\documentclass[12pt]{article}
\pdfoutput=1
\usepackage{amsmath}
\usepackage{amsthm}
\usepackage{amscd}
\usepackage{amsfonts}
\usepackage[psamsfonts]{amssymb}
\usepackage{graphicx}
\usepackage{hyperref}

\numberwithin{equation}{section}

\makeatletter
\renewcommand\section{\@startsection {section}{1}{\z@}%
                                 {-3.5ex \@plus -1ex \@minus -.2ex}
                                   {2.3ex \@plus.2ex}%
                                   {\normalfont\large\bfseries}}
\renewcommand\subsection{\@startsection{subsection}{2}{\z@}%
                                   {-3.25ex\@plus -1ex \@minus -.2ex}%
                                     {1.5ex \@plus .2ex}%
                                     {\normalfont\bfseries}}
\renewcommand\subsubsection{\@startsection{subsubsection}{3}{\z@}%
                                   {-3.25ex\@plus -1ex \@minus -.2ex}%
                                     {1.5ex \@plus .2ex}%
                                     {\normalfont\itshape}}
\makeatother

\begin{document}

\begin{titlepage}
\begin{flushright} 
TAUP-2959/12
\end{flushright}
~
\begin{center}{\Large{\bf Hyperscaling violation : \\~
\vspace{-0.25in}\\
a unified frame for effective holographic theories } }
 ~
 \\ \vspace{0.6in}

{\large \bf Bom Soo Kim}\\ ~\\ \vspace{-0.1in}
\mbox{\normalsize {Raymond and Beverly Sackler School of Physics and Astronomy,}}\\ 
\mbox{\normalsize   Tel Aviv University, 69978 Tel Aviv, Israel} \\ 
\mbox{\normalsize  \it  bskim@post.tau.ac.il}
\end{center}
\bigskip 
\vspace{0.9in}

\begin{abstract}

We investigate systematic classifications of low energy and lower dimensional effective 
holographic theories with Lifshitz and Schr\"odinger scaling symmetries only using metrics 
in terms of hyperscaling violation ($ \theta $) and dynamical ($ z $) exponents. 
Their consistent parameter spaces are constrained by null energy and positive specific 
heat conditions, whose validity is explicitly checked against a previously known result. 
From dimensional reductions of many microscopic string solutions, we observe 
the classifications are tied with the number of scales 
in the original microscopic theories. Conformal theories do not generate 
a nontrivial $ \theta $ for a simple sphere reduction. 
Theories with Lifshitz scaling with one scale are completely fixed by 
$\theta $ and $ z$, and have a universal emblackening factor at finite temperature. 
Dimensional reduction of intersecting M2-M5 requires, we call, spatial anisotropic 
exponents ($ \sharp $), along with $ z=1, \theta = 0 $, because of another scale.
Theories with Schr\"odinger scaling show similar simple classifications at zero temperature, 
while require more care due to an additional parameter 
being a thermodynamic variable at finite temperature.  

\end{abstract}
\end{titlepage}

\tableofcontents

\section{Introduction : hyperscaling violation and effective theories}

The concept of hyperscaling violation \cite{Sachdev:2012dq}, developed in condensed matter 
\cite{Fisher}\cite {SachdevBook}, has attracted much attention recently  
\cite{CGKKM}-\cite{Cadoni:2012ea} in the context of gauge gravity duality 
\cite{Maldacena:1997re}\cite{Aharony:1999ti}. 
For example, hyperscaling violation provides a useful way to realize ``compressible matters'' 
in $ (2+1) $ dimensions \cite{Sachdev:2012dq}\cite{Huijse:2011hp}\cite{Sachdev:2012tj} 
and to pursue the holographic realization of systems with Fermi surfaces 
\cite{Ogawa:2011bz}\cite{Huijse:2011ef}. (See earlier efforts on considering holographic 
Fermi surfaces {\it e.g.} in \cite{Lee:2008xf}-\cite{Faulkner:2009wj}.)
Many physically interesting materials at zero temperature, 
such as high $T_c$ cuprates superconductors, heavy fermion superconductors and organic insulators,  
are compressible, meaning that the ``density'' of their ground states can be dialed by 
some quantum tuning parameters, such as chemical doping, pressure and chemical potentials, respectively. 
As explained nicely in \cite{Sachdev:2012dq}, the ground states of compressible quantum matters can be 
described by a modified Hamiltonian $ \mathcal H' = \mathcal H - \mu \mathcal Q $ 
in the presence of an $ U(1) $ symmetry, where $\mu $ and $\mathcal Q $ are chemical potential 
and charge associated with the $ U(1) $ symmetry, which commutes with the original Hamiltonian 
$ \mathcal H $. The system needs gapless excitations and requires to satisfy  
$\mathrm d \langle \mathrm Q \rangle / \mathrm d \mu \neq 0 $ at $ T=0 $. 
Using scaling arguments, one can get $\mathrm d \langle \mathrm Q \rangle / \mathrm d \mu \sim T^{d-1} $, 
where $ d $ is the number of spatial dimensions. Thus one naturally has compressible matters for $ d=1 $ 
(Luttinger liquids), while it is difficult to realize them for $ d \neq 1 $. 
This difficulty can be overcome in the presence of hyperscaling violation, which effectively gives 
$ d_{\text{eff}} = d -\theta $. Thus the compressible matter at zero temperature in $ 2+1 $ dimensions 
can be achieved with $ \theta =1 $ \cite{Sachdev:2012dq}. 

Hyperscaling violation is realized in holography as a property of the metric, as was first pointed out 
in \cite{Gouteraux:2011ce}, based on the extremal solutions at finite density that were found in \cite{CGKKM}. 
Its implications and importance for strongly coupled condensed matter systems were developed
further in \cite{Huijse:2011ef}.
It has been argued that 
holographic backgrounds with dual Fermi surfaces can be consistently obtained in the systems 
satisfying $\theta = d - 1  $, and thus $ \theta =1 $ for $ d=2 $, 
based on various physical grounds \cite{Ogawa:2011bz}\cite{Huijse:2011ef}.%
\footnote{
This is well fitted with the field theoretic picture outlined in the previous paragraph. 
See \cite{Doiron-Leyraud:2007} for an example of compressible matter with the Fermi surfaces 
present for different doping. }
In particular, this particular system shows a logarithmic violation of the area law 
in the holographic entanglement entropy calculation, 
which is adapted as a working definition of holographic systems with Fermi surfaces 
\cite{Ogawa:2011bz} based on field theoretic results \cite{Eisert:2008ur}. 
The holographic entanglement entropy \cite{HoloEntanglement1}-\cite{Takayanagi:2012kg} 
is further used to identify some novel phases, which interpolates between
the logarithmic violation and extensive volume dependence of entanglement entropy 
\cite{kachru}\cite{kimHyperscaling}. 

\bigskip
{\it Our aim}

While the notion of hyperscaling violation is developed in condensed matter community, 
it can also serve as a useful tool for high energy community as well. 
Here we would like to consider several different classes of microscopic string theory 
solutions and their simplest dimensional reductions along the compact coordinates. 
It becomes clear that hyperscaling violation effects are prevalent upon dimensionally reducing  
the microscopic string solutions. One simple observation is that 
$\theta $ captures the degree of the violation of the conformal / scaling symmetry of 
the original solutions as well as the degree of compactification. 
For example, conformally invariant systems such as D3, M2, M5,  D1-D5 and 
their non-relativistic versions generated by the null Melvin twist show $\theta=0 $ 
upon simple sphere reductions. Motivated by these observations and earlier efforts  
\cite{kachru}\cite{kimHyperscaling}, we investigate the following program. 

\begin{itemize}
\item[A.] The hyperscaling violation exponent $ \theta $ can serve as a unified framework 
for classifying the lower dimensional and low energy effective {\it holographic} theories
with scaling symmetries, especially the Lifshitz and Schr\"odinger scaling.

\item[B.] We only consider gravitational metrics for the classifications with undetermined exponents 
$ (z, \theta) $, without referring to the full solution, such as action and matter contents.%
\footnote{  
One metric can be supported by more than one set of matter sources and action, 
see {\it e.g.} \cite{Balasubramanian:2009rx}. Thus classifying effective theories with full solutions 
would be redundant for investigating the low energy universal properties, which are independent of details, 
not to mention the physical quantities directly related to metric.}

\item[C.] We constrain the parameter space of the consistent solutions using the null energy 
condition, entanglement entropy analysis and the thermodynamic stability, especially the positivity 
of the specific heat, along the line of \cite{Ogawa:2011bz}\cite{kachru}\cite{kimHyperscaling}. 

\end{itemize}
This program for constraining the parameter space of the theories with Lifshitz scaling 
at finite temperature is explicitly compared against the known efforts to do so with 
full solution in the context of the Einstein-Maxwell-Dilaton theories with two parameters 
\cite{CGKKM}\cite{GKMReview}\cite{GKMReview2}. These two results are in very good agreements.
Even though constraining the consistent parameter spaces using the null energy conditions and 
positivity of specific heat constraint is expected to serve as a rough guide, our detailed 
and explicit comparison provides us enough confidences for the validity to do so.%
\footnote{Can we say something about the landscape of the string theory vacua 
and the corresponding classification? While classifying them only with metrics would be 
much simpler, the answer is far from clear, and we have nothing to say about it.  
Our program might be more relevant for classifying general throat geometries with more exponents. 
We are grateful to Piljin Yi for the discussions and comments.} 

\bigskip
{\it Outline and Main points}

We consider the following general metric to classify lower dimensional and low energy 
effective theories, holographically dual to field theories with Lifshitz and Schr\"odinger 
scaling symmetries in $d$ spatial dimensions at zero temperature 
	\begin{align}       \label{BasicMetric}
		\mathrm d s^2 &= r^{-2 +2 \theta /D} \left( - \mathrm b~ r^{-2(z-1)} \mathrm d t^2 
		-2 \mathrm a ~\mathrm d t \mathrm d \xi + \mathrm dr^2
		+ \sum_{i=1}^{\mathrm c} \mathrm d x_i^2 
		+ \sum_{j= \mathrm c+1}^{d} \eta_j (r,t, \vec x) \mathrm d x_j^2  \right) \;, 
	\end{align} 
where $\theta$ and $z$ are the hyperscaling violation and dynamical exponents, respectively.%
\footnote{It is clear that the boundary sits at $r \rightarrow 0 $ from the context. 
One can use different, yet equivalent, coordinate system using $r= \frac{1}{u}$ 
(now the boundary sits at $u \rightarrow \infty $) as 
	\begin{align}        \label{BasicMetricU}
		\mathrm d s^2 &= u^{2 -2 \theta /D} \left( - \mathrm b u^{2(z-1)} \mathrm d t^2 
		-2 \mathrm a \mathrm d t\mathrm d \xi + \frac{\mathrm d u^2}{u^4}
		+ \sum_{i=1}^{\mathrm c}  \mathrm d x_i^2 
		+ \sum_{j=\mathrm c +1}^{d} \eta_j (u, t, \vec x) \mathrm d x_j^2  \right) \;,
	\end{align} 
which provides the identical exponents $ \theta, z$ and $ \sharp $. 
This distinction of the correct coordinate system is important to identify these exponents uniquely. 
Still the distinction can be unclear because 
the metric at hand is valid only for a given range of the radial coordinate.  
Moreover, this distinction can be ambiguous at its fundamental level. One particular example is 
non-commutative super Yang-Mills due to the UV-IR correspondence 
\cite{Hashimoto:1999ut}-\cite{Hubeny:2005qu}. 
Similar features are expected for dipole field theories \cite{Bergman:2001rw}\cite{Bergman:2000cw}
and puff field theories \cite{Ganor:2006ub}\cite{Ganor:2007qh}. 
We are grateful to 
Ori Ganor and Sanny Itzhaki for their comments and discussions on them. 
\label{footnoteUVcoordinate} }  
Note that we split the $ d $ spatial coordinates into two. The first $ \mathrm c $ dimensions 
have been put into a standard form, $\mathrm d x_i^2  $, using coordinate transformations.%
\footnote{
There are some exceptional cases for the Schr\"odinger type solutions to do this successfully. 
Schr\"odinger symmetry require the combination $ -2 \mathrm d t\mathrm d \xi +\mathrm d r^2 $,  
which does not allow further coordinate redefinitions on the radial coordinate. 
One explicit example is considered in detail in \S \ref{sec:SchrConformalCases} and \S \ref{sec:SchrNS5B}. 
\label{footnoteNoSimpleScaling} }  
We call these coordinates as {\it reference} coordinates. 
While $\eta_j (r, t, \vec x)$ can take any general form to provide various different scaling properties 
for each particular spatial direction, here we consider the simplest cases : the parameter 
$\eta $ has only the radial dependence as
	\begin{align}
		\eta_k (r, t, \vec x) \equiv r^{2- 2 \sharp_{k|i }} \;, 
	\end{align} 
where $ \sharp_{k|1}$ denotes ``spatial anisotropic exponent'' of the $k^{th}$ coordinate relative 
to the reference coordinate $ x_1 $. These spatial anisotropic exponents are naturally realized 
in the intersecting Dp-Dq brane systems \S \ref{sec:DpDqSystem} and 
M-brane systems \S \ref{sec:MbrnaeSystem}. We expect that there are more general class of 
systems with the spatial anisotropy. 

The metric is invariant under the following scaling transformations :  
	\begin{align}
		&t \rightarrow \lambda^z t \;, \quad 
		\xi \rightarrow \lambda^{2-z} \xi \;, \quad 
		r \rightarrow \lambda r \;, \quad \\
		&x_i \rightarrow \lambda x_i \;,  \qquad i = 1, \cdots, \mathrm c  \;, \\
		&x_j \rightarrow \lambda^{\sharp_{j|i}} x_j \;,  \qquad j = \mathrm c + 1,  \cdots, d  \;.
	\end{align}  
It is clear that $ z=1 $ and $ \sharp =1 $ represent the usual scaling transformation of the 
relativistic Poincar\'e invariant systems. Thus $ \sharp \neq 1 $ signifies the spatial anisotropy 
and broken rotational symmetries between the reference coordinates and the anisotropic coordinates. 

It turns out that there can be several equivalent sets of the exponent 
for the cases with the nontrivial spatial anisotropic exponents, due to the fact that 
any spatial coordinate is qualified to take the role of the reference coordinate. 
To illustrate this point, let us change the reference coordinate from $x_1 $ to  $x_d $ 
for a simple case $\mathrm a=0, \mathrm b=1 $. Then we get 
	\begin{align}      
		r^{2 \theta /D -2 \sharp_{d|1} } \left( - r^{2 \sharp_{d|1} -2z} \mathrm d t^2 
		+ r^{2 \sharp_{d|1} -2}\mathrm d r^2
		+ r^{2 \sharp_{d|1} -2} \mathrm d x_1^2 + \cdots + \mathrm d x_d^2  \right) \;,		
	\end{align} 
which gives the following new exponents 
	\begin{align}
		\theta' = \frac{\theta}{\sharp_{d|1}} \;, \qquad 
		z' = \frac{z}{\sharp_{d|1}} \;, \qquad 
		\sharp_{1|d}' = \frac{1}{\sharp_{d|1}} \;, \qquad \cdots \;, 
	\end{align}	
after redefining the radial coordinates $ r \rightarrow r^{\frac{1}{\sharp_{d|1}}} $. 	
The physical properties described by this set of exponents are equivalent to those described 
by $\theta, z $	and the original spatial anisotropic exponents $ \sharp_{i|1}, i =2, \cdots, d $. 
If the original anisotropic exponent $ \sharp_{d|1} $ is negative, it is required to use 
the coordinate system described in footnote \ref{footnoteUVcoordinate} and the 
results are the same. The case $ \sharp_{d|1} =0 $ requires a non-polynomial transformation 
to achieve the goal, and eventually one sees that the resulting exponents are not well defined.  
It is plausible that this case can be formulated with the ration $ \theta' / z'  $ fixed, 
while $ \theta' \rightarrow \infty $ and $ z' \rightarrow \infty $ \cite{Hartnoll:2012wm}. 

With these general properties, we consider the physically interesting theories with 
different scaling properties in turn, depending on the parameters 
$D, \mathrm a, \mathrm b, \theta, z $ and $ \eta_i $. 
We also organize our paper accordingly.  
	\begin{itemize}
		\item[I.] Relativistic solutions with full Lorentz symmetry. This case has the parameters 
		$D=d, \mathrm b=1, \mathrm a = 0, z=1$ and $\eta_k =1, k=\mathrm c+1, \cdots, d $, 
		which has the full rotational symmetry on the $d $ spatial directions. 
		This geometry includes the well known AdS spaces with conformal symmetry for $\theta=0$, 
		which is valid for all the energy scales. $\theta=0$ represents the fact that there 
		exists no non-trivial energy scale in the original microscopic string solution. 
		Examples include the dimensional reductions of the D3, M2 and M5 solutions, 
		which are considered in \S \ref{sec:BlackDpLif} \ref{sec:M2Lif} and \ref{sec:M5Lif}, 
		respectively.  
		
		The near extremal black Dp brane solutions are another examples of this type, 
		but now with non-zero hyperscaling violation exponent $\theta$. 
		It is straightforward to see that non-zero hyperscaling violation is directly 
		related to the fact that there is a non-trivial dimensionful parameter in the original 
		Dp brane metrics, which is explicitly demonstrated with dimensional reductions 
		in \S \ref{sec:BlackDpLif}. 
		
		The theories with $ z=1 $ are the special cases of the Lifshitz backgrounds. 
		Thus we consider the relativistic cases with $ z=1 $ as part of the following item [II]. 
		
		\item[II.] Lifshitz type solutions with non-zero $\theta$ and general $z$. The parameters are 
		$D=d, \mathrm b=1, \mathrm a = 0$ and $\eta_k =1, k= \mathrm c +1, \cdots, d$. 
		The physical aspects of the Lifshitz theories with general scaling symmetry $ z $ 
		and the hyperscaling violation exponent $ \theta $ are already considered in \cite{kachru}. 
		It is further considered below in \S \ref{sec:Lifreview} and in \S \ref{sec:LifEntangleToThermal}. 
		Especially we depicted allowed regions (figure \ref{fig:AllowedRegionsLifshitz}) 
		based on the program explained in {\it our aim}. 
		
		This type of gravity solutions
		are constructed for the low energy description of Einstein-Maxwell-Dilaton system \cite{CGKKM}  
		and shown to be the most general IR asymptotic solutions with one gauge and one scalar 
		fields\cite{Gouteraux:2011ce}. It is embedded in higher dimensional theories\cite{Gouteraux:2011ce}.  
		Its finite temperature generalizations are proposed and analyzed in \cite{CGKKM}. 
		This system is explicitly considered in \S \ref{sec:EMD2d} for $ d=2 $ and 
		in \S \ref{sec:EMDgenerald} for arbitrary $ d $, where we re-express the full system 
		in terms of the parameters $ (z, \theta) $ along with the thermodynamic properties. 
		In the figure \ref{fig:AllowedRegionsLifshitzGammaDelta} in \S \ref{sec:ConstraintComparison}, 
		we also present allowed regions of the parameter space of $ (z, \theta) $ from different set 
		of consistency conditions using the full solution \cite{CGKKM}\cite{GKMReview}\cite{GKMReview2}.
		We compare these two different allowed regions and found good agreements.  
		
		We would like to mention that all the worked-out examples of string theory motivated 
		solutions along with the examples \cite{CGKKM} have a particular form of the emblackening factor 
			\begin{align}
				f(r) = 1 - \left( \frac{r}{r_H} \right)^{d+z-\theta} \;, 
			\end{align}
		which reflects the scaling properties and is very interesting. 
		
		\item[III.] Intersecting Dp-Dq and M brane systems with $ \mathrm a =0 $ and $\mathrm b =1 $. 
		The nontrivial spatial anisotropic exponents, $\eta_k \neq 1$, naturally emerge in these systems.%
		\footnote{We are grateful to Yaron Oz and Cobi Sonnenschein for valuable comments on Dp-D(p+4) 
		brane systems and associated anisotropic exponents. $p=0 $ and $p=2 $ cases actually do not generate 
		non-trivial $ \sharp $. $ p=1 $ case can be further compactified down to $ p+2 $ dimensional 
		metric for the compact $ M_4 = T^4 $ or $ K_3 $. Intersecting M2-M5 branes with non-compact directions 
		is presented in \S \ref{sec:MbrnaeSystem}. 
		} 
		In Dp-D(p+4), $ (p=0, 1, 2) $, systems, we get the exponents 
		\begin{align}
			d=p+4 \;, \qquad \theta = \frac{1-p^2}{2-p} \;, \qquad z = 1 \;, \qquad   
			\sharp_{p+4|p} = \frac{1-p}{2(2-p)} \;. 
		\end{align}
		Let us consider D1-D5 system, which is known to have $ AdS_3 \times M_4 $, namely 
		$ AdS_3 $ conformal factor in the dimensionally reduced metric. Correspondingly, 
		we get $ \theta=0 $.  The effective metric is 
		\begin{align}    
			&\mathrm d s_{D1D5}^2 = \sqrt{Q_1 Q_5} u^2 [- f \mathrm d t^2 + \mathrm d x_1^2 
			+ \frac{ 1}{u^4}  \frac{\mathrm d u^2}{f} 
			+ \frac{Q_1}{u^2} \mathrm ds_{M_4}^2 ] \;, 		
		\end{align}
		which shows the non-trivial spatial anisotropy, stemming from the relative scale 
		between these two spaces $ AdS_3 $ and $ M_4 $. 		
		This is described in \S \ref{sec:DpDqSystem} in detail. 
		Intersecting and interpolating M-branes are also considered in \S \ref{sec:MbrnaeSystem} 
		to show similar features. 
		Recently, the hyperscaling violation exponent in the intersecting brane systems are considered 
		in \cite{Dey:2012tg}\cite{Dey:2012rs}\cite{Dey:2012hf}\cite{Dey:2012mc}. 

		\item[IV.]  Schr\"odinger metrics with general $\theta$ and $z$. The parameters are 
		$D=d+1, \mathrm a = 1$ and $\eta_k =1, k=\mathrm c + 1, \cdots, d $. 
		The physical aspects with hyperscaling violation at zero temperature are discussed 
		in \cite{kimHyperscaling}. 	Surprisingly, we find a candidate Sch\"odinger background
		that possesses dual Fermi surfaces according to \cite{Ogawa:2011bz}\cite{Huijse:2011ef} 
		from a simple dimensional reduction of non-relativistic NS5A brane in \S \ref{sec:SchrNS5A}. 
		The background reveals the relation $ \theta = d+1-z $, which is worked out in 
		\cite{kimHyperscaling} based on ``codimension 2'' minimal surface prescription. 
		We further investigate other physical properties of this non-relativistic NS5A brane 
		in \S \ref{sec:SchrNS5A}. Lifshitz background with dual Fermi surfaces has been also claimed 
		in \cite{Narayan:2012hk}. (See also \cite{Perlmutter:2012he}\cite{Sadeghi:2012vv} for the 
		discussions of the hyperscaling violation on Schr\"odinger space.) 
		
		Finite temperature generalizations from effective low energy and lower dimensional point of view 
		turn out to be more difficult than expected, due to the non-trivial 
		asymptotic form and an additional dimensionful parameter $ b$ generated by the 
		null Melvin twist. This parameter $b $ serves as an independent thermodynamic variable, 
		along with $r_H $, and thus can not be ignored. Constructing general low energy 
		effective metrics based on symmetries is a highly non-trivial task. 
		Thus we consider non-relativistic Dp branes generated by the null Melvin twist  
		investigate their physical properties at finite temperature and 
		perform the dimensional reductions of them in \S \ref{sec:DpDimensionalReduction}. 
		Along the way, we identify the exponents $ \theta$ and $z $ 
		in \S \ref{sec:DynamicalExponentSchr}, which turns out to be not straightforward. 
		In \S \ref{sec:StringSolEEntropy}, entanglement entropy at finite temperature is considered 
		to show that it reproduce the thermal entropy at high temperature limit and 
		zero temperature entanglement entropy at low temperature limit. 
		
		Finally, we consider a class of effective Schr\"odigner backgrounds at finite temperature 
		in \S \ref{sec:SchrMetricReduction}. We present a plot for the allowed regions in 
		the figure \ref{fig:AllowedRegionsSpecificHeat} for $ z>0 $ by using the null energy,  
		entanglement entropy conditions and specific heat 
		constraint for fixed chemical potential associated with the null coordinate $ \xi $. 
		
		There exist two different geometric realizations with Schr\"odinger symmetry in holographic 
		approach, Schr\"odinger backgrounds with $\mathrm b =1$, described in \S \ref{sec:Schroedinger},
		and AdS in light-cone frame (ALCF) with $\mathrm b =0$. 
		\S \ref{sec:ALCF} contains the parallel investigations of ALCF compared to 
		the Schr\"odinger background \S \ref{sec:Schroedinger}. In particular, the dynamical 
		exponent $ z $ remains unfixed even after the dimensional reduction, which is one of the 
		reasons we separate our discussions for these two geometric realizations of Schr\"odinger holography. 
		
		Schr\"odinger background and ALCF have the same thermodynamic properties. 
		By rewriting their metrics in a more organized form as 
		equations (\ref{dimenReduSchrMetricADM}) and (\ref{finiteTReducedMetric}), 
		we find an important technical detail responsible for the fact. 
		Basically, the newly introduced $ K $ factor from the null Melvin twist in 
		(\ref{dimenReduSchrMetricADM})	does not enter the thermodynamic analysis. 

	\end{itemize}

The basic properties and the dimensional reductions of several different systems are investigated 
in detail in appendix \S \ref{sec:LifshitzReduction} and \S \ref{sec:SchrReduction}. 
Even though these examples are far from the exhaustive lists of the low energy 
and lower dimensional holographic theories, we hope 
they serve to show the hyperscaling violation exponent $ \theta $ provides 
a unified framework for the low energy and lower dimensional holographic theories.

\section{Theories with Lifshitz scaling}     \label{sec:Lifshitz}

Theories with the Lifshitz scaling symmetries \cite{Kachru:2008yh}\cite{Taylor:2008tg} 
with dynamical exponent $z$ and hyperscaling violation exponent $ \theta$ have been 
considered in \cite{kachru} from the metric point of view without referring to matter contents. 
The full extremal solutions were constructed and analyzed in \cite{CGKKM}. 
Their scaling properties, including their violation of hyperscaling was pointed out later in 
\cite{Gouteraux:2011ce}\cite{Huijse:2011ef}. 
The explicit solution is developed in the context of the holographic bulk solutions with 
the Einstein-Maxwell action with a Dilaton (EMD theories) \cite{Charmousis:2009xr}-\cite{Gouteraux:2011qh}. 

In this section, we first review some salient features 
of the known results in \S \ref{sec:Lifreview}, along with the analysis of the entanglement entropy 
at finite temperature to thermal entropy in \S \ref{sec:LifEntangleToThermal}. 
From our program to classify the low energy and lower dimensional effective holographic theories, 
we would like to check whether the consistency conditions we impose are reasonable guides. 
For this purpose, we make plots of the consistent regions of the parameter spaces $ (z, \theta) $ 
for 2 and 3 spatial dimensions using the null energy condition and positive specific heat constraint. 
We call the allowed regions (I) from these constraints. 

Then we consider the explicit solution \cite{CGKKM}\cite{GKMReview}\cite{GKMReview2}\cite{IKNT} with two parameters 
$ (\gamma, \delta) $ and map it in terms of the parameters $ (z, \theta) $ for $d=2 $ in \S \ref{sec:EMD2d} 
and for general dimensions in \S \ref{sec:EMDgenerald}. 
For the future references, we list also the thermodynamic properties in term of the parameters $ (z, \theta) $ 
in \S \ref{sec:EMD2Dthermo}. The allowed regions of the parameter spaces $ (\gamma, \delta) $ 
at zero temperature are explicitly analyzed by various conditions, such as Gubser's criteria \cite{Gubser:2000nd} 
and well defined fluctuation problems in \cite{CGKKM}\cite{GKMReview}\cite{GKMReview2}\cite{IKNT}. 
In \S \ref{sec:ConstraintComparison}, we plot the allowed regions (II) in terms of $ (z, \theta) $ 
including the positive specific heat constraint at finite temperature. 
These two allowed regions (I) and (II), upon including the positive specific heat constraint, 
are identical, which provides a positive sign for the program to constrain the allowed regions from 
the low energy and lower dimensional point of view.

\subsection{Review and constraint plots}  \label{sec:Lifreview}

The metric we are interested in is described by 
	\begin{align}        \label{HyperscalingLifshitzMetric}
		\mathrm d s^2_{d+2} &= r^{-2(d-\theta)/d} \left( - r^{-2(z-1)} f(r) \mathrm d t^2 
		+ \frac{ \mathrm d r^2}{f(r)} 
		+ \sum_{i=1}^d  \mathrm d x_i^2 \right) \;, \nonumber \\
		f(r) &= 1 - \left(\frac{r}{r_H} \right)^{d+z-\theta} \;.
	\end{align} 
Physical properties of this metric emphasizing the role of the hyperscaling violation exponent 
$\theta $ are analyzed in detail in \cite{kachru}. 

\begin{figure}[!ht]
\begin{center}
	 \includegraphics[width=0.3\textwidth]{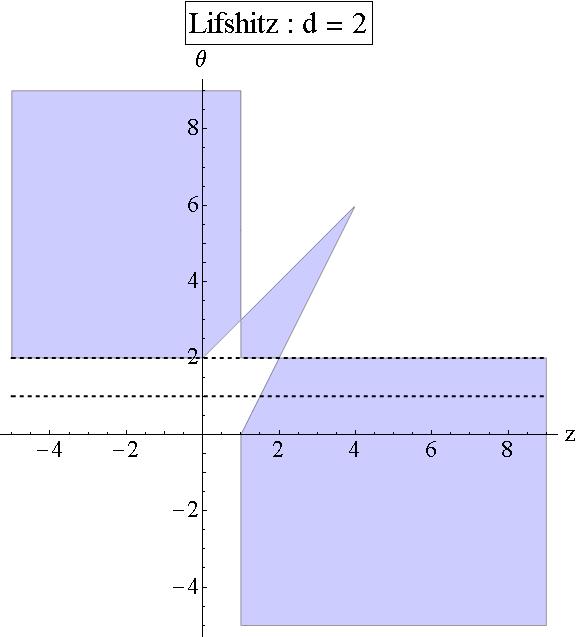}  \quad
	 \includegraphics[width=0.3\textwidth]{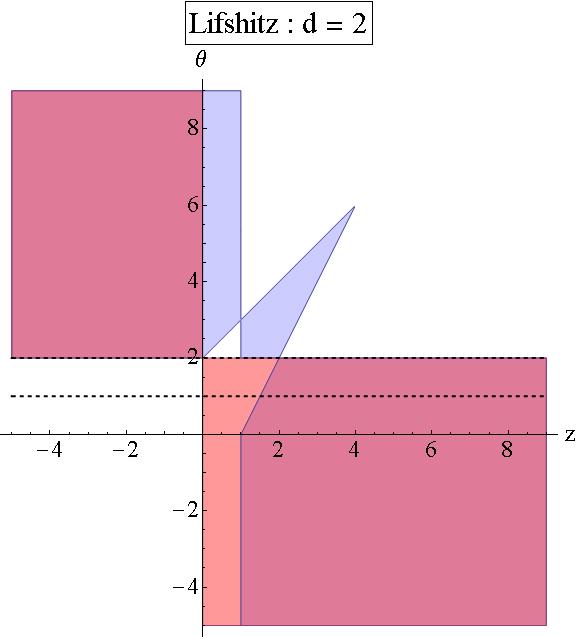}  \quad
	 \includegraphics[width=0.3\textwidth]{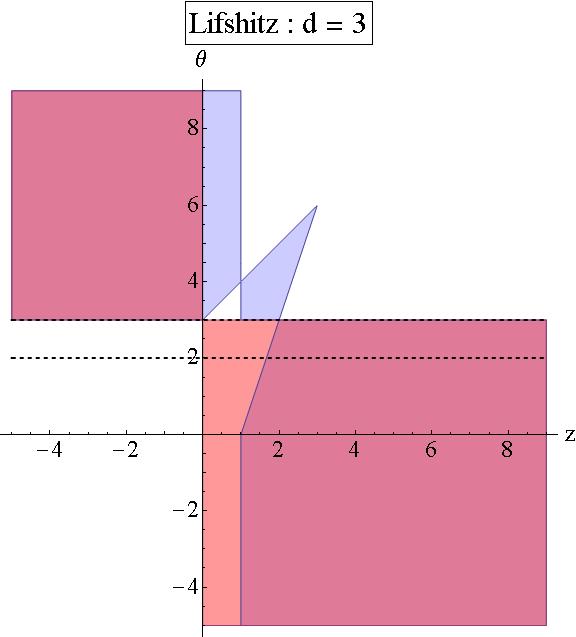}	
	 \caption{\footnotesize The left plot shows the allowed regions for the Lifshitz geometry based on the 
	 null energy conditions, along with the region between two dashed lines indicating 
	 some novel phases from the entanglement entropy analysis \cite{kachru}. The middle plot shows 
	 the allowed (dark purple) regions after taking into account of the specific heat condition 
	 (\ref{SpecificHealConditionLifshitz}). The right plot ($d=3$) shows the regions allowed by 
	 the null energy condition (blue) and the specific heat (red), and thus the dark purple regions, 
	 allowed regions (I), are allowed by both conditions. 
	 These conditions indicate that physically sensible theories are 
	 required to $\theta \geq d$ for $z\leq 0$, while they are required to have 
	 $\theta \leq d$ for positive $z$. It is further suggested that the region $\theta \geq d$ and $z \leq 0$ 
	 are unstable from the entanglement entropy analysis \cite{kachru}. Thus only the bottom right region is 
	 allowed and covers $z \geq 1$ and also $\theta \leq d$, except small top left corner of that region.
	 }
	 \label{fig:AllowedRegionsLifshitz}
\end{center}
\end{figure}
	
Without referring to specific matter contents to support the background, the authors of \cite{kachru} 
restrict the consistent parameter spaces with null energy conditions at zero temperature following 
\cite{Ogawa:2011bz}\cite{Huijse:2011ef}. See also the usage of null energy condition \cite{Hoyos:2010at}. 
The condition is given by 
	\begin{align}
		(d-\theta)(d (z-1) - \theta) \geq 0 \;, \qquad 
		(z-1)(d+z-\theta) \geq 0 \;, 
	\end{align}
which is depicted in the left plot of figure \ref{fig:AllowedRegionsLifshitz} for $d=2$ case.  
The allowed regions are further constrained by some thermodynamic stability, mainly the positivity of 
the specific heat at finite temperature, which is 
	\begin{align}   \label{SpecificHealConditionLifshitz}
		\frac{d -\theta}{z} \geq 0 \;. 
	\end{align}
The result is summarized in the middle plot of figure \ref{fig:AllowedRegionsLifshitz}. 
These constraints are not expected to be completely accurate, but it seems to serve as a good guide 
for all the dimensionally reduced microscopic string solutions considered 
in the appendix \S \ref{sec:LifshitzReduction}.   

Note the specific form of the emblackening factor $f(r)$. The $r$ dependent power is {\it a prior} 
not fixed by any symmetry or dimensional analysis, especially from the view of the resulting effective 
theories which are dimensionally reduced from string theory solutions. Yet, this particular 
combination $f(r) = 1- (r/r_H)^{d+z-\theta} $ is valid for the dimensional reduction 
of the black Dp brane \cite{kachru}, 
black M2 and M5 branes, as well as a large class of exact black hole solutions \cite{CGKKM}\cite{IKNT}. 
Some of them are further studied in the appendix \S \ref{sec:LifshitzReduction}. 
Interestingly, the theories with Schr\"odinger isometries, upon dimensionally reduced along 
the null direction $\xi $, also exhibit this property as explicitly shown in \S \ref{sec:SchrLifReduction} 
and \S \ref{sec:ALCFLifshitzReduction}.   
This combination $d+z-\theta $ is tied with the physical property of the thermal entropy $ S_T $ 
in terms of the temperature $T$ 
	\begin{align}    \label{ThermalEntropyLif}
		S_T \sim T^{(d-\theta)/z} \;, 
	\end{align} 
which can be easily verified from $ T \sim r_H^{-z}$ and $ S_T \sim r_H^{\theta - d}$. 
It will be interesting to check whether these properties are true for all the lower dimensional 
Lifshitz type solutions, which have their microscopic string solutions.

\subsection{Entanglement entropy to thermal entropy}    \label{sec:LifEntangleToThermal}

Entanglement entropy has been a useful tool for classifying the phases of the matter \cite{Eisert:2008ur}, and 
it has been actively discussed in holographic context, see {\it e.g.} 
\cite{HoloEntanglement1}-\cite{Takayanagi:2012kg}. 
 The cross-over from the entanglement entropy to thermal entropy in Lifshitz space is already 
studied in \cite{kachru} motivated by \cite{Swingle:2011mk}. 
Here we review the result for the comparison to the Schr\"odinger geometry 
we study in detail in the following sections. 

The entanglement entropy of the metric (\ref{HyperscalingLifshitzMetric}) with finite temperature 
can be computed for a strip 
	\begin{align}
		-l \le x_1 \le l\;,\quad 0 \le x_i \le L\;,\;\quad i = 2,\cdots \;, d
	\end{align}   
in the limit $l \ll L$. 
The strip is located at $r=\epsilon$, and the profile of the surface in the bulk is given by $r=r(x_1)$. 
The minimal surface has a turning point at $r=r_t$. Thus, to get the entanglement entropy, 
we can evaluate the following expressions 
	\begin{align}    \label{entangleAreaHyprLif}
		l &= \int_0^{r_t}\mathrm d r~\frac{1}{\sqrt{f}} \frac{(r/r_t)^{\alpha_2} }
		{\sqrt{1- (r/r_t)^{2\alpha_2}}} \;, \qquad 
		{\mathcal  A} = L^{d-1} \int_\epsilon^{r_t}\mathrm d r~\frac{1}{\sqrt{f}} \frac{ r^{-\alpha_2} }
		{\sqrt{1-(r/r_t)^{2\alpha_2}}} \;, 
	\end{align}    
where $\alpha_2 = d-\theta $. 	

Assuming $\alpha_2 >0$, we can rewrite the integrals as
	\begin{align}   
		l &= r_t \int_0^{1} \mathrm d \omega ~\frac{\omega^{\alpha_2}}
		{\sqrt{1 - (\gamma \omega)^{\alpha_2 + z}} \sqrt{1- \omega^{2\alpha_2}}} \;,
	\end{align}   
and, following \cite{Swingle:2011mk}
	\begin{align}   
		{\mathcal  A} 
		&= L^{d-1} r_t^{1-\alpha_2} \int_{\epsilon/r_t}^1 \mathrm d \omega~ 
		\left( \left[ \frac{ \omega^{-\alpha_2}}
		{\sqrt{1 - (\gamma \omega)^{\alpha_2 + z}} \sqrt{1-\omega^{2\alpha_2}}} - \frac{1}{\omega^{\alpha_2}} \right] 
		+  \frac{1}{\omega^{\alpha_2}}  \right)	\;, 
	\end{align}   
where the square bracket part is finite for $\alpha_2 >0$ and 
	\begin{align}
		\omega = \frac{r}{r_t} \;, \qquad \gamma = \frac{r_t}{r_H} <  1\;.
	\end{align}
First, we compute the divergent part 
	\begin{align}  
		{\mathcal  A}_{div} 
		&= L^{d-1}  r_t^{1-\alpha_2} \int_{\epsilon/r_t}^1 \mathrm d \omega ~ \frac{1}{\omega^{\alpha_2}} 
		= L^{d-1} \frac{1}{\alpha_2 -1} \frac{1}{\epsilon^{\alpha_2 -1} }	\;, 
	\end{align}   
which is the same as the divergent contribution of the zero temperature background.  	

Now let us compute the finite part. At low temperature, we can evaluate the integral  
	\begin{align}   
		l &= r_t \int_0^{1} \mathrm d \omega ~\frac{\omega^{\alpha_2}}
		{\sqrt{1 - (\gamma \omega)^{\alpha_2 + z}} \sqrt{1- \omega^{2\alpha_2}}}
		\approx \sqrt{\pi} r_t \frac{\Gamma \left(\frac{1+ \alpha_2}{2 \alpha_2} \right)}
		{\Gamma \left(\frac{1}{2 \alpha_2} \right)} \;, 
	\end{align}   
and 
	\begin{align}  
		{\mathcal  A}_{fin} 
		&\approx  L^{d-1} r_t^{1-\alpha_2} \int^1 \mathrm d \omega ~ 
		\frac{ \omega^{-\alpha_2}}{\sqrt{1-\omega^{2\alpha_2}}} 
		\left( 1 + \frac{1}{2} (\gamma \omega)^{\alpha_2 + z} + \cdots  \right)   \;, \nonumber \\
		&= L^{d-1} l^{1-\alpha_2}~ c_\theta  \left(-\frac{1}{\alpha_2 -1}  + 
		\frac{\gamma^{\alpha_2 + z}}{4 \alpha_2 }  \frac{\Gamma \left(\frac{z+1}{2 \alpha_2} \right)}
		{\Gamma \left(\frac{\alpha_2 + z+1}{2 \alpha_2} \right)} 
		\frac{\Gamma \left(\frac{1}{2 \alpha_2} \right)}{\Gamma \left(\frac{1+ \alpha_2}{2 \alpha_2} \right)}
		+ \cdots \right) \;, 				
	\end{align}   	
which serves as an interpolating function from small temperature to high temperature. 
We evaluate it momentarily. At low temperature, the entanglement entropy for a strip 
in the general metric (\ref{HyperscalingLifshitzMetric}) is 
	\begin{align}    
		\mathcal  S 
		&= \frac{(R M_{Pl})^{d} }{4 (\alpha_2-1)} 
		\left(  \left(\frac{\epsilon}{R_\theta}\right)^{\theta} \frac{L^{d-1} }{\epsilon^{d-1}} 
		-  c_{\theta} ~\left(\frac{l}{R_\theta}\right)^{\theta} \frac{L^{d-1} }{l^{d-1}} 
		\left[ 1 - \tilde c_\theta \left( l T^{1/z} \right)^{d+z-\theta} + \cdots  \right] \right) \;,
	\end{align}   
where $R_\theta$ is a scale in which the hyperscaling violation becomes important and 
	\begin{align}
		c_{\theta} = \left( \frac{\sqrt{\pi} \Gamma \left( \frac{1+\alpha_2}{2\alpha_2} \right)}
		{ \Gamma \left( \frac{1}{2\alpha_2} \right) } \right)^{\alpha_2}  \;, \qquad 
		\tilde c_{\theta} = \frac{(\alpha_2 -1) \sqrt{\pi}  }
		{4  \alpha_2 ~c_\theta^{\frac{d+z-\theta +1}{\alpha_2}}} \frac{\Gamma \left(\frac{z+1}{2 \alpha_2} \right)}
		{\Gamma \left(\frac{\alpha_2 +z +1}{2 \alpha_2} \right)} 
		\left( \frac{4\pi}{|d+z-\theta|} \right)^{\frac{d+z-\theta}{z}}\;.
	\end{align}
Thus we check that the entanglement entropy reduces to that of the zero temperature in the low temperature limit. 
	
In the high temperature limit, 
	\begin{align}   
		l &= r_t \int^{1} \mathrm d \omega ~\frac{\omega^{\alpha_2}}
		{\sqrt{1 - (\gamma \omega)^{\alpha_2 + z}} \sqrt{1- \omega^{2\alpha_2}}}  
		= r_t I_+ \left( \gamma \right) \;,
	\end{align}   
and 
	\begin{align}  
		{\mathcal  A} 
		&= L^{d-1}  r_t^{1-\alpha_2} \int^1 \mathrm d \omega~ 
		\frac{ \omega^{-\alpha_2}}{\sqrt{1 - (\gamma \omega)^{\alpha_2 + z}} \sqrt{1-\omega^{2\alpha_2}}} 
		= L^{d-1} r_t^{1-\alpha_2}  I_- \left( \gamma \right)	\;. 
	\end{align}   
When $\gamma =\frac{r_t}{r_H}  \rightarrow 1$, the integrals $I_+ \left( \gamma \right)$ and 	$I_- \left( \gamma \right)$
are dominated by the contribution $\omega \approx 1$, and thus 
$I_+ \left( \gamma \right) \approx I_- \left( \gamma \right) \approx l/r_H$ \cite{kachru}. 
Thus 
	\begin{align}    
		\mathcal  S_{fin} 
		&\propto L^{d-1} l ~T^{(d - \theta)/z} \;,
	\end{align}    	
which agrees with the thermal entropy, given in (\ref{ThermalEntropyLif}).

\subsection{An explicit solution : EMD theories}   \label{sec:EMD2d}

In this section we would like to study an explicit solution with the properties described 
in the previous section. The explicit solution is given in \cite{CGKKM} with two free parameters 
$ (\gamma, \delta) $ in the context of EMD theories \cite{Charmousis:2009xr}-\cite{Gouteraux:2011qh}. 
One of the main motivation of this section 
is to map the solution in terms of new parameters $(z, \theta) $. 
We would like to examine the structures of the action and other fields associated 
with the metric with the hyperscaling violation with emphasis on 2 spatial dimensions in terms of 
those two parameters. In particular, we check whether the program to restrict the 
parameter space of the metric (\ref{HyperscalingLifshitzMetric}) using the null energy condition and 
positive specific heat constraint (allowed region (I)) is reasonable or not, 
by comparing to the previous efforts to 
do so using other means available in the literature \cite{CGKKM}, which include the information 
of the full solution, especially of the matter contents (allowed region (II)).  
We also generalize the solutions for the general dimensions in the following section 
\S \ref{sec:EMDgenerald}. 

The full set of near extremal solution for $d=2$ is obtained with two parameters $(\gamma, \delta)$ 
 \cite{CGKKM} 
  	\begin{align}    \label{EMDsolution}
  		&S =\int \mathrm d ^{3+1}x~\sqrt{-g}\left[R- \frac{e^{\gamma\phi}}{4}F_{\mu\nu}F^{\mu\nu}
  		-\frac{1}{2}(\partial\phi)^2-2\Lambda e^{-\delta\phi}\right]  \;, \nonumber  \\ 
		&d s^2 = -r(r-2m) r^{-4\frac{\gamma(\gamma-\delta)}{wu}} \mathrm d t^2 
		+\frac{ e^{\delta\phi} \mathrm d r^2}{ -w\Lambda r(r-2m)} 
		+ r^{2\frac{(\gamma-\delta)^2}{wu}}  \left( dx^2 + dy^2 \right) \;,   \\
		&e^{\phi} = e^{\phi_0}r^{-4(\gamma-\delta)/(wu)}\;, \quad 
		A_t= 2\sqrt{-v/(wu)}e^{-\frac{\gamma}{2} \phi_0}(r-2m)  \;,  \nonumber \\
		&wu=3\gamma^2-\delta^2 - 2\gamma \delta + 4 \;, 
		\quad u = \gamma^2- \gamma \delta +2 \;, \quad v=\delta^2 - \gamma \delta -2 \;.  \nonumber
	\end{align}
This metric was constructed to get a general scaling solution in IR, and physical properties 
such as energy,  entropy and conductivities show power law behaviors, characteristic features of 
scaling invariant theories. An important difference is that the hyperscaling property is violated.    
Lifshitz solutions are special cases for $\gamma = -\sqrt{4/(z-1)}$ and $\delta=0$. 
This solution provides the most general IR asymptotics at finite density with a single gauge field $A$ 
and a Dilaton field $\phi$, and is embedded in higher dimensional AdS or Lifshitz spacetimes 
\cite{Gouteraux:2011ce}.

Using the coordinate transform 
	\begin{align}
		&r \rightarrow \tilde b r^{\tilde a} \;, \quad  2 m =\tilde b r_H^{\tilde a} \;, \quad 
		\tilde a = - \frac{wu}{\gamma^2 -\delta^2} \;, \quad 
		\tilde b = \left( \frac{-w\Lambda}{\tilde a^2 e^{\delta \phi_0}} \right)^{\tilde a/2}  \;, 
	\end{align}	
the metric can be recast into 
	\begin{align} 
		\mathrm d s^2 
		&= r^{-2 + 2  \theta/d} \left( -  r^{-2(z-1)} f(r) \mathrm d t^2 + \sum_{i=1}^{d}  \mathrm d x_i ^2 
		+ \frac{ \mathrm d r^2}{f(r)}  \right) \;, \\
		f &= 1 - \frac{2m/\tilde b}{r^{\tilde a}}  
		= 1 -\left( \frac{r}{r_H} \right)^{\frac{wu}{\gamma^2 -\delta^2}} 
		= 1 -\left( \frac{r}{r_H} \right)^{d+z-\theta} \;, 
	\end{align}
where we are able to put $\tilde b=1$ with a suitable choice of $ \Lambda $ or 
$\phi_0 $ as $e^{\delta \phi_0} = \frac{-w\Lambda}{\tilde a^2} $, 
and the coefficient of $\mathrm d t^2 $ term becomes unity. 		
The parameters $z$ and $\theta$ are the dynamical and hyperscaling violation exponents, respectively. 
Explicitly, they are identified as follows in terms of $\gamma, \delta$ for $d=2$   	
	\begin{align}    \label{zThetaGD}
		\theta = \frac{4 \delta }{\gamma + \delta} \;, \qquad 
		z = 1 + \frac{2\delta}{\gamma + \delta } +\frac{4}{\gamma^2 -\delta^2} \;. 
	\end{align}
From these we check the relation $d+z-\theta = -\tilde a = \frac{wu}{\gamma^2 -\delta^2}$. 	
Thus we can translate all the results of \cite{CGKKM} in terms of $z$ and $\theta$. Again, we check that 
$\delta=0$ is a Lifshitz solution. In turn, we have the relations 
	\begin{align}    \label{GDzTheta}
		\gamma = \pm \frac{4 - \theta }{\sqrt{-4+4 z-2 z \theta +\theta ^2}} \;, \quad 
		\delta = \pm \frac{\theta }{\sqrt{-4+4 z-2 z \theta +\theta ^2}} \;, \quad 
		\frac{\gamma}{\delta} = \frac{4-\theta}{\theta} \;.
	\end{align}
It is desirable to examine the metric and other fields more closely. For this purpose, 
we completely rewrite this solution in terms of the parameter $z$ and $\theta $. 
  	\begin{align}     \label{EMDsolutionthetaz}
  		&S =\int \mathrm d ^{4}x~\sqrt{-g}\left[R- \frac{Z}{4}F_{\mu\nu}F^{\mu\nu}
  		-\frac{1}{2}(\partial\phi)^2 + V \right]  \;,   \\ 
		&d s^2 = r^{-2 + \theta} \left( -  r^{-2(z-1)} f(r) \mathrm d t^2 + \sum_{i=1}^{2}  \mathrm d x_i ^2 
		+ \frac{ \mathrm d r^2}{f(r)}  \right) \;,  \\
		&e^{\phi} = r^s \;, \qquad s =\pm \sqrt{-4+4 z-2 z \theta +\theta ^2} \;, \\ 
		&V = V_0 e^{-\frac{\theta}{s} \phi} \;, \qquad V_0 = (2+z-\theta)(1+z-\theta)  \;, \\
		&Z = \frac{1}{q^2} e^{\frac{4 - \theta}{s} \phi}  \;, \qquad
		A_t= q \sqrt{\frac{2z-2}{z+2-\theta}} r^{-2-z+\theta} f(r)\;, \\
		&f(r) = 1 -\left( \frac{r}{r_H} \right)^{2+z-\theta}  \;. 
	\end{align}
Note that we set $\tilde b =1 $ using $\Lambda$.%
\footnote{We are grateful to Sang-Jin Sin and Yunseok Seo for the fruitful discussions and comments on 
the metric, gauge field and the associated charge in terms of $(\theta, z) $. 
} 
The remaining constant $ \phi_0 $ serves as an integration constant. 
The two solutions are corresponding to the sign of the relation (\ref{GDzTheta}). 
From the Maxwell equation $ \partial_\mu \left(\sqrt{-g} Z F^{\mu\nu} \right) =0 $, 
we introduce an integration constant $ Q =\sqrt{-g} Z F^{\mu\nu}$, which is identified 
as a charge density. In terms of $q $, it is expressed as $ Q=- \sqrt{2(z-1)(2+z-\theta)}/q $. 
We check explicitly that this satisfy the Einstein equation, 
and the scalar equation is automatically satisfied. 
Under the scaling transformation 
	\begin{align}
		t \rightarrow \lambda^z t \;, \quad x_i \rightarrow \lambda x_i \;, \quad 
		r \rightarrow \lambda r \;,
	\end{align}
the dilaton, the vector potential and metric transform as 
	\begin{align}
		&\phi \rightarrow \pm \sqrt{4 (z-1) +\theta (\theta - 2 z)} \log (\lambda) + \phi \;, \\
		&A \rightarrow \lambda^{\theta -2} A \;, \\
		&\mathrm d s \rightarrow \lambda^{\theta/d}\mathrm d s \;,
	\end{align}
while the action remains invariant under the scaling transformation. 
These properties for the dilaton and vector potential were noticed in \cite{Hartnoll:2012wm}. 

It is interesting to consider various special cases. 
For $\theta=4 $, the gauge field and the dilaton become decoupled. 
If we put $\theta=0 $, the potential $V $ 
becomes a constant. The solution becomes nothing but 
the Lifshitz type supported by a gauge field and a scalar.  
  	\begin{align}     \label{EMDsolutionthetazLifshitz}
  		&S =\int \mathrm d ^{4}x~\sqrt{-g}\left[R- \frac{Z}{4}F_{\mu\nu}F^{\mu\nu}
  		-\frac{1}{2}(\partial\phi)^2 + V_0 \right]  \;,   \\ 
		&d s^2 = r^{-2} \left( -  r^{-2(z-1)} f(r) \mathrm d t^2 + \sum_{i=1}^{2}  \mathrm d x_i ^2 
		+ \frac{ \mathrm d r^2}{f(r)}  \right) \;,  \\
		&V_0  = (2+z)(1+z)  \;, \quad 
		e^{\phi} = r^s \;, \quad s={\pm \sqrt{4 z-4}}\;, \\ 
		&Z = \frac{1}{q^2} e^{4\phi /s}  \;, \quad  A_t= q \sqrt{\frac{2z-2}{z+2}}  r^{-2-z} f(r)\;, \\ 
		&f(r) = 1 -\left( \frac{r}{r_H} \right)^{2+z} \;. 
	\end{align}
In this form, the scalar potential becomes a constant, which can be identified as a cosmological 
constant term for $3+1 $ dimensional gravity system.  
If we further restrict to $z=1$, the metric becomes  
  	\begin{align}     \label{EMDsolutionthetazReducedToAdS}
  		&S =\int \mathrm d ^{4}x~\sqrt{-g}\left[R+ 6 \right]  \;,  \\ 
		&d s^2 = r^{-2} \left( - f(r) \mathrm d t^2 + \sum_{i=1}^{2}  \mathrm d x_i ^2 
		+ \frac{ \mathrm d r^2}{f(r)}  \right) \;,  
	\end{align}
which is the usual AdS solution. 
A particular case $z=1, \theta \neq 0 $ is supported by the dilaton only because the gauge field vanishes.

\bigskip
\bigskip
{\it Alternative}

Using the coordinate transform 
	\begin{align}
		r \rightarrow \tilde b u^{\tilde a} \;, \qquad \tilde a = \frac{wu}{\gamma^2 -\delta^2} \;, \qquad 
		\tilde b = \left( \frac{-w\Lambda}{\tilde a^2 e^{\delta \phi_0}} \right)^{-\tilde a/2}  =1 \;,
	\end{align}
the metric can be recast into 	
	\begin{align} 
		\mathrm d s^2 &= u^{2 -2\theta/d} \left(-u^{2(z-1)} f(u) \mathrm d t^2 
		+ \sum_{i=1}^{d}  \mathrm d x_i ^2 + \frac{du^2}{u^4 f(u)}  \right) \;, \\
		f &= 1 - \frac{2m/\tilde b}{u^{\tilde a}}  
		= 1 -\left( \frac{u_H}{u} \right)^{\frac{wu}{\gamma^2 -\delta^2}} 
		=  1 -\left( \frac{u_H}{u} \right)^{d+z-\theta} \;,  
	\end{align}
where $z$ and $\theta$ are the dynamical and hyperscaling violation exponents, respectively. 
Explicitly, they are identified as follows in terms of $\gamma, \delta$ for $d=2$   	
	\begin{align}
		\theta = \frac{4 \delta }{\gamma + \delta} \;, \qquad 
		z = 1 + \frac{2\delta}{\gamma + \delta } + \frac{4}{\gamma^2 -\delta^2} \;. 
	\end{align}
Thus we explicitly check the alternative coordinate system does not change the dynamical $z $ and 
hyperscaling violation exponent $ \theta $.

\subsubsection{Thermodynamics}    \label{sec:EMD2Dthermo}

As a complete solution is given in (\ref{EMDsolutionthetaz}), we would like to also list thermodynamic 
quantities. These are computed in \cite{CGKKM} using the Euclidean action with appropriate boundary terms. 
We can view this section as a two parameter generalization in terms of $(z, \theta) $ of the AdS$_4$ 
black hole for $(z=1, \theta=0) $. 

The temperature and entropy are given by 
	\begin{align}
		 &T=\frac1{4\pi}\sqrt{-w\Lambda}e^{-\frac\delta2\phi_0}(2m)^{1-2\frac{(\gamma-\delta)^2}{wu}} 
		 = \frac{(2+z-\theta)}{4\pi} r_H^{-z} \;, \\
		 &S= \frac{\Omega_2}{4} (2m)^{2\frac{(\gamma-\delta)^2}{wu}}
		 = \frac{\Omega_2}{4} r_H^{-2+\theta} \;, \qquad \sim T^{\frac{2-\theta}{z}} \;.
	\end{align}
Note the temperature vanishes in the $r_H \rightarrow \infty $ limit if and only if the dynamical exponent 
$ z$ is positive, and the entropy vanishes as $r_H \rightarrow \infty $ for $ \theta <2 $. 
These two observations support further restriction of the allowed regions (I) to the bottom right one 
in the figure \ref{fig:AllowedRegionsLifshitz}, together with the result of the entanglement entropy analysis. 
We observe that $\theta=2 $, for $ d=2 $, is special. The entropy stays finite 
for this case, which corresponds to the case $\gamma=\delta$.  
And the geometry turns out to be a direct product AdS$_2\times S^2$ in the extremal limit $ m=0$. 
This is clear from the geometry given in (\ref{EMDsolution}). 

The energy is 
	\begin{align}
		E = \frac{\Omega_2}{4\pi}\sqrt{\frac{-\Lambda}{wu^2}}(\gamma-\delta)^2m 
		= \frac{\Omega_2}{4\pi} \frac{2-\theta}{4} r_H^{-2-z+\theta} \;. 
	\end{align}
Notice that the mass is identically zero for $\theta=2 $, 
which is again the case with the geometry AdS$_2\times S^2$. 

For the canonical ensemble, where the temperature is allowed to vary and the charge density is fixed, 
the Helmholtz potential is given by 
	\begin{align}
		W = -\frac{\Omega_2}{8\pi}\sqrt{-w\Lambda}e^{-\frac{\delta}{2}\phi_0}\left[1-2\frac{(\gamma-\delta)^2}{wu} \right]m 
		= -\frac{\Omega_2}{16\pi} z ~r_H^{-2-z+\theta} \;.
	\end{align}
Finally the heat capacity is found to be 
	\begin{align}
		C_Q &= \frac{\Omega_2}{4} \frac{2 (\gamma-\delta)^2}{wu} \left[ 1- 2\frac{(\gamma-\delta)^2}{wu} \right]^{-1} 
		 \left(e^{\frac{\delta}{2} \phi_0} \frac{4\pi T}{\sqrt{-w \Lambda}} \right)^
		 {\frac{2(\gamma-\delta)^2}{wu-2(\gamma-\delta)^2}} \;, \\
		&= \frac{\Omega_2}{4} \frac{2-\theta}{ z} \left( \frac{4\pi T}{2+z-\theta} \right)^{\frac{2-\theta}{z}} \;.
	\end{align}
Thus the positive heat capacity gives a condition $\frac{2- \theta}{z} >0 $, which is 
consistent with (\ref{SpecificHealConditionLifshitz}) for $d=2 $.

\subsubsection{Constraints on the parameter space}  \label{sec:ConstraintComparison}

\begin{figure}[!ht]
\begin{center}
	 \includegraphics[width=0.3\textwidth]{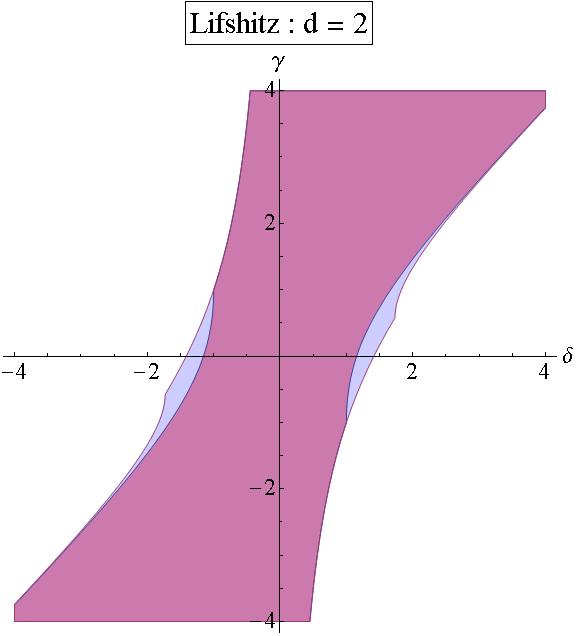}  \quad
	 \includegraphics[width=0.3\textwidth]{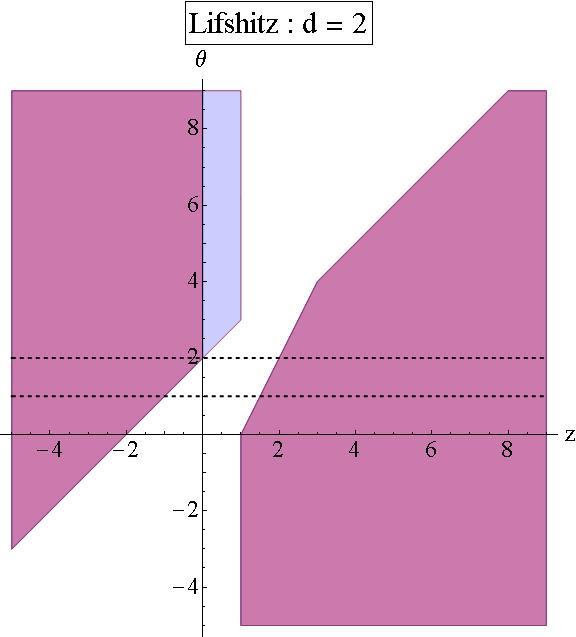}  \quad
	 \includegraphics[width=0.3\textwidth]{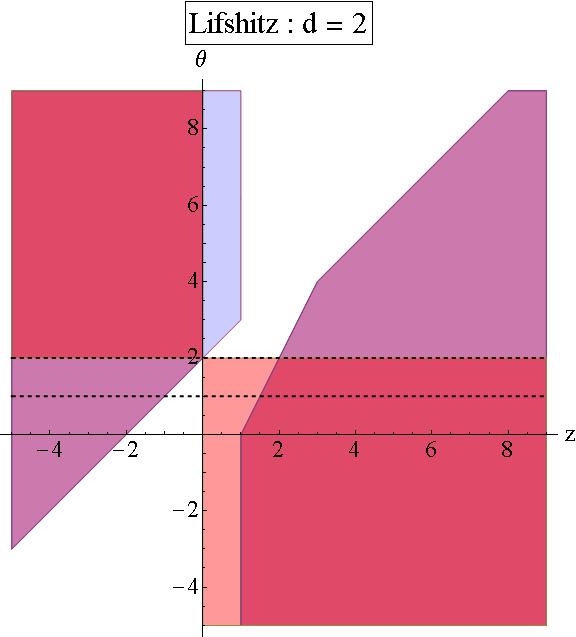}	
	 \caption{\footnotesize 
	 The left plot shows the allowed regions for the Lifshitz geometry in terms of the parameters 
	 $\gamma$ and $\delta $. The outer blue regions are allowed by the Gubser's criteria, while the 
	 inner purple region is allowed after taking into account of the other constraints including 
	 the well defined spectrum \cite{CGKKM}\cite{GKMReview}\cite{GKMReview2}\cite{IKNT}. 
	 The middle plot is the same as the left plot in terms of the parameters $ z$ and $ \theta$, 
	 where the middle blue region is allowed by Gubser's criteria, for example. 
	 The right plot shows the allowed regions after taking into account of the specific heat condition 
	 (\ref{SpecificHealConditionLifshitz}), where the two red regions, top-left and bottom-right regions are 
	 allowed (allowed region (II)). 
	 It is very interesting to confirm that these allowed regions (II) are the same as those of 
	 the middle plot, allowed region (I), in Fig. (\ref{fig:AllowedRegionsLifshitz}). 
	 }
	 \label{fig:AllowedRegionsLifshitzGammaDelta}
\end{center}
\end{figure}

Using the low energy effective holographic approach, consistent regions in the parameter spaces 
$(\gamma, \delta) $ of the solution (\ref{EMDsolution}) are analyzed 
in \cite{CGKKM}\cite{GKMReview}\cite{GKMReview2}\cite{IKNT}. 
Major constraints are the Gubser's bound, excluding naked singularities if the scalar potential is not 
bound below upon evaluating in the solution \cite{Gubser:2000nd}\cite{Gursoy:2007er}\cite{Gursoy:2008za}, 
as well as well defined spin 1 and spin 2 fluctuations around the singularity \cite{Charmousis:2001hg}. 
These constraints on the $(\gamma, \delta) $ parameter spaces are depicted in the left plot of 
figure \ref{fig:AllowedRegionsLifshitzGammaDelta}. Using the relation (\ref{zThetaGD}) or (\ref{GDzTheta}), 
we can re-express this constraints in terms of $(z, \theta) $, which is depicted in the middle plot.  
Finally the allowed regions are severely constrained by the positive specific heat constraint, which 
is the purple regions in the figure \ref{fig:AllowedRegionsLifshitzGammaDelta}. 
This can be compared to that of the middle plot in figure \ref{fig:AllowedRegionsLifshitz}. 
Interestingly, these two results show the same allowed regions.%
\footnote{We are grateful to Elias Kiritsis for discussions and comments on the related subjects.}

This demonstrate the success of the program, which classify the low energy and 
lower dimensional holographic theories in terms of the metrics with the parameters $(z, \theta) $ 
and constrain the allowed parameter spaces using the null energy conditions as well as 
some thermodynamic stability constraints. Allowed region (I) is restricted, mainly, from the metric,  
while allowed region (II) is constrained also from matter contents of the full solution.

\subsection{$d+2$ dimensional EMD solutions}    \label{sec:EMDgenerald}

In this section, we consider the explicit EMD solution for arbitrary number of spatial dimensions 
constructed in \cite{CGKKM} and rewrite the solution in terms of the parameters $(z, \theta) $. 
By doing so, we confirm that the emblackening factor has the same form at finite temperature. 

The full set of near extremal solution for general $d$ spatial dimensions is obtained in \cite{CGKKM}
	\begin{align}  \label{p-EMDSolutions}
	  & S =\int \mathrm d ^{d+2}x~\sqrt{-g}\left[R- \frac{e^{\gamma\phi}}{4}F_{\mu\nu}F^{\mu\nu}
  		-\frac{1}{2}(\partial\phi)^2-2\Lambda e^{-\delta\phi}\right]  \;, \nonumber  \\ 
	  &\mathrm d s^2 = \left(\frac{\tilde r}{ \ell}\right)^{\frac{(\gamma-\delta)^2}{d}}  
	  \left[-\tilde f(\tilde r)  \mathrm d t^2 + \sum_{i=1}^{d}  \mathrm d x_i  \mathrm d x^i \right] 
	  + \frac{ \mathrm d \tilde  r^2}{\tilde f(\tilde r)} \;, \\
	  & \tilde f(\tilde r) = \frac{8d \ell^2 \left(-\Lambda \right) e^{-\delta \phi_0}}{u^2 w}
	  \left(\frac{\tilde r}{ \ell}
	  \right)^{1-\frac{(d+1) (\gamma-\delta)^2}{2d}}
	  \left[\left(\frac{\tilde r}{ \ell}\right)^{\frac{wu}{2d}} -2m \right] \;, \nonumber \\
	  & e^\phi = e^{\phi_0}\left(\frac{\tilde r}{ \ell}\right)^{(\delta-\gamma)} \;, \quad 
	  A_t = \frac{4d}{wu}\sqrt{\frac{\ell^2\Lambda v}{u}}e^{-\frac{(\gamma+\delta)}2\phi_0}
	  \left[\left(\frac{\tilde r}{ \ell}\right)^{\frac{wu}{2d}}-2m\right] \;, \nonumber \\
	  & wu = 2d+(d+1)\gamma^2-2\gamma\delta-(d-1)\delta^2 \;, \quad
	  u = \gamma^2-\gamma \delta+2 \;, \quad
	  v = \delta^2-\gamma \delta -2 \;.  \nonumber
	  \end{align}
One can readily check that this general metric reduces to the $d=2 $ solution (\ref{EMDsolution}), and 
similar properties listed there also apply to this metric. 

Using the coordinate transform 
	\begin{align}
		&\tilde r \rightarrow b r^a \;, \qquad 
		a = - \frac{2d}{(\gamma +(d-1) \delta) (\gamma - \delta)} \;, \\
		&b = \left( \frac{\tilde  f_0}{a^2} \right)^{\frac{a}{2}} \ell^{1-a} \;, \qquad 
		\tilde  f_0 = \frac{8d \ell^2 \left(-\Lambda \right) e^{-\delta \phi_0}}{u^2 w} \;, 
	\end{align}
the metric can be recast into 	
	\begin{align} 
		\mathrm d s^2 &= r^{-2 +2\theta/d} \left(-a^2 \ell^2 r^{-2(z-1)} f(r) \mathrm d t^2 + \sum_{i=1}^{d}  \mathrm d x_i ^2 + \frac{ \mathrm d r^2}{f(r)}  \right) \;, \\
		f &= 1 -  2m \left(\frac{b r^a}{\ell}\right)^{-\frac{wu}{2d}}  
		= 1 -\left( \frac{r}{r_H} \right)^{\frac{wu}{(\gamma +(d-1) \delta) (\gamma - \delta)}} 
		=  1 -\left( \frac{r}{r_H} \right)^{d+z-\theta} \;,  
	\end{align}
where we use $ \Lambda $ or $ \phi_0 $ to set $b/\ell=1 $ as
	\begin{align}
		2\left(-\Lambda \right) e^{-\delta \phi_0} = \frac{u^2 w a^2}{4d }  \;, 
	\end{align}
Explicitly, the exponents are identified as follows in terms of $\gamma, \delta$ for $d+2$ dimensional solution    	
	\begin{align}
		\theta = \frac{d^2 \delta }{\gamma + (d-1) \delta} \;, \qquad 
		z = 1 + \frac{d \delta}{\gamma + (d-1) \delta } + \frac{2 d}{(\gamma -\delta)(\gamma + (d-1) \delta)} \;. 
	\end{align}
In turn we have 
	\begin{align}
		\gamma = \frac{\pm \sqrt{2} \left(d^2-d\theta +\theta \right)}{\sqrt{-d^3+d^3 z-d^2 z \theta +d \theta ^2}} \;, \quad 
		\delta = \frac{\pm \sqrt{2} \theta }{\sqrt{-d^3+d^3 z-d^2 z \theta +d \theta ^2}} \;. 
	\end{align}
Thus we can translate all the parameters in terms of $z$ and  $\theta $. For example, we get 
	\begin{align}
		2\left(-\Lambda \right) e^{-\delta \phi_0} = \frac{u^2 w a^2}{4d } 
		= (d+z-\theta -1) (d+z-\theta )   \;. 
	\end{align}

The resulting solutions has the following form 
	\begin{align} 
	  	&S =\int \mathrm d ^{d+2}x~\sqrt{-g}\left[R- \frac{Z}{4}F_{\mu\nu}F^{\mu\nu}
  		-\frac{1}{2}(\partial\phi)^2 + V \right]  \;,   \\ 
		&\mathrm d s^2 =r^{-2 +2\theta/d} \left(- r^{-2(z-1)} f(r) \mathrm d t^2 +\sum_{i=1}^{d}  
		\mathrm d x_i ^2 +\frac{ \mathrm d r^2}{f(r)}\right) \;, \\
	  	& e^\phi = r^s \;, \qquad s={\pm \sqrt{2 (1-\theta/d ) (d (z-1)-\theta )} }  \;, \\
		&V = V_0 e^{- \frac{2\theta}{s d} \phi} \;, \qquad V_0 =(d+z-\theta -1) (d+z-\theta) \;, \\ 
		&Z = \frac{1}{q^2} e^{ \frac{2(d^2 -(d-1)\theta)}{s d} \phi }\;,  \\
	  	&A_t = q \sqrt{ \frac{2(z-1)}{d+z-\theta }} r^{-d-z+\theta} f \;,  \\
	  	&f= 1 -\left( \frac{r}{r_H} \right)^{d+z-\theta}  \;,
	 \end{align}	 
where we absorb the dimensionful factor by changing coordinate $ a \ell \mathrm d t \rightarrow \mathrm d t $ 
and identify $ q = e^{-\frac{\gamma}2\phi_0}$.	For other useful information, especially for $ d=2 $, 
can be found in the previous section, near the equation (\ref{EMDsolutionthetaz}).

\section{Theories with Schr\"odinger scaling I}  \label{sec:Schroedinger}

Let us consider first the Schr\"odinger backgrounds, whose field theory duals have Schr\"odinger 
symmetry, initiated in \cite{Son:2008ye}\cite{Balasubramanian:2008dm} for the case $ \theta=0 $. 
The holographic dictionary turns out to be drastically different from that of the Lifshitz case. 
There are more than one holographic direction, and $ d+3 $ dimensional gravity backgrounds with 
Schr\"odinger isometry corresponds to $ d+1 $ dimensional field theories. There are two light-cone 
coordinates: one has the role of time, while the other supply the dual particle number $\mathcal M $.%
\footnote{
The role of the light-cone coordinate $ \xi $ has not been fully appreciated in the context of holography.   
See some developments along this line in \cite{Hyun:2011qj}\cite{Hyun:2012fd}, 
where $ \mathcal M $ is generalized to be complex in the context of time dependent setup. 
There one considers a slightly more general metric for $ \theta=0 $ and $\sharp=1 $ as 
	\begin{align}
		\mathrm d s^2 &= r^{-2} \left( - g(r, t) \mathrm d t^2 
		-2 ~\mathrm d t \mathrm d \xi + \mathrm dr^2
		+ \sum_{i=1}^{d} \mathrm d x_i^2 + h(r,t) dr dt  \right) \;. 		
	\end{align}
Two time correlation functions have been constructed to show slow dynamics, power
law decaying behavior in \cite{Hyun:2011qj} for the pure imaginary $ \mathcal M $. 
Their logarithmic extension is investigated in \cite{Hyun:2012fd} to seek connections 
to the Kardar-Parisi-Zhang universality class.  
See also \cite{Guica:2010sw}\cite{Balasubramanian:2010uw} for different considerations.
}
Their finite temperature generalizations are considered in \cite{Herzog:2008wg}-\cite{Ammon:2010eq}
using the null Melvin twist \cite{Alishahiha:2003ru}\cite{Gimon:2003xk}. 
See also further string theory related solutions in \cite{Mazzucato:2008tr}-\cite{Brown:2011av}. 
In this section, we concentrate on a very nice paper \cite{Mazzucato:2008tr}, which worked out many different 
non-relativistic backgrounds using the null Melvin twist. 

At zero temperature, the theories with Schr\"odinger scaling symmetry are described by the metric 
(\ref{BasicMetric}) with $\mathrm b=1, \mathrm a=1 $ and $D=d+1 $ 
	\begin{align}       \label{BasicMetricSchr}
		\mathrm d s^2 &= r^{-2 + \frac{2 \theta }{d+1}} \left( -  r^{-2(z-1)} \mathrm d t^2 
		-2 ~\mathrm d t \mathrm d \xi + \mathrm dr^2
		+ \sum_{i=1}^{\mathrm c} \mathrm d x_i^2 
		+ \sum_{j= \mathrm c+1}^{d} \eta_j (r,t, \vec x) \mathrm d x_j^2  \right) \;. 
	\end{align} 
This metric is extensively analyzed in \cite{kimHyperscaling} for both cases with and without 
hyperscaling violation exponent. 
In particular, a prescription for minimal surfaces of this ``codimension 2 holography,''
is proposed to demonstrate the $(d-1)$ dimensional area law for the entanglement entropy 
starting from $(d + 3)$ dimensional Schr\"odinger backgrounds. 
Surprisingly, the area law is violated for $d + 1 < z < d + 2$ due to the contribution of the 
spectator coordinate $\xi $, even without hyperscaling violation. 
See \cite{Perlmutter:2012he} for the top-down approach for the hyperscaling violation theories 
with Schr\"odinger backgrounds. 

The program to restrict the parameter space of $(z, \theta) $ of the metric (\ref{BasicMetricSchr}) 
for given spatial dimensions $d $ is carried out with the null energy condition. 
These are depicted in figure 1 and 2 of \cite{kimHyperscaling}. We observe a clear distinction, compared to 
the theories with Lifshitz scaling, that the entanglement entropy condition as well as null energy condition 
have non-trivial dependence on $ z$. For example, the entanglement entropy shows an extensive violation of 
the area law, being proportional to volume, for $ \theta = d+2-z $. Thus we expect to exclude the parameter 
spaces for $ \theta > d+2-z $ from the analysis of entanglement entropy. 
We further investigate several different string theory solutions in Appendix \S \ref{sec:SchrReduction}.
Thus, the program equally works for the theories with Schr\"odinger scaling at zero temperature. 

Naively, we desire to find the most general metrics with Schr\"odinger scaling at finite temperature with  
the hyperscaling violation exponent $\theta$ only using the radial coordinates similar to the Lifshitz case 
discussed in \S \ref{sec:Lifshitz}. We quickly realize that this attempt is bound to fail due to the 
dimensionful parameter, $b$, which is present in the known Schr\"odinger black hole solutions 
\cite{Herzog:2008wg}-\cite{Ammon:2010eq}. This parameter is intrinsically different from 
other dimensionful parameters because it actually serves as a thermodynamic variable in the black hole 
thermodynamics. Thus $b $ can not be absorbed in the overall dependence associated with $\theta$ 
or in other coordinates without altering the physical properties of the original solutions. 
It is difficult to constrain the most general Schr\"odinger invariant metric, 
because there are many different scaling invariant combinations one can construct using $ r$ and $b $. 
Because of this, we change our strategy to investigate the theories with a Schr\"odinger isometry 
at finite temperature.  

We concentrate on the non-relativistic Dp branes solutions \cite{Mazzucato:2008tr} at finite temperature 
as our prime examples and dimensionally reduce them to understand their physical properties 
in \S \ref{sec:DpDimensionalReduction}. 
By doing so, we explicitly check whether we can trust the naively constructed effective holographic 
theories with Schr\"odinger isometry. 
In particular, we determine the dynamical exponent of the non-relativistic Dp branes solutions in 
\S \ref{sec:DynamicalExponentSchr}, which turns out to be a non-trivial task. 
We investigate other physical properties including thermodynamics in the rest of the section 
\S \ref{sec:DpDimensionalReduction}. 
While doing this investigation, we find an important 
technical reason for the earlier observation that the thermodynamic properties of the 
Schr\"odinger backgrounds are identical to those of the ALCF. 
We study the finite temperature entanglement entropy in \S \ref{sec:StringSolEEntropy}. 
There we explicitly check that the entanglement entropy cross over to the thermal entropy in the 
high temperature limit, while it reproduces the zero temperature entanglement entropy 
in the small temperature limit. 
Then we consider the effective approach based on the results of the dimensional reduction 
in \S \ref{sec:SchrMetricReduction}.  

We separately consider the other geometries with Schr\"odinger scaling, ALCF, in the 
following section \S \ref{sec:ALCF} because they give fairly different looking physical properties 
at finite temperature.

\subsection{String theory realizations of Schr\"odinger backgrounds}    \label{sec:DpDimensionalReduction} 

Let us consider the following non-relativistic Dp branes solutions \cite{Mazzucato:2008tr}, constructed from 
the black Dp brane solutions \cite{Itzhaki:1998dd}\cite{Boonstra:1998mp} 
using the null Melvin twist \cite{Alishahiha:2003ru}\cite{Gimon:2003xk}. 
The metric can be written in a slightly more organized form as 
	\begin{align}     \label{nonrelBraneBetter}
		\mathrm d s_{Dp}^2 &= \frac{1}{h}
		\left[ - \frac{f}{b^2(1-f)} \mathrm d t^2 + \frac{b^2(1-f)}{K} 
		\left( \mathrm d \xi - \frac{1+f}{(1-f) 2b^2}  \mathrm d t \right)^2 
		+ \sum_{i=1}^{p-1}  \mathrm d x_i^2 \right]  \nonumber \\ 
		&+ h \left[\frac{1}{f} \mathrm d \rho^2 + \rho^2 \left( \frac{1}{K} 
		(\mathrm d \chi +\mathcal A)^2 +\mathrm d s_{\mathbb P}^2 \right) \right] \;, \\
		e^{\Phi} &= \left(\frac{h^{3-p}}{K} \right)^{1/2} \;, \qquad 
		K = 1 + b^2 \rho^2 \left( \frac{\rho_H}{\rho} \right)^{7-p} \;,  \\
		B &= \frac{\rho^2}{ K} (\mathrm d \chi + \mathcal A) \wedge 
		\left( \frac{1+f}{2} \mathrm d t - b^2 (1-f)\mathrm d \xi \right) \;,  \\
		h^2 &= 1 + \left(\frac{\rho_p}{\rho} \right)^{7-p}  \;, \qquad
		f(\rho) = 1 - \left( \frac{\rho_H}{\rho} \right)^{7-p}  \;. 
	\end{align}

We consider the near horizon geometry with $ h \rightarrow   \left(\frac{\rho_p}{\rho} \right)^{(7-p)/2} $, 
followed by the change of coordinate $\rho = 1/u$, followed by the compactification of the solution 
down to $(p+2)$ dimensions. 
Dimensional reduction to $p+2$ dimensions and going to Einstein frame give 
	\begin{align}   \label{dimenReduSchrMetricADM}
		\mathrm d s^2 &= r^{-\frac{2(9-p)}{p(5-p)}} K^{\frac{1}{p}}
		\left[ \frac{- f}{b^2(1-f)} \mathrm d t^2 + \frac{b^2(1-f)}{K} 
		\left( \mathrm d \xi - \frac{1+f}{2b^2(1-f)} \mathrm d t \right)^2 \right. \nonumber \\
		&\left. +  \sum_{i=1}^{p-1}  \mathrm d x_i^2 + \frac{ \mathrm d r^2}{f} \right]  \;,  \\
	 	K &= 1 + c~ b^2 ~ r_H^{-2\frac{7-p}{5-p}}~ r^2  \;, \qquad 
	 	f = 1 - \left( \frac{r}{r_H} \right)^{\frac{2(7-p)}{5-p}} \;,
	\end{align}
where we use $ u = \left(\frac{2}{5-p} \right)^{-\frac{2}{5-p}} u_p^{\frac{7-p}{5-p}} r^{\frac{2}{5-p}}$ and 
$c= \left(\frac{2}{5-p} \right)^{\frac{4}{5-p}} u_p^{-2\frac{7-p}{5-p}}$. 
With this form at hand, we realize that the thermodynamic properties, related to the horizon such as 
temperature and entropy, of	the Schr\"odinger type solutions are independent of $K $. 
This is an important technical detail that verify the earlier claims that the thermodynamics of the 
Schr\"odinger type theories are identical to those of the ALCF 
\cite{Herzog:2008wg}\cite{Maldacena:2008wh}\cite{Kim:2010tf}. 
This is checked explicitly in \S \ref{sec:ThermodynamicsSchr} below. 

Thus the non-relativistic Dp brane solutions (\ref{nonrelBraneBetter}) 
gives the Schr\"odinger type theories with hyperscaling violation with the identifications  
	\begin{align}
		\theta = p - \frac{9-p}{5-p}  =  - \frac{(p-3)^2}{5-p} \;, 
	\end{align}
where we use $D=p=d+1 $. 
Note that the overall dimensionful parameter only contains $u_p$, which is important to 
provide a unique physical meaning to the hyperscaling violation exponent, without 
being mixed with the other thermodynamic parameters, $b $ and $r_H $.

\subsubsection{Determination of the dynamical exponent}  \label{sec:DynamicalExponentSchr}

Here we would like to identify the dynamical exponent in terms of the parameter $p$ by 
taking the zero temperature limit carefully. There is actually a subtle point. 

At first glance, 
it is tempting to use the ADM form of the metric (\ref{dimenReduSchrMetricADM}) to identify the dynamical 
exponent from the first term with $ \mathrm d t^2 $.  
	\begin{align}     \label{zprimeIden}
		\frac{-f}{b^2(1-f)} \mathrm d t^2 = \left[ b^{-2} r_H^{\frac{2(7-p)}{5-p}} \right]
		r^{\frac{-2(7-p)}{5-p}} (-f \mathrm d t^2) \;.
	\end{align}
Naively, we can identify the dynamical exponent in terms of the radial dependence by absorbing 
all the dimensionful parameters into $t $ by redefining $t \rightarrow   b r_H^{-\frac{(7-p)}{5-p}} t$. 
Then we get 
	\begin{align}    \label{zPrimeIndetificationSchr}
		z' = \frac{12-2p}{5-p} \;, 
	\end{align}
Seemingly, it is consistent with the fact that 
$K\approx 1 $ is not important because of the physically relevant range $ 0\leq r \leq r_H$.
But this identification is not consistent with that of the zero temperature 
Schr\"odinger background due to a subtle cancellation involved with a subleading term in $ K $.%
\footnote{If one include the scaling dimension of $ b $ for the identification of $ z $ 
from (\ref{zprimeIden}), he or she would get the correct value. 
Our motivation starts from the low energy effective description without knowing precise information 
on $ b$ and thus handling the parameter would be an important task.  
\label{footnote:zidentification}}

Let us identify the dynamical exponent using a slightly different form of the 
metric (\ref{dimenReduSchrMetricADM})  
	\begin{align}     \label{dimenReduSchrMetricUsualFrom}
		\mathrm d s^2  &= r^{-2+ \frac{2\theta}{p}} \left( \frac{(1+f)^2 - 4 f K }{4 b^2(1-f) K} \mathrm d t^2 
		+ \frac{b^2(1-f)}{K} \mathrm d \xi^2 
		- \frac{1+f}{K} \mathrm d t \mathrm d \xi \right.  \nonumber  \\
		&\left. + \sum_{i=1}^{p-1}  \mathrm d x_i^2 + \frac{ \mathrm d r^2}{f} \right)  \;.
	\end{align}
Let us take a zero temperature limit, $r_H \rightarrow \infty $, or asymptotic form at the 
boundary, $r \rightarrow 0 $. 
	\begin{align}	
		\mathrm d s^2  &=  r^{-2+ 2\theta/p} \left( - c r^{\frac{-4}{5-p}} \mathrm d t^2 - 2 \mathrm d t \mathrm d \xi 
		+ \sum_{i=1}^{p-1}  \mathrm d x_i^2 + \frac{ \mathrm d r^2}{f} \right) \;.
	\end{align}
The constant $c $ in front of the term $\mathrm d t^2 $ can be absorbed by redefining $t $ and $\xi $. 
Thus we should identify the dynamical exponent  
	\begin{align}    \label{zIdentificationSchr}
		2-2 z = \frac{-4}{5-p} \qquad  \longrightarrow  \qquad 
		z = \frac{7-p}{5-p} \;, 
	\end{align}
to be in consistent with the zero temperature background. 
In turn, we can identify $ z' $ in terms of $ z $ as 
	\begin{align}
		z' = \frac{12-2p}{5-p}  = z+1  \;.
	\end{align}
Thus $ z $ is the correct dynamical exponent we are going to use in this section because it is consistent with 
the zero temperature limit. 	
As a byproduct, we identify the parameter $ \beta $ in terms of $ c $ as 
	\begin{align}
		c = \left(\frac{2}{5-p} \right)^{\frac{4}{5-p}} u_p^{-2\frac{7-p}{5-p}} \;,
	\end{align}
which is independent of the parameters $b $ and the horizon radius related to the temperature. 	

Note there are three different dimensionful parameters, $u_p $, $b $ and $u_H $, in the 
Schr\"odinger metric at finite temperature. 
We can pull out $u_p $ to be an overall factor by redefining coordinates, and it contributes  
to the part related to the hyperscaling violation. 
The other two are physical parameters, being two independent 
thermodynamic variables, which are important to keep track. 
In this sense, $z $ in equation (\ref{zIdentificationSchr}) is well established 
because the physical parameters $b $ and $u_H $ are not involved in the identification process.  
On the other hand, we need to be careful for the identification of (\ref{zPrimeIndetificationSchr}) 
because it requires to redefine  $t \rightarrow   b r_H^{-\frac{(7-p)}{5-p}} t$. 
The latter case would change or lose some physical information because the coordinate redefinition 
is temperature dependent. (see related motivation in footnote \ref{footnote:zidentification}.)

\subsubsection{Thermodynamic Properties}    \label{sec:ThermodynamicsSchr}
	
Thermodynamic properties of the metric (\ref{dimenReduSchrMetricUsualFrom}) are more transparent in the ADM form 
(\ref{dimenReduSchrMetricADM}). We can read off the lapse function $N$, the shift function
$V^i$ and the horizon coordinate velocity in the $x^-$ direction $\Omega_H$, 
which can be interpreted as chemical potential associated with a conserved quantity along the $x^-$ direction.
	\begin{align}
	  N = r^{-\frac{9-p}{p(5-p)}}  \sqrt{\frac{f}{(1-f) b^2}}  \;,\quad
	  V^- = - \frac{1}{2b^2}\frac{1+f}{1-f}  \;, \quad
	  \Omega_H = \frac{1}{2b^2} \;.
	\end{align}

Some of the thermodynamic properties can be directly analyzed using the horizon properties. 
Temperature, entropy and chemical potential along the $\xi$ direction are given by 
	\begin{align}    \label{thermoDpBraneSchr}
		&T = \frac{1}{2 \pi b r_H} \bigg| \frac{7-p}{5-p} \bigg| \;, \qquad 
		S_T \approx b r_H^{-\frac{9-p}{5-p}} V_{p-1} V_\xi \;, \qquad 
		\Omega = \frac{1}{2b^2} \;.
	\end{align}
Note that the entropy is independent of $K $, as explained above.	
There is a hidden dimensionful parameters $R_\theta $ in the entropy expression. 
Assuming the front factor in metric (\ref{dimenReduSchrMetricADM}) being   
$ \left( \frac{r}{R}\right)^{-2} \left(\frac{r}{R_\theta} \right)^{2\theta/p} $, 
the entropy becomes 
$ S_T = \frac{(M_{pl} R)^p}{4} R_\theta^{-\theta} V_{p-1} V_\xi b r_H^{-\frac{9-p}{5-p}} $.

\subsubsection{Reduction to Lifshitz type theories}    \label{sec:SchrLifReduction}

It is interesting to connect to the Schr\"odinger type theories to Lifshitz type theories 
in one way or another because the properties of the latter might shed some light to the former. 
Here we just formally identify the exponents of the Schr\"odinger theories to that of the Lifshitz 
theories by dimensionally reducing (\ref{dimenReduSchrMetricADM}) along $\xi $ coordinate. 

The resulting $p+1 $ dimensional Lifshitz type solution has the following metric in Einstein frame 
	\begin{align}    \label{dimenReduSchrMetricToLifshitz}
		\mathrm d s^2 &= r^{-\frac{4}{(p-1)(5-p)}} \left( b r_H^{-\frac{7-p}{5-p}} \right)^{\frac{2}{p-1}} 
		\left[ \frac{- f}{b^2 r_H^{-\frac{2(7-p)}{5-p}}} r^{-\frac{2(7-p)}{5-p}} \mathrm d t^2 
		+  \sum_{i=1}^{p-1}  \mathrm d x_i^2 + \frac{ \mathrm d r^2}{f} \right]  \;, 
	\end{align}
where the metric has the same emblackening factor $f = 1 - \left( \frac{r}{r_H} \right)^{\frac{2(7-p)}{5-p}} $, 
which still has the pivotal role for the identification of the dynamical exponent as in the Schr\"odinger case. 
Note that the front factor $K^{\frac{1}{p}} $ cancels out the contribution from reduced $\xi $ coordinate. 
The dimensionful parameters in the term $\mathrm d t^2 $ can be absorbed by the coordinate redefinition.%
\footnote{We further assume that $ b r_H^{-\frac{7-p}{5-p}} =1 $ upon taking the dimensional reduction along 
the coordinate $ \xi $. 
The thermodynamic properties of the Lifshitz space are expected to be independent of the parameter $ b $, 
and the resulting Lifshitz metric is also expected to be independent of $ r_H $. 
\label{footnote:LifReduction}
} 
Thus we identify 
	\begin{align}
		\theta_{LF} = -\frac{p^2-6p+7}{5-p} \;, \qquad z_{LF} = \frac{12-2p}{5-p}   \;.
	\end{align}
We would like to point out that the emblackening factor becomes  
	\begin{align}
		d+z-\theta = \frac{2(7-p)}{5-p} \;, \qquad f = 1 - \left( \frac{r}{r_H} \right)^{d+z-\theta} \;, 
	\end{align} 
which shares the same property of the general theories with Lifshitz isometry advertised in \S \ref{sec:Lifreview}.

\subsection{Entanglement entropy}      \label{sec:StringSolEEntropy}

We would like to evaluate the entanglement entropy for the Schr\"odinger backgrounds 
at finite temperature described by the metric (\ref{dimenReduSchrMetricADM}) 
	\begin{align}    \label{EntangleMetricSchrZTheta}
		\mathrm d s^2 &= r^{\frac{2\theta}{p}-2} K^{\frac{1}{p}} 
		\left[\frac{-f~ \mathrm d t^2}{(1-f)b^2} 
	  + \frac{(1-f)b^2}{K} \bigg( \mathrm d \xi - \frac{1+f}{1-f} \frac{\mathrm d t}{2b^2} \bigg)^2 
 	  + \sum_{i=1}^{p-1}  \mathrm d x_i^2 + \frac{ \mathrm d r^2}{f} \right] \;,  \nonumber \\		 
	 	f &= 1 - \left( \frac{r}{r_H} \right)^{2z}  \;, \quad
	 	K = 1 + \beta ~ b^2 ~ r_H^{-2z}~ r^2 \;, 
	\end{align}	
where $D=p=d+1 $ and $ \beta = \left(\frac{2}{5-p} \right)^{\frac{4}{5-p}} u_p^{-2z} 
= (z-1)^{2(z-1)} u_p^{-2z}$. 
Note that we use $\theta, z $ just for the notational simplicity 
in this section. They are given by 
	\begin{align}
		\theta = - \frac{(p-3)^2}{5-p} \;,  \qquad z = \frac{7-p}{5-p} \;. 
	\end{align}

The entanglement entropy of the metric (\ref{EntangleMetric}) with finite temperature can be computed 
for a strip with $\xi$ direction 
	\begin{align}
		0 \le \xi \le L_\xi\;,\quad -l \le x_1 \le l\;,\quad 0 \le x_i \le L\;,\;\quad i = 2,\cdots \;, p-1
	\end{align}   
in the limit $l \ll L, L_\xi$. 
The strip is located at $r=\epsilon$, and the profile of the surface in the bulk is given by $r=r(x_1)$. 
The minimal surface has turning point at $r=r_t$. Thus, to get the entanglement entropy, 
we evaluate the following expression 
	\begin{align}   
		l &= \int_0^{r_t}\mathrm d r~\frac{1}{\sqrt{f}} \frac{(r/r_t)^{\alpha_2} ~(K(r)/K(r_t))^{p/2p-1/2}}
		{\sqrt{1- (r/r_t)^{2\alpha_2} ~(K(r)/K(r_t))^{p/p-1}}}  \;,
	\end{align}   
and 
	\begin{align}  
		{\mathcal  A} 
		&= L^{p-2} L_\xi \int_\epsilon^{r_t}\mathrm d r~\frac{1}{\sqrt{f}} \frac{\left(b r_H^{-z} \right)
		 r^{-\alpha_2} ~K(r)^{-p/2p+1/2}}
		{\sqrt{1-(r/r_t)^{2\alpha_2}~(K(r)/K(r_t))^{p/p-1}}} \;,
	\end{align}   
where $\alpha_2 = p-z -\theta$. Note that the expression is independent of $K$ because the front factor 
$ K^{1/p}$ multiplied by $p$ times cancel $ 1/K $ in front of $ \mathrm d \xi^2 $. Thus the entanglement entropy 
of the Schr\"odinger black hole is identical to that of the ALCF, which is discussed in the following section.  
Thus we get 
	\begin{align}   
		l &= \int_0^{r_t}\mathrm d r~\frac{1}{\sqrt{f}} \frac{(r/r_t)^{\alpha_2}}
		{\sqrt{1- (r/r_t)^{2\alpha_2}}} \;, 
	\end{align}   
and 
	\begin{align}  
		{\mathcal  A} 
		= L^{p-2} L_\xi \left(b r_H^{-z} \right) \int_\epsilon^{r_t}\mathrm d r~\frac{1}{\sqrt{f}} \frac{ r^{-\alpha_2}}
		{\sqrt{1-(r/r_t)^{2\alpha_2}}} 	\;. 
	\end{align}    
Note the extra dimensionless factor 
$ \left(b r_H^{-z} \right)$, which comes from the contribution $(1-f)b^2 $ in front of $ \mathrm d \xi^2 $ term. 
This factor gives us a chance that this entanglement entropy can be equal to the thermal entropy 
given in (\ref{thermoDpBraneSchr}). 

Assuming $\alpha_2 >0$, we can rewrite the integrals as, following \cite{Swingle:2011mk}
	\begin{align}   
		l &= r_t \int_0^{1} \mathrm d \Omega ~\frac{\omega^{\alpha_2}}
		{\sqrt{1 - (\gamma \omega)^{2z}} \sqrt{1- \omega^{2\alpha_2}}} \;,
	\end{align}   
and 
	\begin{align}  
		{\mathcal  A} 
		&= L^{d-1} L_\xi \left(b r_H^{-z} \right) r_t^{1-\alpha_2} \int_{\epsilon/r_t}^1 \mathrm d \Omega~ 
		\left( \left[ \frac{ \omega^{-\alpha_2}}
		{\sqrt{1 - (\gamma \omega)^{2z}} \sqrt{1-\omega^{2\alpha_2}}} - \frac{1}{\omega^{\alpha_2}} \right] 
		+  \frac{1}{\omega^{\alpha_2}}  \right)	\;, 
	\end{align}   
where the square bracket part is finite for $\alpha_2 >0$ and 
	\begin{align}
		\omega = \frac{r}{r_t} \;, \qquad \gamma = \frac{r_t}{r_H} <  1\;.
	\end{align}
The last term can be computed as 
	\begin{align}  
		{\mathcal  A}_{div} 
		&= L^{p-2} L_\xi \left(b r_H^{-z} \right) r_t^{1-\alpha_2} \int_{\epsilon/r_t}^1 \mathrm d \Omega~  \frac{1}{\omega^{\alpha_2}} 
		= L^{p-2} L_\xi \left(b r_H^{-z} \right) \frac{1}{\alpha_2 -1} \frac{1}{\epsilon^{\alpha_2 -1} }	\;, 
	\end{align}   
which is the divergent contribution of the zero temperature background.  	

At small temperature, we can evaluate the integral as 
	\begin{align}   
		l &= r_t \int_0^{1} \mathrm d \Omega ~\frac{\omega^{\alpha_2}}
		{\sqrt{1 - (\gamma \omega)^{2z}} \sqrt{1- \omega^{2\alpha_2}}}
		\approx \sqrt{\pi} r_t \frac{\Gamma \left(\frac{1+ \alpha_2}{2 \alpha_2} \right)}
		{\Gamma \left(\frac{1}{2 \alpha_2} \right)}
	\end{align}   
and 
	\begin{align}  
		{\mathcal  A}_{fin} 
		&\approx  L^{p-2} L_\xi \left(b r_H^{-z} \right) r_t^{1-\alpha_2} \int^1 \mathrm d \Omega~ 
		\frac{ \omega^{-\alpha_2}}{\sqrt{1-\omega^{2\alpha_2}}} 
		\left( 1 + \frac{1}{2} (\gamma \omega)^{2z} + \cdots  \right)   \;, \nonumber \\
		&= L^{p-2} L_\xi \left(b r_H^{-z} \right) l^{1-\alpha_2}~ c_\theta  \left(-\frac{1}{\alpha_2 -1}  + 
		\frac{\gamma^{2z}}{4 \alpha_2 }  \frac{\Gamma \left(\frac{-\alpha_2 + 2 z+1}{2 \alpha_2} \right)}
		{\Gamma \left(\frac{2z +1}{2 \alpha_2} \right)} 
		\frac{\Gamma \left(\frac{1}{2 \alpha_2} \right)}{\Gamma \left(\frac{1+ \alpha_2}{2 \alpha_2} \right)}
		+ \cdots \right) \;, 				
	\end{align}   	
Thus, for the low temperature limit, 	
the entanglement entropy for a strip in the general metric (\ref{EntangleMetricSchrZTheta}) is 
given in terms of $(b, r_H) $ as 
	\begin{align}      \label{lowTentangleEntropyMicro}
		\mathcal  S 
		= \frac{(R M_{Pl})^{p} }{4 (\alpha_2-1)} \frac{b}{r_H^{z}}
		&\left(  \left(\frac{\epsilon}{R_\theta}\right)^{\theta} \frac{L^{p-2} L_\xi }{\epsilon^{p-z-1}} 
		\right. \nonumber \\
		&\left. -  c_{\theta} ~\left(\frac{l}{R_\theta}\right)^{\theta} \frac{L^{p-2} L_\xi }{l^{p-z-1}} 
		\left[ 1 - \tilde c_\theta \left(\frac{ l}{r_H} \right)^{2z} + \cdots  \right] \right) \;,
	\end{align}   
where $R_\theta$ is a scale in which the hyperscaling violation becomes important and 
	\begin{align}
		c_{\theta} = \left( \frac{\sqrt{\pi} \Gamma \left( \frac{1+\alpha_2}{2\alpha_2} \right)}
		{ \Gamma \left( \frac{1}{2\alpha_2} \right) } \right)^{\alpha_2}  \;, \qquad 
		\tilde c_{\theta} = \frac{(\alpha_2 -1) \sqrt{\pi}  }{4 \alpha_2 }  
		c_\theta^{-\frac{2z+1}{\alpha_2}}
		\frac{\Gamma \left(\frac{-\alpha_2 + 2 z+1}{2 \alpha_2} \right)}
		{\Gamma \left(\frac{2z +1}{2 \alpha_2} \right)} \;.
	\end{align}
Note that the scaling mismatches of the factor $ \frac{L^{p-2} L_\xi }{\epsilon^{p-z-1}}$ and 
$ \frac{L^{p-2} L_\xi }{l^{p-z-1}}$ are made up by that of $ \frac{b}{r_H^{z}} $. 

In the high temperature limit, 
	\begin{align}   
		l &= r_t \int^{1} \mathrm d \Omega ~\frac{\omega^{\alpha_2}}
		{\sqrt{1 - (\gamma \omega)^{2z}} \sqrt{1- \omega^{2\alpha_2}}}  
		= r_t I_+ \left( \gamma \right) \;,
	\end{align}   
and 
	\begin{align}  
		{\mathcal  A} 
		&= L^{p-2} L_\xi \left(b r_H^{-z} \right) r_t^{1-\alpha_2} \int^1 \mathrm d \Omega~ 
		\frac{ \omega^{-\alpha_2}}{\sqrt{1 - (\gamma \omega)^{2z}} \sqrt{1-\omega^{2\alpha_2}}} \nonumber \\
		&= L^{p-2} L_\xi \left(b r_H^{-z} \right) r_t^{1-\alpha_2}  I_- \left( \gamma \right)	\;. 
	\end{align}   
When $\gamma \rightarrow 1, r_t \approx r_H$ , the integrals $I_+ \left( \gamma \right)$ and 
$I_- \left( \gamma \right)$ are dominated by the contribution $\omega \approx 1$ and thus 
$I_+ \left( \gamma \right) \approx I_- \left( \gamma \right) \approx l/r_t$ \cite{kachru}. 
Thus 
	\begin{align}    \label{highTentangleEntropyMicro}
		\mathcal  S_{fin} \approx \frac{(M_{pl} R)^p}{4} R_\theta^{-\theta} L^{p-2} L_\xi l ~b~ r_H^{-p+\theta} 
		= \frac{(M_{pl} R)^p}{4} R_\theta^{-\theta} L^{p-2} L_\xi l ~b~ r_H^{-\frac{9-p}{5-p}} \;,
	\end{align}    	
which agrees with the thermal entropy evaluated in (\ref{thermoDpBraneSchr}).

\subsection{Searching for effective theories}       \label{sec:SchrMetricReduction}

In this section, we would like to answer a question, 
{\it ``Can we construct the most general low energy and lower dimensional effective holographic metrics 
with Schr\"odinger scaling at finite temperature?''} The answer seems to be partial due to the special 
form we consider.

For now, we can think about the class of metrics of the form with two parameters $(\theta, z) $
	\begin{align}    \label{EntangleMetricSchrEff}
		\mathrm d s^2 &= r^{\frac{2\theta}{p}-2} K^{\frac{1}{p}} 
		\left[\frac{-f~ \mathrm d t^2}{(1-f)b^2} 
	  + \frac{(1-f)b^2}{K} \bigg( \mathrm d \xi - \frac{1+f}{1-f} \frac{\mathrm d t}{2b^2} \bigg)^2 
 	  + \sum_{i=1}^{p-1}  \mathrm d x_i^2 + \frac{ \mathrm d r^2}{f} \right] \;,  \nonumber \\		 
	 	f &= 1 - \left( \frac{r}{r_H} \right)^{2z}  \;, \quad
	 	K = 1 + \beta ~ b^2 ~ r_H^{-2z}~ r^2 \;.
	\end{align}	
Note the emblackening factor $ f$, whose form is fixed only with dynamical exponent $ z $. 
Let us comment the scaling symmetry of the metric (\ref{dimenReduSchrMetricADM}), 
which is described by the transformations 
	\begin{align}
		&t \rightarrow \lambda^z t \;, \quad 
		\xi \rightarrow \lambda^{2-z} \xi \;, \quad 
		r \rightarrow \lambda r \;, \quad 
		x_i \rightarrow \lambda x_i \;, \\
		&b \rightarrow \lambda^{z-1} b \;, \quad 
		\beta \rightarrow \lambda^{0} \beta \;, \quad 
		r_H \rightarrow \lambda r_H \;.
	\end{align}  
The corresponding zero temperature limit is explicitly considered in \S \ref{sec:DynamicalExponentSchr}.  
We would like to investigate the physical properties of this metric from the spirit of 
\S \ref{sec:Lifreview} for $ z>0 $ and postpone an important question whether this class of metrics can 
be further generalized. 

Can we restrict further the allowed regions of the parameter space of $ (\theta, z)$ using 
specific heat constraint? We already calculate them in (\ref{thermoDpBraneSchr}) 
of \S \ref{sec:ThermodynamicsSchr}.  
	\begin{align}    \label{thermoDpBraneSchrEff}
		&T \sim \frac{1}{b r_H}  \;, \qquad  
		S_T \sim b r_H^{-p + \theta} \;, \qquad 
		\Omega = \frac{1}{2b^2} \;.
	\end{align}
From these thermodynamic quantities, we calculate the specific heat for fixed $\Omega $ for $ z>0 $, 
even though the result is less restrictive than that for fixed particle number $ N $.%
\footnote{
It is expected to constrain the parameter space more strictly 
if we use the specific heat with fixed particle number. 
The over all sign of the $\Omega N $ term is negative similar to the pressure and volume $PV $, where 
$C_V $ is more constraining than $C_P $ at least for the ideal thermodynamics. 
It would be interesting to evaluate all the thermodynamic quantities including energy $E $ and 
dual particle number $N $. Simply these are not available yet.  
Thus we would like to consider a specific heat with fixed $\Omega $ to constrain the parameter spaces. 
We thanks to Carlos Hoyos for the discussions related to this point. 
\label{footnote:specificHeat}
}
The condition $ T \frac{\partial S}{\partial T} |_{\Omega} >0 $ gives $ \theta < p=d+1$ for $ z>0 $. 
This condition is actually less restrictive than the entanglement entropy constraint, 
$ \theta < d+2-z $. Thus we are not able to constrain more than what we did with the zero temperature 
metric once we include the constraint entanglement entropy analysis as well as null energy conditions, 
which are analyzed in \cite{kimHyperscaling}. 
In particular, the null energy condition gives two independent constraints
	\begin{align}   \label{nullECondition}
		& (z-1) (d+2 z) (d+1) - z (d+1) \theta + \theta^2 \geq 0  \;, \nonumber\\ 
		& (z-1) ( d+2 z - \theta ) \geq 0\;,
	\end{align}  
where we use $D=p=d+1$. 

\begin{figure}[!ht]
\begin{center}
	 \includegraphics[width=0.31\textwidth]{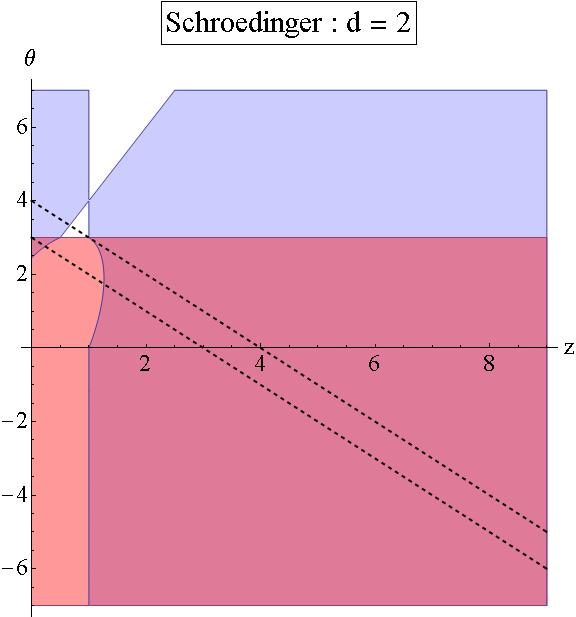} \quad 
	 \includegraphics[width=0.31\textwidth]{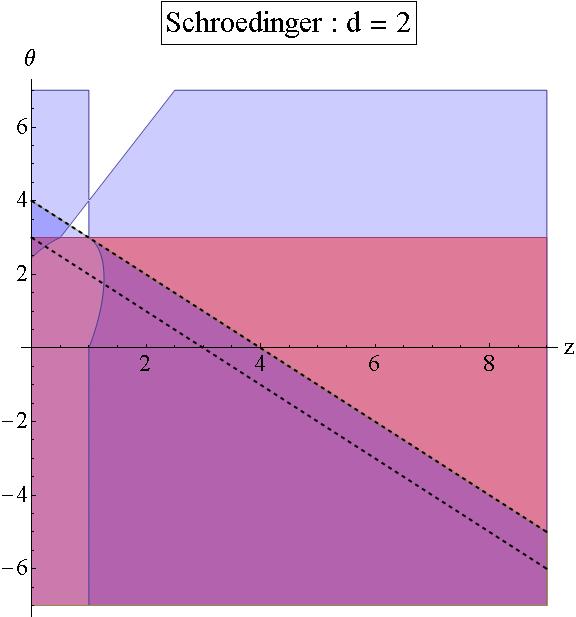}	 \quad 
	 \includegraphics[width=0.31\textwidth]{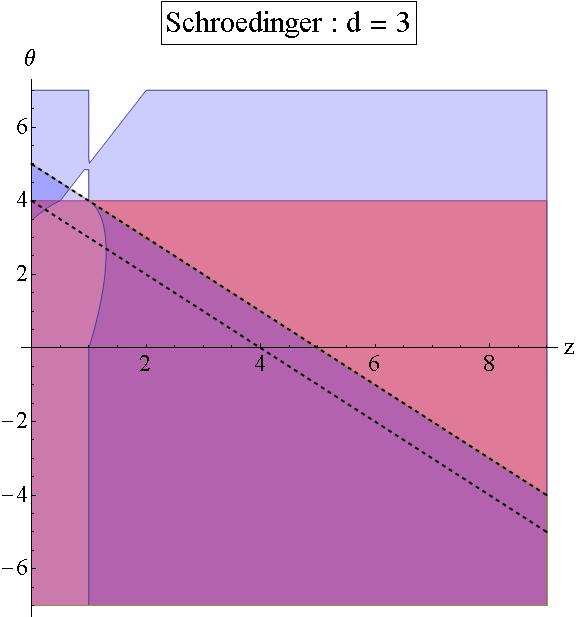}
	 \caption{\footnotesize 
	 The left plot shows the allowed region (dark red) for the Schr\"odinger background 
	 for $ d=2$ after taking into account of the specific heat condition for fixed $\Omega $, 
	 $ \theta < d+1 $ in addition to the null energy condition. 
	 The region between two dashed lines indicates some novel phases 
	 from the entanglement entropy analysis \cite{kimHyperscaling}. 
	 The left plot can be further constrained by the entanglement entropy analysis, $ \theta < d+2-z $, 
	 which is given in the middle plot. Thus the purple region is allowed for $ z>0 $.
	 The right plot shows the regions allowed by the null energy condition (blue), 
	 the specific heat (red), and the entanglement analysis. The purple region is allowed 
	 after taking into account these conditions for $d=3$. 
	 }
	 \label{fig:AllowedRegionsSpecificHeat}
\end{center}
\end{figure}

Can we remove the parameter $ b $ for the low energy description? The answer is no. 
The reason is already stressed above that the physical parameter $ b $ has a crucial role 
in black hole thermodynamics. Thus any change would result in changing its thermodynamics, 
which is not allowed. Furthermore, $ b $ is relevant for all energy scales because it has zero 
mass dimension, while it carries definite scaling dimensions.

Let us finish this section by commenting the case $ z<0 $. Seeming there are several different 
solutions with {\it negative} dynamical exponents, which can be obtained from the consistent string 
theory backgrounds \cite{Bobev:2009mw}\cite{Bobev:2011qx}\cite{Costa:2010cn}.%
\footnote{We thank to Nikolay Bobev and especially Yaron Oz 
for the references and related discussions and comments.
} 
While it is not clear that these theories are consistent or useful, it is interesting to 
have careful investigations for them.    
Some holographic backgrounds with negative dynamical exponents are also reported in different 
contexts, see {\it e.g.} \cite{Singh:2012un}. 
We also find that the null Melvin twisted version of the seemingly non-singular KK monopole solution 
exhibits a negative dynamical exponent \S \ref{sec:KKMonolope}.

\section{Theories with Schr\"odinger scaling II}    \label{sec:ALCF} 

As we already point out, there are two different realizations of holographic theories 
with Schr\"odinger scaling. Here we consider a simpler realization, so-called AdS in light-cone (ALCF). 
The case without hyperscaling violation, $ \theta=0 $, is introduced in 
\cite{Goldberger:2008vg}\cite{Barbon:2008bg}, and further 
generalized to finite temperature in \cite{Maldacena:2008wh}\cite{Kim:2010tf}. 
The transport properties of this background is analyzed in \cite{Kim:2010tf}\cite{Kim:2010zq}. 
The magneto-transport properties analyzed in \cite{Kim:2010zq} are very similar to 
those of the high $T_c $ cuprates at very low temperature.
See also \cite{Kim:2011zd}. The transport properties are further generalized using higher derivative 
corrections in \cite{Fadafan:2012hr}. 

At zero temperature, ALCF are described by the metric (\ref{BasicMetric}) 
with $\mathrm b=0, \mathrm a=1 $ 
	\begin{align}       \label{BasicMetricALCF}
		\mathrm d s^2 &= r^{-2 + \frac{2 \theta }{d+1}} \left(	-2 ~\mathrm d t \mathrm d \xi + \mathrm dr^2
		+ \sum_{i=1}^{\mathrm c} \mathrm d x_i^2 
		+ \sum_{j= \mathrm c+1}^{d} \eta_j (r,t, \vec x) \mathrm d x_j^2  \right) \;, 
	\end{align} 
where $D=d+1 $. This metric is also analyzed in \cite{kimHyperscaling} along with the 
Schr\"odinger backgrounds (\ref{BasicMetricSchr}) 
for both cases with and without hyperscaling violation exponent. 
There similar results on the minimal surface prescriptions and thus entanglement entropy are obtained. 
Recently, hyperscaling violation of the R-charged black holes are analyzed in \cite{Sadeghi:2012vv}. 

The program to restrict the parameter space of $(z, \theta) $ of the metric (\ref{BasicMetricALCF}) 
for $d $ spatial dimensions is carried out with the null energy condition, which are depicted 
in figure 3 for $ d=2$ and $3 $ \cite{kimHyperscaling}. For this case, the null energy condition is similar to 
the Lifshitz case, while the entanglement entropy condition is similar to Schr\"odinger backgrounds 
with non-trivial dependence on $ z$. Thus the classification at zero temperature works as the other cases. 

The program of classifying ALCF at finite temperature has the similar difficulties as 
the Schr\"odinger background. See the introduction of \S \ref{sec:Schroedinger}. 
Thus we concentrate on the ALCF version of the relativistic black Dp branes solutions 
as our prime examples and dimensionally reduce them to understand their physical properties 
in \S \ref{sec:DpDimensionalReductionAdSLC}. We find the dynamical exponent of the ALCF at 
finite temperature is not fixed by the parameters of the microscopic string theories.
In the rest of \S \ref{sec:DpDimensionalReductionAdSLC}, we study the thermodynamics ALCF and 
their dimensional reduction to Lifshitz theories along the $ \xi $ coordinate.  
We study the finite temperature entanglement entropy in \S \ref{sec:StringSolEEntropy}, followed by 
the effective approach based on the results of the dimensional reduction 
in \S \ref{sec:SchrMetricReduction}.

\subsection{String theory realizations for ALCF}    \label{sec:DpDimensionalReductionAdSLC}

We are primarily interested in the black Dp branes solutions 
\cite{Itzhaki:1998dd}\cite{Boonstra:1998mp}\cite{Aharony:1999ti}, whose metric is given by 
	\begin{align}     
		&\mathrm d s_{Dp}^2 = h^{-1/2} \left(- f d\tau^2 + \sum_{i=1}^p  \mathrm d x_i^2 \right) 
		+ \frac{h^{1/2}}{\tilde r^4}  \left( \frac{1}{f}   \mathrm d \tilde  r^2 
		+ \tilde r^{2} \mathrm d \Omega_{8-p}^2  \right) \;,  \\ 
		&f =  1 - \left(\frac{\tilde r}{\tilde r_H} \right)^{7-p} \;, \quad 
		h = 1 +  \left( \frac{\tilde r}{\tilde r_0} \right)^{7-p} \;, 
		\quad e^{-2(\Phi - \Phi_\infty)} = h^{(p-3)/2} \;, 
	\end{align}
where we omit RR fields and the corresponding type II supergravity action. 
We compactify this solution on $S^{8-p}$ down to $(p+2)$ dimensions to get  	
	\begin{align}     
		&\mathrm d s_{p+2}^2 = h^{-1/2} \left(- f d\tau^2 + \sum_{i=1}^p  \mathrm d x_i^2 \right) 
		+  \frac{h^{1/2}}{\tilde r^{4} f}   \mathrm d \tilde  r^2  \;,  
	\end{align}	
and the action acquire additional radial dependence from the radial dependence of $S^{8-p}$
	\begin{align}
		S = \frac{V(\Omega_{8-p})}{2 \kappa_{10}^2} \int \mathrm d ^{p+2} x 
		\sqrt{-g_{p+2}} e^{-2 \Phi_{p+2}}  \mathcal R_{p+2} + \cdots \;,  
	\end{align}	
where $e^{-2\Phi_{p+2}} =  h^{\frac{p+2}{4}} {\tilde r}^{p-8}  $ represents the radial dependence, 
not a dilaton. 
Using $\mathrm d s^2  \rightarrow \left( e^{-2\Phi_{p+2}} \right)^{\frac{2}{p}}\mathrm d s^2 $, 
we go to the Einstein frame in $p+2$ dimensions 
	\begin{align} 
		\mathrm d s_{Dp}^2 &= r^{-\frac{2(9-p)}{p(5-p)}}  
		\left(- f d\tau^2 + \sum_{i=1}^p  \mathrm d x_i^2 + \frac{1}{f}\mathrm d r^2\right) \;, 
		\qquad f = 1 - \left( \frac{r}{r_H} \right)^{\frac{2(7-p)}{5-p}}  \;, 
	\end{align}
where we use $ h \rightarrow \left( \frac{\tilde r}{\tilde r_0} \right)^{7-p}$ 
for the near horizon limit and additional change of a variable $ \tilde r \sim  r^{\frac{2}{5-p}}$. 	
The overall dimensionful factor was not carried over. 

The operation of the dimensional reduction commutes with changing into the light-cone frame, which   
is defined by   
	\begin{align}
	  t =& b(\tau+x) \;,\qquad  \xi  = \frac{1}{2b}(\tau-x)  \;.
	\end{align}
We assign the scaling dimension of $b$ as $[b]=1-z$ in the mass unit, 
and thus $[t] = -z$ and $[\xi] = z-2$ to have manifest dynamical exponent $z$.  
The metric takes the form
	\begin{align}     \label{finiteTReducedMetric}
		\mathrm d s^2 &= r^{-\frac{2(9-p)}{p(5-p)}}  \left(\frac{1-f}{4b^2}d t^2
  		-(1+f)\mathrm d t \mathrm d \xi + (1-f)b^2 \mathrm d \xi^2 + \sum_{i=1}^{p-1}  \mathrm d x_i^2 + \frac{1}{f}\mathrm d r^2\right) \;. 
	\end{align}
At zero temperature, $f=1 $, this metric is reduced to 
	\begin{align}    \label{zeroTReducedMetric}
		\mathrm d s^2 &= r^{-\frac{2(9-p)}{p(5-p)}}  \left(-2 \mathrm d t \mathrm d \xi 
		+ \sum_{i=1}^{p-1}  \mathrm d x_i^2 + \mathrm d r^2\right) \;. 
	\end{align}	
The basic properties of this metric was analyzed in \cite{kimHyperscaling}.  
We have the following identification for $\theta $
	\begin{align}   \label{thetaValueALCF}
		\theta_{LC} = - \frac{(p-3)^2}{5-p} \;,
	\end{align} 
where we use $ D=p=d+1$ for the Schr\"odinger case. 	

For later use, we would like to explicitly write the physical ADM form 
	\begin{align}    \label{finiteTADMMetricAdSLC}
		\mathrm d s^2 &= \left(\frac{r}{R_\theta}\right)^{-\frac{2(9-p)}{p(5-p)}}  
		\left(\frac{-f~ \mathrm d t^2}{(1-f)b^2} 
		+ (1-f)b^2 \bigg( \mathrm d \xi - \frac{1+f}{1-f} \frac{\mathrm d t}{2b^2} \bigg)^2
 		+ \sum_{i=1}^{p-1}  \mathrm d x_i^2 + \frac{ \mathrm d r^2}{f} \right) \;, 
	\end{align}	
where $R_\theta $ represent a scale where the hyperscaling violation becomes significant. 
Let us comment the scaling symmetry of the metric (\ref{finiteTADMMetricAdSLC}), 
which is described by the transformations 
	\begin{align}   \label{scalingTrALCF}
		&t \rightarrow \lambda^z t \;, \quad 
		\xi \rightarrow \lambda^{2-z} \xi \;, \quad 
		r \rightarrow \lambda r \;, \quad 
		x_i \rightarrow \lambda x_i \;, \quad 
		b \rightarrow \lambda^{z-1} b \;, \quad 
		r_H \rightarrow \lambda r_H \;.
	\end{align}  

Similar to the Schr\"odinger background, we would like to determine the dynamical exponent for 
the ALCF  metric (\ref{finiteTADMMetricAdSLC}). 
We would like to remind the reader that the identification of the dynamical exponent $z$ is not 
clear for the zero temperature metric (\ref{zeroTReducedMetric}). Actually the metric 
(\ref{zeroTReducedMetric}) describes ALCF system for general $z$ at zero temperature 
\cite{kimHyperscaling}. 
Furthermore, the scaling transformation (\ref{scalingTrALCF}) does not fix the dynamical 
exponent in terms of other parameters, such as $ p $. 
Thus we leave the dynamical exponent undetermined for the case here, too.%
\footnote{ 	
From this metric (\ref{finiteTADMMetricAdSLC}), we consider the first term in the parenthesis 
$ \frac{-f}{(1-f)b^2} = -  b^{-2} r_H^{\frac{2(7-p)}{5-p}} r^{-\frac{2(7-p)}{5-p}}  f $
where the dimensionful parameters $ b^{-2} r_H^{\frac{2(7-p)}{5-p}} $ can be absorbed by 
redefining the coordinate $t $. It is tempting to identify the dynamical exponent $z_{LC} $ as 
$z_{LC} = 1 + \frac{7-p}{5-p}= \frac{12-2p}{5-p} $. But this is dangerous thing to do due to the 
fact that physical thermodynamic parameters are absorbed in time. 
}

\subsubsection{Thermodynamic Properties}
	
Thermodynamic properties of the metric (\ref{finiteTReducedMetric}) are more transparent in the ADM form 
(\ref{finiteTADMMetricAdSLC}). 
From this ADM form, we can read off the lapse function $N$, the shift function
$V^i$ and the horizon coordinate velocity in the $x^-$ direction
$\Omega_H$, which can be interpreted as chemical potential associated with the 
conserved quantities along the $x^-$ direction,  as
	\begin{align}
	  N = r^{-\frac{9-p}{p(5-p)}}  \sqrt{\frac{f}{(1-f) b^2}}  \;,\quad
	  V^- = - \frac{1}{2b^2}\frac{1+f}{1-f}  \;, \quad
	  \Omega_H = \frac{1}{2b^2} \;.
	\end{align}

Some of the thermodynamic properties can be directly analyzed using the horizon properties. 
The temperature, entropy and chemical potential along the $\xi$ direction are given by 
	\begin{align}    \label{thermoDpBraneAdSLC}
		T = \frac{1}{2 \pi b r_H} \bigg| \frac{7-p}{5-p} \bigg| \;, \qquad 
		S_T = b \left( \frac{r_H}{R_\theta} \right)^{-\frac{9-p}{5-p}} V_{p-1} V_\xi  \;, \qquad 
		\Omega = \frac{1}{2b^2} \;.
	\end{align}
Due to the presence of the parameter $b$, it is not straightforward to identify the 
precise relation between the thermal entropy $ S_T$ and the temperature $T$.

\subsubsection{Reduction to theories with Lifshitz scaling}    \label{sec:ALCFLifshitzReduction}

As done in the Schr\"odinger type theories, here we just formally identify the exponents of the ALCF theories 
to that of the Lifshitz theories by dimensionally reducing (\ref{finiteTADMMetricAdSLC}) along $\xi $ coordinate. 

The resulting $p+1 $ dimensional Lifshitz type solution has the following metric in Einstein frame 
	\begin{align}    \label{finiteTADMMetricAdSLCToLifshitz}
		\mathrm d s^2 &= r^{-\frac{4}{(p-1)(5-p)}}  
		\left[ - f r^{-\frac{2(7-p)}{5-p}} \mathrm d t^2 
		+  \sum_{i=1}^{p-1}  \mathrm d x_i^2 + \frac{ \mathrm d r^2}{f} \right]  \;, 
	\end{align}
where we also take $ b r_H^{-\frac{7-p}{5-p}} =1 $ as we comment in footnote \ref{footnote:LifReduction}. 
Now the metric has the same emblackening factor $f = 1 - \left( \frac{r}{r_H} \right)^{\frac{2(7-p)}{5-p}} $ 
as that of (\ref{dimenReduSchrMetricToLifshitz}) considered in \S \ref{sec:SchrLifReduction}.
We note that this form is the same as that of the Schr\"odinger case.  
Thus we identify 
	\begin{align}    \label{ALCFdynamicalExponentLifshitzReduction}
		\theta_{LF} = -\frac{p^2-6p+7}{5-p} \;, \qquad z_{LF} = \frac{12-2p}{5-p}   \;.
	\end{align}
We would like to emphasize that this case also belongs to the typical case of emblackening factor, which can 
be identified as 
	\begin{align}
		d+z_{LF}-\theta_{LF} = \frac{2(7-p)}{5-p} \;, \qquad 
		f = 1 - \left( \frac{r}{r_H} \right)^{d+z_{LF}-\theta_{LF}} \;, 
	\end{align} 
which seems to be persisting for this case also, as advertised in \S \ref{sec:Lifreview}.

\subsection{Entanglement entropy}      \label{sec:StringSolEEntropyALCF}

We would like to evaluate the entanglement entropy for the ALCF background at finite temperature 
described by the metric (\ref{finiteTADMMetricAdSLC}) following \cite{Swingle:2011mk}\cite{kachru}.  
	\begin{align}    \label{EntangleMetric}
		\mathrm d s^2 &= \left(\frac{r}{R_\theta}\right)^{-2+2\theta/p} 
		\left[\frac{-f~ \mathrm d t^2}{(1-f)b^2} 
	  + (1-f)b^2 \bigg( \mathrm d \xi - \frac{1+f}{1-f} \frac{\mathrm d t}{2b^2} \bigg)^2
 	  + \sum_{i=1}^{p-1}  \mathrm d x_i^2 + \frac{ \mathrm d r^2}{f} \right] \;,  \nonumber \\		 
	 	f &= 1 - \left( \frac{r}{r_H} \right)^{2\frac{7-p}{5-p}}  \;, 
	\end{align}	
where $D=p = d+1 $. Note that we did not specify the emblackening factor in terms of $ z $. 

The entanglement entropy of the metric (\ref{EntangleMetric}) with finite temperature can be computed 
for a strip with $\xi$ direction 
	\begin{align}
		0 \le \xi \le L_\xi\;,\quad -l \le x_1 \le l\;,\quad 0 \le x_i \le L\;,\;\quad i = 2,\cdots \;, p-1
	\end{align}   
in the limit $l \ll L, L_\xi$. 
The strip is located at $r=\epsilon$, and the profile of the surface in the bulk is given by $r=r(x_1)$. 
The minimal surface has turning point at $r=r_t$. Thus, to get the entanglement entropy, 
we evaluate the following expressions 
	\begin{align}   
		l &= \int_0^{r_t}\mathrm d r~\frac{1}{\sqrt{f}} \frac{(r/r_t)^{\alpha_2}}
		{\sqrt{1- (r/r_t)^{2\alpha_2}}} \;, 
	\end{align}   
and 
	\begin{align}  
		{\mathcal  A} 
		= L^{p-2} L_\xi \left(b r_H^{-\frac{7-p}{5-p}} \right) \int_\epsilon^{r_t} 
		\mathrm d r~\frac{1}{\sqrt{f}} \frac{ r^{-\alpha_2}}
		{\sqrt{1-(r/r_t)^{2\alpha_2}}} 	\;, 
	\end{align}   
where $\alpha_2 = p-\frac{7-p}{5-p} - \theta$. 
Note that there is a clear difference compared to (\ref{entangleAreaHyprLif}), the extra dimensionless factor 
$ \left(b r_H^{-\frac{7-p}{5-p}} \right)$, which comes from the contribution $(1-f)b^2 $ 
in front of $ \mathrm d \xi^2 $ term. 
This factor gives us a chance that this entanglement entropy can be equal to the thermal entropy 
(\ref{thermoDpBraneAdSLC}). 
	
The rest of the calculations are similar to the previous case \S \ref{sec:StringSolEEntropy}, 
and we present the final results. 
For the low temperature limit, 	
the entanglement entropy for a strip in the general metric (\ref{EntangleMetric}) is 
given by 
	\begin{align}      \label{lowTentangleEntropyMicroALCF}
		\mathcal  S 
		= \frac{(R M_{Pl})^{p} }{4 (\alpha_2-1)} \frac{b}{r_H^{\frac{7-p}{5-p}}}
		&\left(  \left(\frac{\epsilon}{R_\theta}\right)^{\theta} \frac{L^{p-2} L_\xi }{\epsilon^{p-1-\frac{7-p}{5-p}}} 
		\right. \nonumber \\
		&  \left. - c_{\theta} ~\left(\frac{l}{R_\theta}\right)^{\theta} 
		\frac{L^{p-2} L_\xi }{l^{p-1-\frac{7-p}{5-p}}} 
		\left[ 1 - \tilde c_\theta \left( \frac{l}{r_H} \right)^{2\frac{7-p}{5-p}} + \cdots  \right] \right) \;,
	\end{align}   
which has the same form as (\ref{lowTentangleEntropyMicro}) with suitable identification of $ z $, 
which we do not specify due to the reason explained above. 
	
In the high temperature limit, 
	\begin{align}   
		l &= r_t \int^{1} \mathrm d \Omega ~\frac{\omega^{\alpha_2}}
		{\sqrt{1 - (\gamma \omega)^{2z-2}} \sqrt{1- \omega^{2\alpha_2}}}  
		= r_t I_+ \left( \gamma \right) \;,
	\end{align}   
and 
	\begin{align}  
		{\mathcal  A} 
		&= L^{d-1} L_\xi \left(b r_H^{-\frac{7-p}{5-p}} \right) r_t^{1-\alpha_2} \int^1 \mathrm d \Omega~ 
		\frac{ \omega^{-\alpha_2}}{\sqrt{1 - (\gamma \omega)^{2z-2}} \sqrt{1-\omega^{2\alpha_2}}} \nonumber \\
		&= L^{d-1} L_\xi \left(b r_H^{-\frac{7-p}{5-p}} \right) r_t^{1-\alpha_2}  I_- \left( \gamma \right)	\;. 
	\end{align}   
When $\gamma \rightarrow 1, r_t \approx r_H$ , the integrals $I_+ \left( \gamma \right)$ and 
$I_- \left( \gamma \right)$ are dominated by the contribution $\omega \approx 1$ and thus 
$I_+ \left( \gamma \right) \approx I_- \left( \gamma \right) \approx l/r_t$ \cite{kachru}. 
Thus 
	\begin{align}   
		\mathcal  S_{fin} \approx R_\theta^{-\theta} L^{d-1} L_\xi l ~b~ r_H^{-p+\theta} 
		= R_\theta^{-\theta} L^{d-1} L_\xi l ~b~ r_H^{-\frac{9-p}{5-p}} \;,
	\end{align}    	
which agrees with the thermal entropy evaluated in (\ref{thermoDpBraneAdSLC}).

\subsection{Searching for ALCF effective theories}       \label{sec:ALCFEff}

In this section, we would like to ask a similar question, 
{\it ``Can we construct the most general low energy and lower dimensional ALCF effective holographic metrics 
with Schr\"odinger scaling at finite temperature?''} 
The answer seems to be more unclear for this case compared to Schr\"odinger case at finite temperature. 

For now, we can think about the class of metrics of the form with a parameter $\theta $ for arbitrary $ z $
	\begin{align}    \label{EntangleMetricALCF2}
		\mathrm d s^2 &= r^{\frac{2\theta}{p}-2} \left[\frac{-f~ \mathrm d t^2}{(1-f)b^2} 
	  + (1-f)b^2 \bigg( \mathrm d \xi - \frac{1+f}{1-f} \frac{\mathrm d t}{2b^2} \bigg)^2 
 	  + \sum_{i=1}^{p-1}  \mathrm d x_i^2 + \frac{ \mathrm d r^2}{f} \right] \;,  \nonumber \\		 
	 	f &= 1 - \left( \frac{r}{r_H} \right)^{\frac{7-p}{5-p}} \;.
	\end{align}	
Note the emblackening factor $ f$ does not have a definite dependence on $ z $, due to the fact 
that this metric is valid for arbitrary dynamical exponent $ z $ \cite{kimHyperscaling}. 
See also the discussion around (\ref{finiteTADMMetricAdSLC}). 
With this form, specific heat constraint from (\ref{thermoDpBraneAdSLC}) does not give us 
further constraint because the expressions are well defined for given $ p $. 

Now we change our attention a little and try to understand the differences between the Schr\"odinger 
backgrounds (\ref{EntangleMetricSchrEff}) and those of ALCF (\ref{EntangleMetricALCF2}). For this 
purpose we fix the dynamical exponent of the ALCF metric as $ z_{LC} = \frac{7-p}{5-p} $. 
Thus for the rest of the section we consider the metric 
	\begin{align}    \label{EntangleMetricALCFEff}
		\mathrm d s^2 &= r^{\frac{2\theta}{p}-2} \left[\frac{-f~ \mathrm d t^2}{(1-f)b^2} 
	  + (1-f)b^2 \bigg( \mathrm d \xi - \frac{1+f}{1-f} \frac{\mathrm d t}{2b^2} \bigg)^2 
 	  + \sum_{i=1}^{p-1}  \mathrm d x_i^2 + \frac{ \mathrm d r^2}{f} \right] \;,  \nonumber \\		 
	 	f &= 1 - \left( \frac{r}{r_H} \right)^{z} \;.
	\end{align}	
Thermodynamical properties of this metric is given by 
	\begin{align}    \label{thermoDpBraneAdSLCEff}
		T \sim \frac{1}{b r_H}  \;, \qquad 
		S_T \sim b r_H^{\theta - p}  \;, \qquad 
		\Omega = \frac{1}{2b^2} \;.
	\end{align}
Similar to the discussion around the equation (\ref{thermoDpBraneSchrEff}) and footnote 
\ref{footnote:specificHeat}, we would like to  restrict further the allowed regions of the parameter 
space of $ (\theta, z)$ using specific heat constraint at fixed $ \Omega $. 
The condition $ T \frac{\partial S}{\partial T} |_{\Omega} >0 $ gives $ \theta < p=d+1$ for $ z>0 $. 
This condition is actually less restrictive than the entanglement entropy constraint, 
$ \theta < d+2-z $, which might be improved once we can use the specific heat constraint for 
fixed particle number. 

\begin{figure}[!ht]
\begin{center}
	 \includegraphics[width=0.31\textwidth]{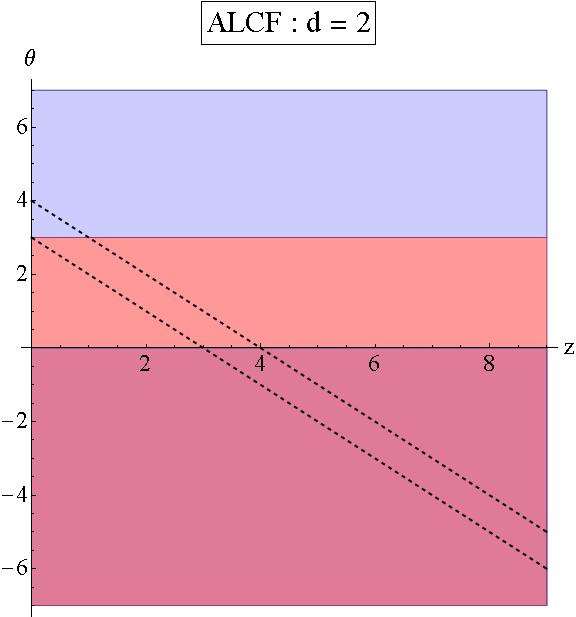} \quad 
	 \includegraphics[width=0.31\textwidth]{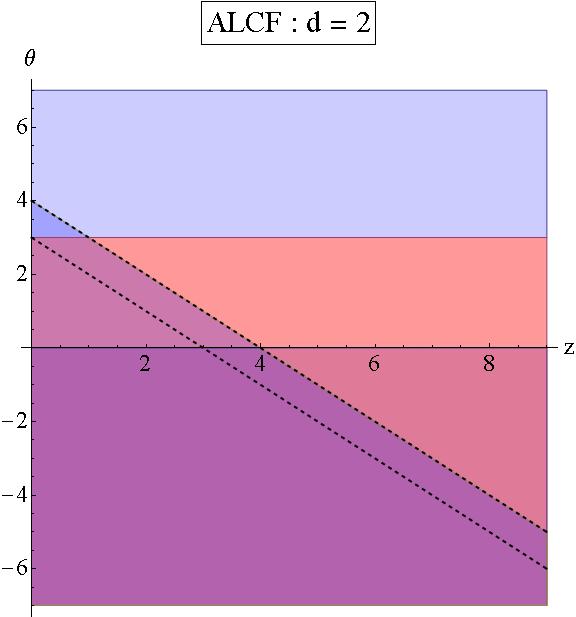}	 \quad 
	 \includegraphics[width=0.31\textwidth]{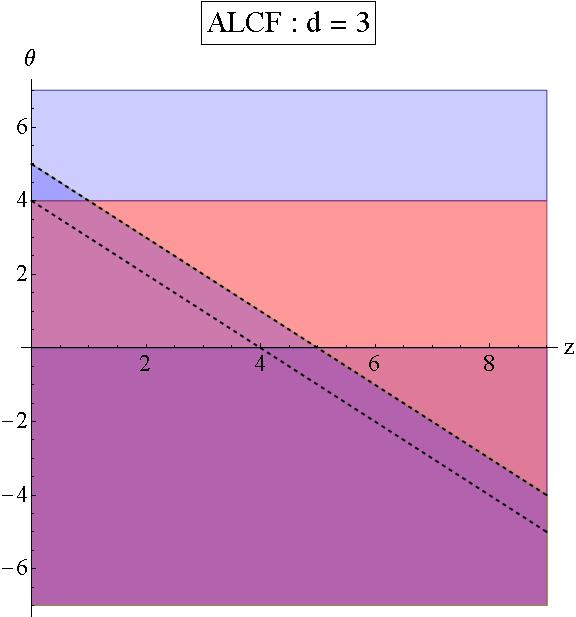}
	 \caption{\footnotesize 
	 The left plot shows the allowed region (dark red) for the Schr\"odinger background 
	 for $ d=2$ after taking into account of the specific heat condition for fixed $\Omega $, 
	 $ \theta < d+1 $ in addition to the null energy condition. 
	 The region between two dashed lines indicates some novel phases 
	 from the entanglement entropy analysis \cite{kimHyperscaling}. 
	 The left plot can be further constrained by the entanglement entropy analysis, $ \theta < d+2-z $, 
	 which is given in the middle plot. Thus the purple region is allowed for $ z>0 $.
	 The right plot shows the regions allowed by the null energy condition (blue), 
	 the specific heat (red), and the entanglement analysis. The purple region is allowed 
	 after taking into account these conditions for $d=3$.   
	 }
	 \label{fig:AllowedRegionsSpecificHeatALCF}
\end{center}
\end{figure}

Thus we have the following constraints. At zero temperature, we have the entanglement entropy analysis
$ \theta < d+2-z $ as well as the null energy conditions 
\cite{kimHyperscaling}. 
	\begin{align}   \label{nullEConditionAdSLC}
		& \theta (\theta - d-1) \geq 0  \qquad \rightarrow \qquad  
		\theta \leq 0 \quad \text{or} \quad \theta \geq  d+1  \;.
	\end{align}
At finite temperature, we add $ \theta < p=d+1$ for $ z>0 $ from specific heat constraint at 
fixed chemical potential $ \Omega $. This is summarized in figure \ref{fig:AllowedRegionsSpecificHeatALCF}.

\section{Summary and Outlook} 

The main theme we would like to put forward 
is that {\it the hyperscaling violation exponent $\theta$ along with the dynamical $z$ 
and the spatial anisotropic $ \sharp $ exponents can be 
viewed a unified framework for the low energy and lower dimensional effective holographic theories.} 
One clear observation is that the hyperscaling violation exponent captures the degree of violation of 
conformal symmetry of the microscopic string theory solutions. 
Conformal invariant solutions would give $\theta=0$ upon simple dimensional reduction, 
while nonconformal solutions would produce non zero $\theta$. 

We initiate a simple classification of these effective holographic theories with definite scaling symmetries.  
Our examples include the theories with Lifshitz, including the relativistic $ z=1 $ case, and Schr\"odinger 
scaling. Our classification is simple because we only care about the metric, discarding the corresponding 
action and matter contents. This can be justified, partially at least, by the fact that one metric 
can be supported by more than single set of action with corresponding matter contents 
\cite{Balasubramanian:2009rx}. Thus classifying theories with full solutions might be redundant, 
certainly for the physical properties directly related to the metric.  

As shown in the main body, this program is successful for the theories with Lifshitz scaling, 
which is described by the simple metric (\ref{HyperscalingLifshitzMetric})
	\begin{align}        
		\mathrm d s^2_{d+2} &= r^{-2(d-\theta)/d} \left( - r^{-2(z-1)} f(r) \mathrm d t^2 
		+ \frac{ \mathrm d r^2}{f(r)} 
		+ \sum_{i=1}^d  \mathrm d x_i^2 \right) \;, \nonumber \\
		f(r) &= 1 - \left(\frac{r}{r_H} \right)^{d+z-\theta} \;, 
	\end{align} 
where $\theta $ and $z $ are the hyperscaling violation and dynamical exponents. 
This metric is extensively studied in the context of Einstein-Maxwell system with a dilaton 
\cite{CGKKM}\cite{IKNT} and in the context of systems with hyperscaling violation 
\cite{Huijse:2011ef}\cite{kachru}. 
Several observations are in order. First, many different microscopic string solutions reduce to this 
simple and universal form upon sphere reduction, as we have checked here. 
The metric is universal even at finite temperature with a fixed emblackening factor $ f $.  
Second, the information of the dimensionful parameter in the string theory solution is directly 
transferred to the front factor, captured by the hyperscaling violation exponent $ \theta $. 
Third, this metric has been known to be the most general IR scaling solution for Einstein-Maxwell system 
with one scalar \cite{Gouteraux:2011ce}.
Fourth, if there are more than one dimensionful parameters active in the effective theories, 
additional exponents, called spatial anisotropic exponents, would come into play.   

Once one accepts this, further procedures are required to restrict the allowed space of 
the parameters $ (z, \theta) $ \cite{Ogawa:2011bz}\cite{kachru}. 
This can be done by using null energy conditions and the condition 
from entanglement entropy analysis at zero temperature. 
The allowed parameter space is further restricted by the positive specific heat constraint. 
The result for the theories with Lifshitz scaling is summarized in the figure 
\ref{fig:AllowedRegionsLifshitz}. 
This result is tested against the earlier, similar in spirit, program to constrain the allowed 
parameter space of the full solution \cite{CGKKM}\cite{GKMReview}, using the Gubser's constraint 
\cite{Gubser:2000nd} as well as the well defined fluctuation problems around the horizon. 
The allowed regions of the latter is summarized in the figure \ref{fig:AllowedRegionsLifshitzGammaDelta}.
After taking into account of the positive specific heat constraint, these two programs to 
restrict the allowed region of the parameter space $(z, \theta) $ are identical, signalling 
success of our program at least for the theories with Lifshitz isometry. 

We also consider two different theories with Schr\"odinger scaling : Schr\"odinger backgrounds 
and ALCF (AdS in light-cone). The classification goes well at zero temperature similar to that of 
the Lifshitz isometry. 
	\begin{align}       
		\mathrm d s^2 &= r^{-2 +2 \theta /D} \left( - \mathrm b~ r^{2-2z} \mathrm d t^2 
		-2 \mathrm d t \mathrm d \xi + \mathrm dr^2
		+ \sum_{i=1}^{\mathrm c} \mathrm d x_i^2 
		+ \sum_{j= \mathrm c+1}^{d} r^{2- 2 \sharp_j}  ~\mathrm d x_j^2  \right) \;, 
	\end{align} 
where $ D=d+1 $ and $ \sharp_j $ are the spatial anisotropic exponents. 
We consider here two different theories, Schr\"odinger backgrounds with $ \mathrm b=1 $ 
and ALCF with $ \mathrm b=0 $.
At finite temperature, there exist additional dimensionful parameter $ b $ entering our story, 
which also has role in their thermodynamics.  We can not simply remove or absorb this $ b $ 
without changing their physical properties. Thus we should carry over this parameter. 
The classification becomes complicated in the description of the effective theories 
because there are many different scaling invariant combinations one can come up with 
the parameter $ b $ along with the radial direction and other parameters in the theory. 
Thus the program need much more care compared to that of the Lifshitz case. The 
results are summarized in the figure \ref{fig:AllowedRegionsSpecificHeat} for the Schr\"odinger case 
and in the figure \ref{fig:AllowedRegionsSpecificHeatALCF} for the ALCF.

We are observing intricate and active interplay between many different disciplines,  
especially the string theory and condensed matter, via gauge gravity duality. 
Our program would be useful for the high energy community to provide useful guides for the 
available theories with various scaling symmetries, not to mention for the condense matter applications. 
We investigate the systematic studies of the holographic backgrounds with scaling symmetries 
using the simplest sphere reduction. It is interesting to generalize this program by 
incorporating non-trivial internal geometries and various higher modes in the dimensional reduction 
as well as by tackling less symmetric holographic backgrounds. Generally, it is expected to 
introduce more exponents as the reduction becomes more complicated. 
Theories with Schr\"odinger isometry are more involved due to the presence of a new thermodynamic 
parameter at finite temperature and requires more investigations.  
It will be also interesting to investigate further the backgrounds with the negative dynamical exponents.

\section*{Acknowledgments}

We are grateful to N. Bobev, C. Charmousis, O. Ganor, B. Gout\'eraux, C. Hoyos, 
S. Hyun, N. Itzhaki, J. Jeong, E. Kiritsis, 
B.-H. Lee, R. Meyer, K. Narayan, Y. Oz, Y. Seo, S.-J. Sin, C. Sonnenschein and P.-j. Yi 
for discussions, correspondences and comments. 
We are indebted to Yaron Oz for the collaboration at the early stage and 
for valuable comments on the draft.  
Various parts of the paper have been developed through his numerous advices. 
We thank to the members of the CQUeST for their warm hospitality and to the organizers of 
``APCTP Focus Program : Holography at LHC," Aug. 1-10, 2012 at APCTP, Pohang, Korea.  
Some of the results were presented there, and the author received valuable comments. 
We are supported in part by the Centre of Excellence supported by the Israel Science 
Foundation (grant number 1468/06).

\appendix 

\section*{Appendix}

\section{Dimensional reduction of theories with Lifshitz scaling}    \label{sec:LifshitzReduction} 

In this section, we consider several different string theory solutions and their dimensional reductions 
(simple sphere reduction) of their compact coordinates. 
We compare the resulting backgrounds to the general form given 
in (\ref{BasicMetric}) to check whether we can get some useful information, which is universal 
over several different examples. In particular, all the worked out examples in this section 
has the metric structure given in (\ref{HyperscalingLifshitzMetric}) 
at finite temperature (if the finite temperature generalizations are available). 
For the convenience, we write the metric here again 
	\begin{align}       
		\mathrm d s^2_{d+2} &= r^{-2(d-\theta)/d} \left( - r^{-2(z-1)} f(r) \mathrm d t^2 
		+ \frac{ \mathrm d r^2}{f(r)} + \sum_{i=1}^d  \mathrm d x_i^2 \right) \;, \qquad 
		f(r) = 1 - \left(\frac{r}{r_H} \right)^{d+z-\theta} \;.  \nonumber 
	\end{align} 
The examples we consider here include the Dp branes, M2, M5 branes, Dp-Dq and 
intersecting M-brane systems.

\subsection{Near extremal Black Dp branes}   \label{sec:BlackDpLif}

Dimension reduction of Dp branes in the context of hyperscaling violation are analyzed in \cite{kachru}. 
This serves as our prime examples. Here we add some more details.  
The 10 dimensional non-extremal Dp brane metric in string frame is 
	\begin{align} 
		&\mathrm d s_{Dp, str}^2 = H^{-1/2} \left( -f \mathrm d t^2 + \sum_{i=1}^{p}  \mathrm d x_i ^2  \right) 
		+ H^{1/2} \left( \frac{ \mathrm d \tilde  r^2}{f} + \tilde r^2 \mathrm d \Omega_{8-p}^2 \right) \;,  \\
		&H = 1 + \sinh^2 \beta \frac{\tilde r_H^{7-p}}{\tilde r^{7-p}}  \;, \qquad
		f = 1 - \frac{\tilde r_H^{7-p}}{\tilde r^{7-p}} \;, \nonumber \\
		&e^{\phi-\phi_0} = g_s H^{(3-p)/4} \;, \qquad 
		C_{01 \cdots p} = \coth \beta ~g_s^{-1} (1 - H^{-1}) \;. \nonumber 
	\end{align}
To analyze the thermodynamic properties and dimensional reduction later, 
we rewrite the metric in Einstein frame using $ \mathrm d s_E^2 
= e^{-\phi/2}\mathrm d s_{str}^2  = H^{\frac{p-3}{8}}\mathrm d s_{str}^2 $. 
	\begin{align} 
		\mathrm d s_{Dp, E}^2 &= H^{-(7-p)/8} \left( -f \mathrm d t^2 + \sum_{i=1}^{p}  \mathrm d x_i ^2  \right) 
		+ H^{(p+1)/8} \left( \frac{ \mathrm d \tilde  r^2}{f} + \tilde r^2 \mathrm d \Omega_{8-p}^2 \right) \;. 
	\end{align}	
The temperature and entropy of the geometry are given by
	\begin{align}     \label{TSforDpBrane}
		T = \frac{7-p}{4\pi \cosh \beta} \frac{1}{\tilde r_H} \;, \qquad 
		S = \frac{\Omega_{8-p}}{4G_{10}} V_p \cosh \beta  ~\tilde r_H^{8-p}    \;, 
	\end{align} 	
where $\Omega_{8-p} $ is the volume of the $S^{8-p} $, and we use 
$ H = 1 + \sinh^2 \beta \frac{\tilde r_H^{7-p}}{\tilde r^{7-p}} = \cosh^2 \beta $. 

We also consider the near extremal limit for a small temperature, 
$ H = 1 + \sinh^2 \beta \frac{\tilde r_H^{7-p}}{\tilde r^{7-p}} 
\rightarrow \frac{\tilde r_0^{7-p}}{\tilde r^{7-p}}$. 
The temperature and entropy of the geometry for the near extremal limit are given by
	\begin{align}     \label{TSforDpBraneNearExtremal}
		T &= \frac{f' (\tilde r_H)}{4\pi H(\tilde r_H)^{1/2}} 
		= \frac{7-p}{4\pi} \frac{\tilde r_H^{(5-p)/2}}{\tilde r_0^{(7-p)/2}}  
		\;, \\
		S &= \frac{V_p V(S^{8-p}) }{4G_{10}} H(\tilde r_H)^{1/2} \tilde r_H^{8-p} 
		= \frac{\Omega_{8-p}}{4G_{10}} V_p  \tilde r_0^{(7-p)/2} \tilde r_H^{(9-p)/2} \;, 
	\end{align} 
where we use the volume of unit $n$ sphere (surface area of unit $n+1$ dimensional ball)
$\Omega_n = \frac{2 \pi^{(n+1)/2}}{\Gamma ((n+1)/2)} $. 
Thus we check the temperature dependence of the entropy $S \sim T^{\frac{9-p}{5-p}}$ 
in the near extremal limit. 
	
With these basic information, we would like to compactify this theory on $S^{8-p} $ down to $p+2$ dimensions. 
Here the action has the form 
	\begin{align}
	S &= \frac{1}{2\kappa_{10}^2} \int \mathrm d ^{10} x \sqrt{-g}  \mathcal R + \cdots   \nonumber \\
	&=\frac{V(S^{8-p}) R^{8-p}}{2\kappa_{10}^2} \int \mathrm d ^{p+2} x \sqrt{-g_{p+2}} e^{-2 \Phi_{p+2}} 
	\left( \mathcal R_{p+2} + \cdots \right) \;,  
	\end{align}
where $R$ is some dimensionful constant to match the dimension. Here it is natural to take $R = \tilde r_0 $. 
In the main body, we assume that there is a dimensionful parameter to match the dimension after the 
dimensional reduction.  
$e^{-2 \Phi_{p+2}} $ represents $\tilde r$ dependence coming from the compactification, 
and the lower dimensional newton constant $\kappa_{p+2} $ can be expressed explicitly. 
They are 
	\begin{align}
		e^{-2 \Phi_{p+2}} = H^{(p+1)(8-p)/16} \left( \frac{\tilde r}{R} \right)^{8-p}   \;, \qquad
		\frac{1}{2\kappa_{p+2}^2} = \frac{V(S^{8-p}) R^{8-p}}{2\kappa_{10}^2} \;.
	\end{align}	
Note that $\Phi_{p+2}$ is not a dilaton. 	
Finally, using $\mathrm d s^2 \rightarrow  \left(e^{-2\Phi_{p+2}} \right)^{\frac{2}{p}}\mathrm d s^2 $, 
the Einstein frame metric has the following form 
	\begin{align} 
		\mathrm d s_{p, E}^2 
		&= H^{\frac{1}{p}} \left( \frac{\tilde r}{R} \right)^{\frac{2(8-p)}{p}} 
		\left( -f \mathrm d t^2 + \sum_{i=1}^{p}  \mathrm d x_i ^2 + H \frac{ \mathrm d \tilde  r^2}{f} \right) \;,
	\end{align}
By changing a variable $H^{1/2} \mathrm d \tilde r =\mathrm d r$, we can get the standard form. 

We explicitly consider the near extremal limit, $H \approx \frac{\tilde r_0^{7-p}}{\tilde r^{7-p}}$.  
Then we get  
	\begin{align} 
		\mathrm d s_{p, E}^2 &= \left( \frac{\tilde r_0^{7-p}}{\tilde r^{7-p}}\right)^{\frac{1}{p}}  \left( \frac{\tilde r}{R} \right)^{\frac{2(8-p)}{p}} 
		\left( -f \mathrm d t^2 + \sum_{i=1}^{p}  \mathrm d x_i ^2+\frac{\tilde r_0^{7-p}}{\tilde r^{7-p}}\frac{ \mathrm d \tilde  r^2}{f}\right) \;. 
	\end{align}
Finally, by changing a variable 
$\tilde r = \left(\frac{ (p-5)^2 \tilde r_0^{p-7}}{4} \right)^{1/(p-5)} r^{\frac{2}{p-5}} $, we arrive at 
	\begin{align}	
		\mathrm d s_{p, E}^2  &= a~ r^{\frac{2(9-p)}{p(p-5)}} 
		\left( -f \mathrm d t^2 + \sum_{i=1}^{p}  \mathrm d x_i ^2 + \frac{\mathrm d r^2}{f} \right)  \;, \qquad
		f = 1 - \left( \frac{r}{r_H} \right)^{-\frac{2(7-p)}{p-5}}  \;, 
	\end{align}
where $ a= \left(\frac{p-5}{2}\right)^{\frac{2}{p-5}} R^{-\frac{2(8-p)}{p}} 
\left( \tilde r_0 \right)^{\frac{5(p-7)}{p(p-5)}} $ 
and the $r$ dependent factor in emblackening factor can be identified as $ -2(7-p)/(p-5) = {d+z-\theta}$. 	
Compared to the metric (\ref{HyperscalingLifshitzMetric}), we conclude 
	\begin{align}    \label{ExponentsForDp}
		d=p\;, \qquad \theta = \frac{(p-3)^2}{p-5} \;, \qquad z = 1 \;. 
	\end{align}
This case has a positive dynamical exponent $z=1$ as we expected.  
The corresponding temperature and entropy of the reduced effective geometry are 
	\begin{align}     \label{TSforDpReduction}
		T &= \frac{f' (r_H)}{4\pi} = \frac{|d+z-\theta |}{4\pi r_H} 
		= \frac{1}{2\pi r_H} \bigg|\frac{7-p}{5-p} \bigg| \;,  \nonumber \\
		S &= \frac{1}{4G_{p}} a^{p/2} V_p r_H^{-\frac{9-p}{5-p}} 
		\sim  T^{\frac{9-p}{5-p}} = T^{\frac{d-\theta}{z}} \;. 
	\end{align} 
The temperature and entropy (\ref{TSforDpReduction}) are the same as those evaluated in 
(\ref{TSforDpBraneNearExtremal}) if one correctly identifies the parameters before and 
after the dimensional reduction.

\subsection{Black M2 brane}   \label{sec:M2Lif}

Similarly we consider the dimensional reduction of the black M2 brane solution and 
its dimensional reduction to 4 dimensions. The 11 dimensional M2 brane metric is 
	\begin{align}    \label{11dM2Metric}
		&S = \frac{1}{2\kappa_{11}^2} \int \mathrm d ^{11} x \sqrt{-g} \mathcal R  
		- \frac{1}{4\kappa_{11}^2} \int \left(F_4 \wedge * F_4 +\frac{1}{3} A_3 \wedge F_4 \wedge F_4 \right) \;, 
		\nonumber \\
		&\mathrm d s_{M2}^2 = H^{-2/3} \left( -f \mathrm d t^2 + \sum_{i=1}^{2}  \mathrm d x_i ^2  \right) 
		+ H^{1/3} \left( \frac{ \mathrm d \tilde  r^2}{f} + \tilde r^2 \mathrm d \Omega_7^2 \right) \;, \\
		&H = 1 + \frac{R^6}{\tilde r^6}  \;, \quad
		f = 1 - \frac{\tilde r_H^6}{\tilde r^6} \;, \quad *F_4 = F_7 = 6 R^6 ~V (S^7)   \;. \nonumber 
	\end{align}
This metric is explicitly given in \cite{Aharony:1999ti}. 
The quantization condition is $ R^9 \pi^5 = \sqrt{2} N^{3/2} \kappa_{11}^2 $. 
The temperature and entropy of the geometry are 
	\begin{align}
		T = \frac{3}{2\pi} \frac{\tilde r_H^{2}}{R^{3}} \;, \qquad 
		S = \frac{8\sqrt{2} \pi^2}{27} V_2 N^{3/2} T^2 \;, 
	\end{align} 	
where we use the explicit expression $V (S^7) = \frac{\pi^4}{3} $. 
	
We compactify the solution on $S^{7} $. Typically for the conformal cases, the dimensional reduction 
does not introduce any extra radial dependence on the field theory coordinates.  
Thus the resulting action becomes 
$S =\frac{1}{2\kappa_{4}^2} \int \mathrm d ^{4} x \sqrt{-g_4} \mathcal R_4 + \cdots $ with 
$\frac{1}{2\kappa_4^2} = \frac{R^7 V(S^7)}{2\kappa_{11}^2} $. 
And the corresponding Einstein frame metric is nothing but the original metric (\ref{11dM2Metric}) 
without the last term $\mathrm d \Omega_7^2 $. 
By changing a variable $\tilde r = \frac{1}{\sqrt{2}} R^{3/2} r^{-1/2} $, we get  
	\begin{align} 
		\mathrm d s_{E}^2 &= \frac{R^2}{4 r^2} \left(- f \mathrm d t^2 + \sum_{i=1}^{2}  \mathrm d x_i^2 +   \frac{\mathrm d r^2}{f}  \right) \;, 
		\qquad f = 1 - \frac{r^3}{r_H^3} \;.  
	\end{align}
Compared to the metric (\ref{HyperscalingLifshitzMetric}), we conclude 
	\begin{align}
		d=2\;, \qquad \theta = 0 \;, \qquad z = 1 \;, 
	\end{align}
which show $z=1$ and $\theta=0$ as we expected.  
The corresponding temperature and entropy of the reduced effective geometry are 
	\begin{align}
		T = \frac{3}{4\pi r_H} \;, \qquad 
		S = \frac{\pi^2 R^2}{9 G_4} V_2  T^2  \;, 
	\end{align} 
where the temperature and the entropy are the same as those of the original M2 brane 
if we consider the relation $\tilde r_H = \frac{1}{\sqrt{2}} R^{3/2} r_H^{-1/2} $ 
and use the reduction $\frac{1}{2\kappa_4^2} = \frac{1}{16 \pi G_4} = \frac{R^7 V_7}{2\kappa_{11}^2} $.

\subsection{Black M5 brane}   \label{sec:M5Lif}

The 11 dimensional M5 brane action and metric are 
	\begin{align} 
		&S = \frac{1}{2\kappa_{11}} \int \mathrm d ^{11} x \sqrt{-g} \mathcal R  
		- \frac{1}{4\kappa_{11}} \int \left(F_4 \wedge * F_4 
		+\frac{1}{3} A_3 \wedge F_4 \wedge F_4 \right) \;, \\
		&\mathrm d s_{M5}^2 = H^{-1/3} \left( -f \mathrm d t^2 + \sum_{i=1}^{5}  \mathrm d x_i ^2  \right) 
		+ H^{2/3} \left(\frac{ \mathrm d \tilde  r^2}{f} + \tilde r^2 \mathrm d \Omega_4^2 \right) \;, \\
		&H = 1 + \frac{R^3}{\tilde r^3} \quad \rightarrow \quad  \frac{R^3}{\tilde r^3} \;, \quad 
		f = 1 -\frac{\tilde r_H^3}{\tilde r^3} \;, \quad F_4 = d A_3 = 3 R^3 V(S^4)  \;. \nonumber 
	\end{align}
The temperature and entropy of the geometry are 
	\begin{align}
		T = \frac{3}{4\pi} \frac{\tilde r_H^{1/2}}{R^{3/2}}  \;, \qquad 
		S = \frac{2^7 \pi^3}{3^6} V_5 N^3 T^5 \;, 
	\end{align} 	
where we use $ V (S^4) = \frac{8 \pi^2}{3} $ and the quantization condition 
$ R^9 \pi^5 2^7 =  N^{3} \kappa_{11}^2 $. 

Upon compactifying down to $7$ dimensions, the action becomes 
$S =  \frac{1}{2\kappa_{7}^2}  \int \mathrm d ^{7} x \sqrt{-g_7} \mathcal R_7 + \cdots $ with
$ \frac{1}{2\kappa_{7}^2} = \frac{R^4 V(S^4)}{2\kappa_{11}^2}$. Similar to the M2 brane, 
the resulting metric becomes 
	\begin{align} 
		\mathrm d s_{E}^2 &= \frac{4 R^2}{r^2} \left(-f \mathrm d t^2 + \sum_{i=1}^{5}  \mathrm d x_i^2 
		+ \frac{\mathrm d r^2}{f}  \right) \;, 
		\qquad f = 1 -\frac{r^6}{r_H^6} \;, 
	\end{align}
after changing a variable $\tilde r = 4R^3 r^{-2} $. 
Compared to the standard form of the metric (\ref{HyperscalingLifshitzMetric}), we conclude 
	\begin{align}
		d=5 \;, \qquad \theta = 0 \;, \qquad z = 1 \;. 
	\end{align}
This case has a positive dynamical exponent $z=1$ and $\theta=0$ as we expected.  
The corresponding temperature and entropy of the reduced effective geometry are 
	\begin{align}
		T = \frac{6}{4\pi r_H} \;, \qquad 
		S = \frac{2^8 \pi^5 R^5}{3^5 G_7} V_5  T^5  \;, 
	\end{align} 
where the temperature and the entropy are the same as those of the original M5 brane 
if we consider the relation $\tilde r_H = 4R^3 r_H^{-2}  $ 
and use the reduction $\frac{1}{2\kappa_7^2} =\frac{R^5 V(S^5)}{16 \pi G_{11}} $.

\subsection{Dp-D(p+4) branes}    \label{sec:DpDqSystem}

In this section we would like to consider the intersecting Dp-Dq, $ q=p+4, p=0,1,2 $ solutions 
\cite{Horowitz:1996ay}\cite{Tseytlin:1997cs}\cite{Maldacena:2001km}, 
studied recently in the context of the hyperscaling violation solution \cite{Dey:2012rs}. 
	\begin{align} 
		&\mathrm d s_{pq}^2 = \frac{1}{ \sqrt{H_pH_q} } [- \mathrm d t^2 + \sum_{i=1}^{p} \mathrm d x_i^2 ]  
		+ \sqrt{\frac{H_p}{H_q}} \sum_{j = p+1}^{p+4} \mathrm d x_j^2 
		+ \sqrt{H_p H_q}\left[ \mathrm d \rho^2 + \rho^2 \mathrm d \Omega_{4-p}^2 \right] \;,  \\		 
		&e^{\Phi} = H_p^{\frac{3-p}{4}} H_q^{-\frac{p+1}{4}}   \;, \quad 
		H_{p,q} = 1- \frac{Q_{p,q}}{\rho^{3-p}} \;, \\
		&A_{p+1} = (1- H_p^{-1}) \mathrm d t \wedge \mathrm d x^1 \wedge \cdots \wedge \mathrm  d x^p \;, \\
		&A_{p+5} = (1- H_q^{-1}) \mathrm d t \wedge \mathrm d x^1 \wedge \cdots \wedge \mathrm  d x^{p+4} \;,  
	\end{align}
We would like to take the near horizon limit first, $ H_p H_q =H^2 = \frac{Q^2}{\rho^{2(3-p)}}, Q^2 =Q_p Q_q$, 
and dimensionally reduce to the Einstein metric to $ p+6 $ dimensions, using 
$ \mathrm d s_E^2=  H^{\frac{p+2}{p+4}} \rho^{\frac{2(4-p)}{p+4}} \mathrm d s_{str}^2 $. 
	\begin{align} 
		\mathrm d s_{pq, E}^2 
		&= Q_p^{\frac{-3}{p+4}} Q_p^{\frac{2}{p+4}} \rho^{\frac{14-4p}{p+4}} 
		\left( - \mathrm d t^2 + \sum_{i=1}^{p} \mathrm d x_i^2   
		+  \frac{Q_p}{\rho^{3-p}} \sum_{j = p+1}^{p+4} \mathrm d x_j^2 
		+  \frac{Q_p Q_q}{\rho^{2(3-p)}} \mathrm d \rho^2 \right)		\;. 
	\end{align}	
Using $ \rho = \left( \frac{Q}{2-p} \right)^{\frac{1}{2-p}} u^{\frac{1}{2-p}} $ for $p<2 $, we obtain 
	\begin{align} 
		\mathrm d s_{pq, E}^2 
		&=  u^{\frac{14-4p}{(p+4)(2-p)}} 
		\left( - \mathrm d t^2 + \sum_{i=1}^{p} \mathrm d x_i^2   
		+ u^{\frac{p-3}{2-p}} \sum_{j = 1}^{4} \mathrm d x_j^2 + \frac{\mathrm d u^2}{u^4} \right) \;, 
	\end{align}	
where we omit the overall factor $Q_p^{\frac{-3}{p+4}} Q_p^{\frac{2}{p+4}} 
\left( \frac{Q}{2-p} \right)^{\frac{14-4p}{(2-p)(p+4)}} $ and absorb 
$  Q_p \left( \frac{Q}{2-p} \right)^{\frac{p-3}{2-p}} $ factors to $ x_j $. 
This metric reveals that spatial anisotropic exponent is necessary to describe the string theory 
solutions with two dimensionful parameters. 
Compared to the standard metric (\ref{BasicMetricU}), we conclude 
\begin{align}
		d=p+4 \;, \qquad \theta = \frac{1-p^2}{2-p} \;, \qquad z = 1 \;, \qquad   \sharp = \frac{1-p}{2(2-p)} \;. 
\end{align}
This is one of the main example to reveal the spatial anisotropic exponent advertised in the introduction. 
Surely this example has two independent dimensionful parameters. Effectively, one can be absorbed in the 
front factor to the hyperscaling violation exponent, while the other one is pushed into some part of the 
spatial coordinates with spatial anisotropic exponent. 
Until now we consider the dimensional reduction to $ p+6 $ dimensions to show the manifestation 
of the advertised spatial anisotropic exponent. 

Now for the coordinates $ (p+1) - (p+4) $ are compact, 
this metric can be further dimensionally reduced down to $ p+2 $ dimensions. 
Similarly, we obtain 
	\begin{align} 
		\mathrm d s_{p, E}^2 
		&= (Q_p Q_q)^{\frac{2}{p(2-p)}} \left( \frac{1}{2-p} \right)^{\frac{2}{p(2-p)}} u^{\frac{2}{p(2-p)}} 
		\left( - \mathrm d t^2 + \sum_{i=1}^{p} \mathrm d x_i^2   
		+  \frac{\mathrm d u^2}{u^4} \right) \;. 
	\end{align}	
Comparing to the standard metric (\ref{BasicMetricU}), we conclude 
\begin{align}
		d=p \;, \qquad \theta = -\frac{(1-p)^2}{2-p} \;, \qquad z = 1 \;. 
\end{align}
Considering the finite temperature generalization with the emblackening factor
\begin{align}
	f = 1 - \frac{\rho_H^{3-p}}{\rho^{3-p}} = 1 - \left( \frac{u_H}{u} \right)^{\frac{3-p}{2-p}} \;, 
\end{align}
we again check that $ d+z-\theta = \frac{3-p}{2-p} $.

\smallskip
{\it D1D5 system}

Let us concentrate on the conformal case with $ p=1 $.
This geometry shows an interesting property, which can be clearly seen in the following form  
	\begin{align} 
		&\mathrm d s_{D1D5}^2 = \frac{\rho^2}{ Q } [- f \mathrm d t^2 + \mathrm d x_1^2 ]
		+ \frac{ Q }{\rho^2}  \frac{\mathrm d \rho^2}{f} 
		+ Q  \mathrm d \Omega_3^2   
		+ \sqrt{\frac{Q_1}{Q_5}} \mathrm ds_{M_4}^2 \;, 
	\end{align}
which is $ AdS_3 \times S^3 \times M_4 $. 	
We check that there is no hyperscaling violation when we take dimensional reductions 
along the directions $S^3 $. 
	\begin{align}   \label{D1D5metricEffective}
		&\mathrm d s_{D1D5}^2 = Q \rho^2 [- f \mathrm d t^2 + \mathrm d x_1^2 
		+ \frac{ 1}{\rho^4}  \frac{\mathrm d \rho^2}{f} 
		+ \frac{Q_1}{\rho^2} \mathrm ds_{M_4}^2 ] \;, 		
	\end{align}
which is already in a standard form because the boundary sits at $ u \rightarrow \infty $.
Thus we get 
\begin{align}
		d=5 \;, \qquad \theta = 0 \;, \qquad z = 1 \;, \qquad   \sharp_{M4|1} = 0 \;. 
\end{align}
Note that normally we expect to get $ \sharp = 1 $ to have rotationally invariant metric. 
Thus $ \sharp_{M4|1} = 0 $ reveals the spatial anisotropy. For the effective metric 
$ AdS_3 \times M_4 $, we observe a different scale between the two spaces $ AdS_3 $ and $ M_4 $. 
This is manifested in the prefactors $ 1 $ and $ \frac{Q_1}{\rho^2} $ in (\ref{D1D5metricEffective}). 
This picture would be applicable for non-compact $ M_4 $ as well as compact $ M_4 $ for 
the physical cases requiring compact coordinates. 

Now let us take the dimensional reductions further along the directions $M_4 $ 
for the compact $M_4 =T^4 $. Then we arrive ($u=\rho $) 
	\begin{align} 
		\mathrm d s_E^2 
		&\propto \rho_1 \rho_5 u^2\left( - f \mathrm d t^2 + \mathrm d x_1^2 
		+ \frac{ 1 }{u^4}  \frac{\mathrm d u^2}{f} \right)  \;,
	\end{align}
where a single scale $\rho_1 \rho_5 $ is absorbed into the coordinates $t, x_1 $. 	
It is clear that the boundary is at $\rho \rightarrow \infty $ from the coordinate $x_1 $, 
and thus the corresponding exponents are $\theta=0, z=1 $ for $d=1 $. 
It is worthwhile to mention that this case falls into an interesting case 
$\theta =d-1 $ for the Lifshitz type theories trivially. 
As already mentioned, this case falls into the conformal case and we don't 
expect to have any pysically relevant scale. 

This system still satisfy the relation $d+z-\theta =2 $, which is the power behavior of the 
emblackening factor. Note that this system is related to a conformal case, which is signified by 
$\theta =0 $, in contrast with the case of D1 brane $\theta=-1, z=1$ for $ d=1 $ 
given in (\ref{ExponentsForDp}).

\subsection{Intersecting and interpolating M-branes}    \label{sec:MbrnaeSystem}
 
In this section we would like to consider an intersecting M2-M5 solution \cite{Boonstra:1997dy} and 
an interpolating background between $ 2\bot 2 $ and $ 2 \bot 5 $ \cite{Tseytlin:1997cs}. 
Some of them are considered recently in \cite{Dey:2012tg}. 
Here we observe that these intersecting and interpolating solutions reveal the necessity of the 
spatial anisotropic exponent(s) $ \sharp $. 

\smallskip 
{\it Intersecting solution}

Let us first start with the intersecting M2-M5 solution \cite{Boonstra:1997dy}. 
	\begin{align} 
		\mathrm d s_{M2M5}^2 &= H_2^{-2/3} H_5^{-1/3} \left(- \mathrm d t^2 + \mathrm d x_1^2 \right) 
		+ H_2^{-2/3} H_5^{2/3} \left( \mathrm d x_2^2 \right)    \nonumber \\
		&+ H_2^{1/3} H_5^{-1/3} \left( \mathrm d x_3^2 + \cdots  +  \mathrm d x_6^2 \right) 
		+H_2^{1/3} H_5^{2/3} \left[ \mathrm d \rho^2 + \rho^2 \mathrm d \Omega_{3}^2 \right] \;,  \\		
		H_{p,q} &= 1 + \frac{Q_{p,q}}{\rho^{2}} \;, \quad 
		F_{r012} = \pm \partial_\rho H_2^{-1} \;, \quad 
		F_{r\alpha\beta\gamma} = \pm \epsilon_{\alpha\beta\gamma} \partial_\rho H_5\;. 
	\end{align}
After taking the near horizon limit, we get 
	\begin{align} 
		\mathrm d s_{M2M5}^2 &= Q_2^{-2/3} Q_5^{-1/3} \rho^2 \left(- \mathrm d t^2 + \mathrm d x_1^2 \right) 
		+ Q_2^{-2/3} Q_5^{2/3} \left( \mathrm d x_2^2 \right)    \nonumber \\
		&+ Q_2^{1/3} Q_5^{-1/3} \left( \mathrm d x_3^2 + \cdots  +  \mathrm d x_6^2 \right) 
		+ Q_2^{1/3} Q_5^{2/3} \rho^{-2} \left[ \mathrm d \rho^2 + \rho^2 \mathrm d \Omega_{3}^2 \right] \;.  
	\end{align}
Here the geometry is factorized into the product of an AdS$_3$ spacetime, a three-sphere $ S^3 $, 
and a flat Euclidean five-dimensional space $ E^5 $ \cite{Boonstra:1997dy}. 	
	
We dimensionally reduce the compact coordinates to the Einstein metric in $ 8 $ dimensions. 
Using $ \mathrm d s_E^2=  (Q_2^{1/2} Q_5)^{1/3} \mathrm d s_{str}^2 $, we get   
	\begin{align}    \label{IntersectingBrane}
		\mathrm d s_{25,E}^2 
		&= 	Q_2^{1/2} Q_5  \rho^2 \left(- \mathrm d t^2 + \mathrm d x_1^2 
		+ \frac{1}{\rho^{2}} \left( \mathrm d x_2^2 + \mathrm d x_3^2 + \cdots  +  \mathrm d x_6^2 \right) 
		+ \frac{ \mathrm d \rho^2 }{\rho^{4} } \right) \;,		  
	\end{align}	
where we absorb $ Q $ factors in the spatial coordinates $ x_2, \cdots x_6 $ and redefine 
$ \rho \rightarrow \sqrt{Q_2 Q_5} \rho $. 
This is already in a standard form and we can read off the exponents as 
\begin{align}
		d=7 \;, \qquad \theta = 0 \;, \qquad z = 1 \;, \qquad   \sharp = 0 \;. 
\end{align}
The fact $ \theta = 0 $ is consistent with the observation that the near horizon geometry is 
AdS$_3 \times E^5 \times S^3$. Note that $ \sharp =0 $ is non-trivial, and provide an example 
to generate the spatial anisotropic exponent. 
There are other examples listed in \cite{Boonstra:1997dy}. We expect to get the similar results.

\smallskip 
{\it Interpolating background}

Now, let us turn to the interpolating background between $ 2\bot 2 $ and $ 2 \bot 5 $ 
\cite{Tseytlin:1997cs}. 
	\begin{align} 
		\mathrm d s^2 &= \tilde H_3^{1/3} H_3^{1/3} H_1^{1/3} 
		\left[ - H_1^{-1} H_3^{-1} \mathrm d t^2 + H_1^{-1} \mathrm d x_1^2 
		+ H_3^{-1} ( \mathrm d x_2^2 + \mathrm d x_3^2 )   \right. \nonumber \\
		&\left. + \tilde H_3^{-1} H_3^{-1}  \mathrm d x_5^2 
		+ \tilde H_3^{-1} ( \mathrm d x_4^2 + \mathrm d x_{11}^2 ) 
		+ \mathrm d \rho^2 + \rho^2 \mathrm d\Omega_3^2 \right] \;, 
	\end{align}
where $ H_i = 1 + \frac{Q_i}{\rho^{2}} $ and  $ \tilde H_3 = 1 + \frac{\tilde Q_3}{\rho^{2}} $. 
After taking the near horizon limit, we get 
	\begin{align} 
		\mathrm d s^2 &= \frac{\tilde Q_3^{1/3} Q_3^{1/3} Q_1^{1/3} }{\rho^2}
		\left[ - \frac{\rho^4}{Q_1 Q_3} \mathrm d t^2 + \frac{\rho^2}{Q_1} \mathrm d x_1^2 
		+ \frac{\rho^2}{Q_3} ( \mathrm d x_2^2 + \mathrm d x_3^2 )   \right. \nonumber \\
		&\left. +  \frac{\rho^4}{\tilde Q_3 Q_3} \mathrm d x_5^2 
		+ \frac{\rho^2}{\tilde Q_3} ( \mathrm d x_4^2 + \mathrm d x_{11}^2 ) 
		+ \mathrm d \rho^2 + \rho^2 \mathrm d\Omega_3^2 \right] \;, 
	\end{align}
We dimensionally reduce the compact coordinates to the Einstein metric in $ 8 $ dimensions. 
Using $ \mathrm d s_E^2=  (Q_1 Q_3 \tilde Q_3)^{1/6} \mathrm d s_{str}^2 $, we get   
	\begin{align} 
		\mathrm d s_{25,E}^2 
		&= 	(Q_1 Q_3 \tilde Q_3)^{1/2} \rho^{2} 
		\left[ -  \mathrm d t^2 +  \mathrm d x_5^2 
		+ \frac{1}{\rho^2} ( \mathrm d x_2^2 + \mathrm d x_3^2 
		+ \mathrm d x_4^2 + \mathrm d x_{11}^2 ) 
		+ \frac{ \mathrm d \rho^2}{\rho^4}  \right]  \;,		  
	\end{align}	
where we absorb $ Q $ factors in the coordinates $t, x_1, \cdots x_5, x_{11} $. 
Thus we check this metric is the same as (\ref{IntersectingBrane}) and thus 
\begin{align}
		d=7 \;, \qquad \theta = 0 \;, \qquad z = 1 \;, \qquad   \sharp_{i|1} = 0 \;, 
\end{align}
where $ i=2, \cdots, 6 $. 
Again, note that $ \sharp =0 $ is non-trivial, and provide an example 
to generate the spatial anisotropic exponent.

\section{Dimensional reduction of solutions with Schr\"odinger scaling}  \label{sec:SchrReduction}

Let us consider the dimensional reduction of the theories with Schr\"odinger scaling. 
These are mostly at zero temperature. 
They are constructed by null Melvin twist \cite{Alishahiha:2003ru}\cite{Gimon:2003xk}, 
which serves as an effective way to generate non-relativistic solutions with Schr\"odinger isometry, 
starting from the known relativistic solutions. 
Most of the solutions we consider are listed in \cite{Mazzucato:2008tr}.%
\footnote{We are grateful to Yaron Oz for his numerous advices and valuable comments 
especially on this section.  
} 
Non-relativistic Dp branes are analyzed in detail in the main body, including their finite 
temperature generalizations \S \ref{sec:Schroedinger}.  

There exists an interesting solution, type IIA NS5 brane in \S \ref{sec:SchrNS5A}, 
among the solutions we consider, which satisfies the following condition  
	\begin{align}
		\frac{\theta}{D} = \frac{d+1-z}{d+1} \;, \qquad D=d+1=5 \;, \qquad  \theta = 3 \;. 
	\end{align} 
This solution is expected to posses Fermi surfaces in the dual field theory in the context of 
Schr\"odinger type theories \cite{kimHyperscaling}, similar to the Lifshitz type theories 
\cite{Ogawa:2011bz}\cite{Huijse:2011ef}\cite{kachru}. This is the first explicit example belong 
to this class from the top down string theory solutions with Schr\"odinger isometry.

\subsection{`Conformal' cases}   \label{sec:SchrConformalCases}

We consider the null Melvin twist of the conformal branes including M2, M5, D3, D1D5, F1NS5 systems  
\cite{Mazzucato:2008tr}. All the non-relativistic conformal branes share the property that 
the part of the metric along the internal directions are independent of the radial coordinates, and 
the associated dilaton is either not present or constant.  
Thus we can write a general form of the metric at zero temperature as  
	\begin{align}    \label{NRConformalMetric}
		\mathrm d s^2 &= \left(\frac{\rho_{p}}{r} \right)^2 \left( - \frac{\tilde \Delta^2}{r^{2z-2}} 
		\mathrm d t^2 -2 \mathrm d t \mathrm d \xi 
		+ \sum_{i=1}^{p-1} \eta_i \mathrm d x_i^2 +  \mathrm d r^2 \right) + \rho_p^2 \mathrm d \Omega^2 \;, 
	\end{align} 
where $d=p-1 $. For example, 
$\rho_{p} = \rho_2, ~\tilde \Delta = 2 \beta \rho_2$, $p=2 $, $\eta_i=1 $ and 
$ d \Omega^2 = d \Omega_7^2 $ for non-relativistic M2 brane, 
while  $\rho_{p}^2 = \rho_1 \rho_5, \tilde \Delta = \beta \rho_1 \rho_5$, 
$p=5 $, $\eta_i= r^2/\rho_5^2  $ and  $ d \Omega^2 = d \Omega_3^2  $ 
for D1D5 system. We observe a clear distinction between these two cases: 
D1D5 has explicit radial dependence in $\eta_i $, which can not be removed by redefining the 
radial coordinate. 
	
The dimensional reduction along the coordinates $ d \Omega^2$ does not produce any radial dependence. 
Thus the resulting metric is the same as before, which is (\ref{NRConformalMetric}) without the last term. 
We list various exponents of these cases by comparing to the standard form (\ref{BasicMetric}) 
with $\mathrm a =1 $, $\mathrm b=1 $ and $ D=d+1 $ 
(after absorbing the dimensionful parameter $\tilde \Delta $ into $dt $). 
	\begin{align}
		\text{D3 : }& \qquad  d=2\;, \qquad\qquad \theta = 0 \;, \qquad z = 2 \;, \\
		\text{M2 : }& \qquad  d=1\;, \qquad\qquad \theta = 0 \;, \qquad z = 3/2 \;, \\		
		\text{M5 : }& \qquad  d=4\;, \qquad\qquad \theta = 0 \;, \qquad z = 3 \;, \\		
		\text{D1D5 : }& \qquad  d=0 + 4\;, \qquad~ \theta = 0 \;, \qquad z = 2 \;, \label{D1D5Schr} \\		
		\text{F1NS5 : }& \qquad d= 0 + 4\;, \qquad~ \theta = 0 \;, \qquad z = 2 \;. \label{F1D5Schr}
	\end{align}
All the cases reveal $\theta = 0 $. M2 and M5 cases reveal interesting variations 
in dynamical exponents and do not have special non-relativistic conformal symmetry, which is clear 
from the non-trivial dynamical exponent. It will be interesting to investigate these cases further 
from this point of view.  

The last two cases, (\ref{D1D5Schr}) and (\ref{F1D5Schr}), are different 
because the spatial dimensions have explicit radial dependence. 
After dimensional reduction, the metric is  
	\begin{align} 
		\mathrm d s_{D1D5}^2 &= \left(\frac{\rho_{p}}{r} \right)^2 \left( - \frac{\tilde \Delta^2}{r^{2}} \mathrm d t^2 -2 \mathrm d t \mathrm d \xi 
		+  \mathrm d r^2 \right) +\mathrm d s_{M_4}^2  \;.
	\end{align}
To make this metric in the form (\ref{BasicMetric}) with $\eta_i =1 $, it is required to change coordinate 
with non-polynomial type, $r \sim e^{y}$. Thus dynamical exponent is not determined. 
Thus this metric has neither the Galilean boost, nor the special conformal transformations.  
To have nontrivial $ \sharp $, we need more general intersecting D-branes. 

The finite temperature generalizations for D1D5 and F1NS5 systems can be considered similarly. 
These metrics generate the `$K $' factor similar to (\ref{dimenReduSchrMetricADM}), 
which might spoil some of the properties of the zero temperature. 
It turns out that the K-factors become constant in the near horizon limit \cite{Mazzucato:2008tr}, 
which is special for this case.

\subsection{NS5A brane}    \label{sec:SchrNS5A}

In this section, we consider the non-relativistic version of the type IIA NS5 brane solution at 
zero temperature in some detail. This case turns out to be a very interesting case, 
being expected to possess Fermi surface according to recent conjecture 
\cite{Ogawa:2011bz}\cite{Huijse:2011ef}. 

The metric and dilaton are given in \cite{Mazzucato:2008tr}
	\begin{align} 
		\mathrm d s^2 &= - \frac{2 \Delta^2}{ \tilde r^2} \mathrm d t^2 -2 \mathrm d t \mathrm d \xi 
		+ \sum_{i=1}^{4}  \mathrm d x_i^2 + \frac{\rho_5}{\tilde r} \left(  \mathrm d \tilde  r^2 
		+ \tilde r^2 \mathrm d \Omega_3 ^2 \right) \;, 	 
		\qquad e^{\Phi} = \left( \frac{\tilde r}{\rho_5}\right)^{3/2}   \;, 
	\end{align}
where we omit the other fields. 
We compactify this solution on $S^{3}$ and get the Einstein metric in $7$ dimensions 
	\begin{align}   \label{NS5AReducedMetric}
		\mathrm d s_{E}^2 &= 4^{3/5} \rho_5 ^{12/5} r^{-6/5} \left( - \frac{\tilde \Delta^2}{r^4} \mathrm d t^2 
		-2 \mathrm d t \mathrm d \xi 
		+ \sum_{i=1}^{4}  \mathrm d x_i^2 +  \mathrm d r^2 \right) \;, 
	\end{align}
where we use $d s_E^2 = \rho_5 ^{9/5} \tilde r^{-3/5} \mathrm d s^2 $, 
$\tilde r = \frac{r^2}{4 \rho_5} $ and $\tilde \Delta = 32 \rho_5^2 \Delta^2$. 
Compared to the standard metric (\ref{BasicMetric}), we get  
	\begin{align}
		d=4 \;, \qquad \theta = 2 \;, \qquad z = 3 \;,
	\end{align}
which belongs to the category of logarithmic violation case with the condition 
	\begin{align}
		\frac{\theta}{D} = \frac{d+1-z}{d+1} \;, 
	\end{align}
where $ D=d+1 $.	
What is so special about this metric? It turns out that the holographic stress energy tensor has 
some distinctive signature along with the entanglement entropy. 
We further analyze this case in some detail here.

\smallskip  
{\it  Entanglement entropy}

Entanglement entropy has become a new useful tool to classify and understand different phases of
field theory. For example, it differentiates the fermionic models from the bosonic ones 
\cite{Eisert:2008ur}. 
This is extensively investigated in the context of holography \cite{HoloEntanglement1}. 

The entanglement entropy can be computed using the minimal surface prescription of the `codimension 2 
holography' utilizing the stationary ADM form, developed in \cite{kimHyperscaling} 
	\begin{align} 
		\mathrm d s_{E}^2 &= 4^{3/5} \rho_5 ^{12/5} r^{-6/5} \left( - \frac{\tilde \Delta^2}{r^4} 
		\left(\mathrm d t +  \frac{r^4}{\tilde \Delta^2} \mathrm d \xi \right)^2  + \frac{r^4}{\tilde \Delta^2} \mathrm d \xi ^2
		+ \sum_{i=1}^{4}  \mathrm d x_i^2 +  \mathrm d r^2 \right) \;, 
	\end{align}
with the condition $\mathrm d t +  \frac{r^4}{\tilde \Delta^2} \mathrm d \xi =0$. 
The entanglement entropy is given by (see the details in \S 4 of \cite{kimHyperscaling}) 
	\begin{align}     
		\mathcal  S 
		&= \frac{M_{Pl}^{5}}{4 } \frac{8 \rho_5^6}{\tilde \Delta} 
		\left(\frac{L^{3}}{M_\xi} \right) \log \left(\frac{2l}{\epsilon} \right)  \;.
	\end{align}   
This result shows the logarithmic violation of entanglement entropy. 
This important property is associated with the conjecture \cite{Ogawa:2011bz}\cite{Huijse:2011ef} 
that ``the entanglement entropies of the system with Fermi surface show the logarithmic violation 
of the area law."

\smallskip  

{\it  Holographic stress energy tensor}

Holographic stress energy tensor in the context of Schr\"odinger holography is computed 
using Brown-York method in \S 2.3 of \cite{kimHyperscaling}. 
We summarize the result here (using the same notations of \cite{kimHyperscaling}).  
	\begin{align}     \label{HSETNS5ALog}
		\langle \hat \tau_{tt}\rangle = -\frac{1}{r_c^{1+z}} h_{00} \;,  \qquad
		\langle \hat \tau_{t\xi}\rangle =\langle \hat \tau_{\xi t}\rangle =-\frac{z}{r_c^{1+z}} h_{t\xi}  \;, \qquad	
		\langle \hat \tau_{ij}\rangle = -\frac{z}{r_c^{1+z}}  h_{ij}  \;.
	\end{align} 
In particular, the coefficient of $\langle \hat \tau_{tt}\rangle $ becomes unity and 
is independent of parameters, $z, \theta, d$ and $D $. 
Let us compare this to the extensive violation case $\theta = d+2-z $ for $D=d+1 $. Then 
	\begin{align}     \label{HSETNS5AExten}
		\langle \hat \tau_{tt}\rangle = 0 \;,   \qquad
		\langle \hat \tau_{t\xi}\rangle =\langle \hat \tau_{\xi t}\rangle =-\frac{z-1}{r_c^{z}} h_{t\xi} \;, \qquad		
		\langle \hat \tau_{ij}\rangle = -\frac{z-1}{r_c^{z}}  h_{ij}  \;.
	\end{align}   
We also observe the coefficient of $\langle \hat \tau_{tt}\rangle $ is independent of the parameters. 

These coefficients of the holographic stress energy tensor are independent of the the normalization 
and thus universal. Thus we expect that these provide some important properties in further investigating 
the physical significances of the theories for the range 
$d+1-z \leq \theta \leq d+2-z $ for $D=d+1 $.

\smallskip 

{\it  Scalar correlation functions}

The equation of motion for a scalar field with mass $m$ in the background (\ref{NS5AReducedMetric}) is 
given (in the momentum space) as 
	\begin{align}     \label{scalarFieldEqMom}
		\left(\partial_r^2-\frac{3}{r}\partial_r - \vec k^2+ 2 M \omega - \beta \frac{M^2}{ r^{4}} 
		-\frac{m^2}{r^{6/5}}\right)\phi=0 \;,
	\end{align}    
where $\vec k$ and $ \omega $ are Fourier transform of $\vec x$ and $t$, respectively. We treat the $\xi$ direction 
special and replace $\partial_\xi = i M$ for the scalar field following \cite{kimHyperscaling}. 

It is clear that the asymptotic expansion is not well posed with the polynomial form 
due to the term proportional to $M^2$. Many of the interesting cases have similar difficulties 
for computing the correlation functions. 
This can be interpreted as a signal that the geometry is not valid all the way to the boundary, 
thus the asymptotic expansion can not be trusted. Thus we would like to evaluate the semiclassical 
propagator.

\smallskip 
{\it  Semiclassical propagators}

Semiclassical propagators of the Schr\"odinger type backgrounds are studied in detail 
in \S 2.2 of \cite{kimHyperscaling}. 
For the Schr\"odinger type metric, there are three different types of propagator. 
All the results are valid in the limit where the second exponential factors are suppressed. 
 
One can evaluate the static semiclassical propagator as  
	\begin{align}    \label{staticPropagatorNS5A}
		G(\Delta x_i) &\sim  \exp \left[2m \frac{D}{\theta} ~ \epsilon^{\theta/D}\right]  
		 \exp \left[-m \frac{D}{\theta} c_{\theta/D} |\Delta x_i|^{\theta/D}\right] \;, 
	\end{align}     
where $ c_{\theta/D} $	only depends on the combination $\theta/D $ and given in equation 
(2.22) of \cite{kimHyperscaling}.
The static case only depends on the combination $\theta/D=2/5$, independent of other parameters. 
Note the non-trivial dependence of the propagator on the hyperscaling violation exponent $\theta $. 

Due to the cross term present in the metric (\ref{NS5AReducedMetric}), there is also a stationary 
propagator. In general, this case is rather complicated. This simplifies when we constrain the 
travel distance along the $\xi $ coordinate to be the total length $ L_\xi $, which is the defining 
length associated with the dual particle number $M_\xi$. Then  
	\begin{align}    \label{stationaryPropagatorNS5A}
		G(L_\xi) &\sim  \exp \left[2m \frac{D}{\theta} ~ \epsilon^{\theta/D}\right]  
		 \exp \left[-2m \frac{D}{\theta} \hat c_{L_\xi} L_\xi^{\frac{\theta}{(z-2) D}}\right] \;, 
	\end{align}
where $ \hat c_{L_\xi}$ is given in equation (2.33) of \cite{kimHyperscaling}. 
The correlation function can not decay faster than this because there exist the maximum distance 
$ L_\xi $ in $ \xi $ direction. This is a unique property of the Schr\"odinger type theories.

The timelike propagator is also complicated in general, which can be found in \S 2.2 of \cite{kimHyperscaling}. 
Thus we consider the special case, where the constant of motion along $\xi $ coordinate is small. 
The result of the timelike propagator is 
	\begin{align}        \label{timelikePropagatorNS5A}
		G(\Delta t) &\sim  \exp \left[2m \frac{D}{\theta} ~ \epsilon^{\theta/D}\right]  
		 \exp \left[-2m \frac{D}{\theta} c_{\xi} |\Delta t|^{\frac{\theta}{2D-\theta}}\right] \;, 
	\end{align}   
where $c_{\xi}$ is given in equation (2.43) of \cite{kimHyperscaling}. 

Typically the semiclassical propagators are exponentially suppressed in the valid regime, accordingly 
for $-2m |\Delta x_i|^{\theta/D} \gg 1$, $ -2m L_\xi^{\frac{\theta}{(z-2) D}} \gg 1 $ and 
$-2m |\Delta t|^{\frac{\theta}{2D-\theta}} \gg 1$. 
Still it is possible to get the standard form 
$ G \sim \exp \left[-m \left(\frac{|\Delta x_i|^2}{2 |\Delta t|} \right)^{\frac{\theta}{(z-2) D}} \right] $
for the general case \cite{kimHyperscaling}.  

Further investigations of this case would be very interesting to figure out 
whether these systems posses some physically distinctive properties.

\subsection{NS5B brane}    \label{sec:SchrNS5B}

Let us consider type IIB NS5 brane solution in 10 dimensions \cite{Mazzucato:2008tr}. 
The metric and dilaton are  
	\begin{align} 
		\mathrm d s_{NS5A}^2 &= -\frac{2 \Delta^2}{ \tilde r^2} \mathrm d t^2 -2 \mathrm d t \mathrm d \xi 
		+ \sum_{i=1}^{4}  \mathrm d x_i^2 + \frac{\rho_5}{\tilde r^2} \left(  \mathrm d \tilde  r^2 
		+ \tilde r^2 \mathrm d \Omega_3 ^2 \right) \;, \qquad 		 
		e^{\Phi} = \frac{\tilde r}{\rho_5}   \;, 
	\end{align}
where we omit other irrelevant fields again. 
We compactify this solution on $S^{3}$ down to $7$ dimensions. The resulting metric in Einstein frame is 
given by  
	\begin{align} 
		\mathrm d s_{E}^2 &\sim \tilde r^{-4/5} \left( - \frac{2 \Delta^2}{ \tilde r^2} \mathrm d t^2 
		-2 \mathrm d t \mathrm d \xi 
		+ \sum_{i=1}^{4}  \mathrm d x_i^2 + \frac{\rho_5}{\tilde r^2}   \mathrm d \tilde  r^2 \right) \;, 
	\end{align}
where we use $\mathrm d s_E^2 \sim \tilde r^{-4/5} \mathrm d s^2 $. 
Apparently, we can not put this metric in the standard form (\ref{BasicMetric}) with the reference coordinate, 
because it is necessary to make a coordinate transformation with exponential form $\tilde r \sim e^{a r}$. 
Thus the power law scaling properties do not hold anymore. 
This case is briefly outlined in footnote \ref{footnoteNoSimpleScaling} in general context. 
 
Let us consider this case more closely. 
One can pull $\rho_5 $ out of the parenthesis and absorb it into other field theory coordinates appropriately, 
without modifying the radial dependence of the metric. There seems no preferred scaling for the radial coordinate, 
yet we would take it as a reference scale $r \rightarrow \lambda r $. 
With these discussion, we can put the metric in, yet, another standard form as 
	\begin{align}       \label{BasicMetricNS5B}
		\mathrm d s^2 &= r^{2 \theta /D} \left( - r^{-2 z} \mathrm d t^2 
		-2 \mathrm d t \mathrm d \xi + \frac{\mathrm dr^2}{r^2}
		+ \sum_{i=1}^{d} \mathrm d x_i^2   \right) \;, 
	\end{align} 
where the scaling symmetry of this metric is given by 
	\begin{align}
		&t \rightarrow \lambda^z t \;, \quad 
		\xi \rightarrow \lambda^{-z} \xi \;, \quad 
		r \rightarrow \lambda r \;, \quad 
		x_i \rightarrow  x_i \;. 
	\end{align}  
From this metric, it is not clear whether there is precise meaning of dynamical exponent 
because the spatial directions do not scale. 
Thus this types of metric does not possess scaling symmetries of the kind we consider here.

\subsection{F1 brane}

F1 brane solution in 10 dimensions is described by the metric and dilation \cite{Mazzucato:2008tr}
	\begin{align} 
		\mathrm d s^2 &= \left( \frac{\rho_0}{\tilde r}\right)^{6}  
		\left[ - \frac{2 \Delta^2}{ \tilde r^2} \mathrm d t^2 -2 \mathrm d t \mathrm d \xi \right]
		+ \left( \frac{\rho_0}{\tilde r} \right)^4 \left(  \mathrm d \tilde  r^2 
		+ \tilde r^2 \mathrm d \Omega_7 ^2 \right) \;, \qquad 		 
		e^{\Phi} = \left( \frac{\rho_0}{\tilde r} \right)^3   \;, 
	\end{align}
where we omit other irrelevant field contents. 	
We compactify this solution on $S^{7}$ down to $3$ dimensional Einstein metric 
	\begin{align} 
		\mathrm d s_{E}^2 &\sim r^{-4} \left( - \frac{\tilde \Delta^2}{r} \mathrm d t^2 
		-2 \mathrm d t \mathrm d \xi +  \mathrm d r^2 \right) \;, 
	\end{align}
where we use $\tilde r = \sqrt{r} $. Note that the naive identification of the dynamical exponent 
shows $z = \frac{1}{2} $, which actually violate the null energy condition. A moment of thought 
tells that the meaning of the dynamical exponent is not well posed because 
there is no spatial directions left after the dimensional reduction. 
Furthermore, there can not be symmetries for the Galilean boost and special conformal transformation.

\subsection{KK monopole}  \label{sec:KKMonolope}

Let us consider the metric of the KK monopole in 11 dimensional Einstein frame \cite{Mazzucato:2008tr} 
	\begin{align} 
		\mathrm d s_{KK}^2 &= - \frac{2 \Delta^4}{ \tilde r^4} \mathrm d t^2 -2 \mathrm d t \mathrm d \xi 
		+ \sum_{i=1}^{5}  \mathrm d x_i^2 
		+ \left( \frac{\rho_0}{\tilde r} \right)^4 \left(  \mathrm d \tilde  r^2 +
		 \tilde r^2 \mathrm d \Omega_3 ^2 \right) 	\;.
	\end{align}	
One might prefer to view that the metric approaches to the boundary for $\tilde r \rightarrow 0 $. 
Yet the situation is not clear once we perform the dimensional reduction. 
	
We compactify this on $S^{3}$. The resulting $8 $ dimensional Einstein metric is 
	\begin{align}    \label{KKMonopoleMetricTildeR}
		\mathrm d s_{E}^2 &\sim \tilde r^{-1} \left(
		- \frac{2 \Delta^4}{ \tilde r^4} \mathrm d t^2 -2 \mathrm d t \mathrm d \xi 
		+ \sum_{i=1}^{5}  \mathrm d x_i^2 
		+ \left( \frac{\rho_0}{\tilde r} \right)^4   \mathrm d \tilde  r^2 
		 \right) \;,
	\end{align}
where we use $\mathrm d s_E^2 = \tilde r^{-1} \mathrm d s^2 $. 
To get the standard form (\ref{BasicMetric}), we change a variable $\tilde r = \frac{1}{r} $. 
Then we get  
	\begin{align}  \label{KKMonopoleMetricR}
		\mathrm d s_{E}^2 &\sim r \left( - r^4 \mathrm d t^2 -2 \mathrm d t \mathrm d \xi 
		 + \sum_{i=1}^{5}  \mathrm d x_i^2 + \mathrm d r^2 \right) \;. 
	\end{align}
Compared to the standard metric (\ref{BasicMetric}), we obtain 
	\begin{align}   \label{KKMonopoleExponent}
		d=5\;, \qquad \theta = 9 \;, \qquad z = -1 \;, 
	\end{align}
where we use $D=6 $.
Note that the dynamical exponent is negative and the hyperscaling exponent is a relative large positive number. 

To be more careful, we also try to put the metric in another form using the $ u $ coordinate as in 
footnote \ref{footnoteUVcoordinate}. It turns out that the metric (\ref{KKMonopoleMetricTildeR}) 
is already in a standard form given in footnote \ref{footnoteUVcoordinate}. Thus we put $u =\tilde r $, 
then we get 
	\begin{align}    \label{KKMonopoleMetricU}
		\mathrm d s_{E}^2 &\sim u^{-1} \left(
		- \frac{1}{u^4} \mathrm d t^2 -2 \mathrm d t \mathrm d \xi + \sum_{i=1}^{5}  \mathrm d x_i^2 
		+ \frac{1}{u^4}   \mathrm d u^2 \right) \;,
	\end{align}
where we absorb various constants into the appropriate coordinates. According to the standard form there, 
we obtain the same exponents as (\ref{KKMonopoleExponent}).

Thus we conclude that the KK monopole solution has a negative dynamical exponent. 
Can we understand the negative dynamical exponent, which means that space and time scale in an opposite way? 
Some thoughts give an observation that the standard forms given in (\ref{KKMonopoleMetricR}) and 
(\ref{KKMonopoleMetricU}) may not appropriately reflect the location of the boundary, meaning that 
the $ r $ coordinate in (\ref{KKMonopoleMetricR}) describe the boundary at $r \rightarrow \infty $ 
and thus can be described more appropriately by $u $ coordinate, and vice versa. 
But there is no way to put the dimensionally reduced metric into this `more appropriate' form. 
The metrics with the negative dynamical exponent means that we are forced to put them in the wrong 
form with opposite boundary with respect to time.   
The KK monopole solution with null Melvin twist seems to be well defined. 
Thus it is reasonable to take this case seriously to investigate whether this makes sense, 
along with other systems with a negative dynamical exponent. 
Here we calculate a few physical properties associated with this KK metric, postponing 
more serious investigations to the future. 

\smallskip 

{\it  Entanglement entropy}

The entanglement entropy of the metric can be computed using the prescription given in \cite{kimHyperscaling}. 
Using a strip geometry, the entanglement entropy of the metric (\ref{BasicMetric}) with 
$\mathrm a=1 $, $\mathrm b=1 $ and $ \eta_j =1 $ can be computed as  
	\begin{align}   
		l &= \int_0^{r_t}\mathrm d r~ \frac{(r/r_t)^{\alpha}}{\sqrt{1- (r/r_t)^{2\alpha}}} 
		= - \frac{i \pi }{2} r_t  \;, 
	\end{align}   
and 
	\begin{align}  
		{\mathcal  A} 
		&= L^{d-1} L_\xi \int_\epsilon^{r_t}\mathrm d r~ \frac{ \beta^{-1/2} r^{-\alpha}}{\sqrt{1-(r/r_t)^{2\alpha}}} 
		= \beta^{-1/2} L^{d-1} L_\xi  \left( -\frac{i \pi r_t^2}{4} + \mathcal O (\epsilon^3) \right)	\;, 
	\end{align}   
where $\alpha =-1$ for the KK monopole metric (\ref{KKMonopoleMetricR}). 	
The entanglement entropy for a strip in the general metric (\ref{BasicMetricSchr}) (with $ \eta_i =1 $) is 
	\begin{align}      \label{entanglementEntropyKKMonopole}
		\mathcal  S 
		&= \frac{(R M_{Pl})^{6} }{4 } 
		\left( i \frac{l^2}{\pi} \frac{L^{4} L_\xi}{R_\theta^{9}} -   \mathcal O (\epsilon^3)  \right) \;,
	\end{align}   
where $R_\theta$ is a scale in which the hyperscaling violation becomes important. 
The result is rather unusual. The length $l$ is imaginary and the entanglement entropy is 
also imaginary.

\smallskip  

{\it  Holographic stress energy tensor}

Let us calculate the stress energy tensor for this case. 
	\begin{align}     \label{HSETKKMonopole}
		&\langle T_{tt}\rangle = -\frac{d + 2 -z -(d+1) \theta/D}{r_c^{d+2-(d+1)\theta/D}} h_{00} 
		= -\frac{-1}{r_c^{-2}} h_{00} \;,  \\ 
		&\langle T_{t\xi}\rangle =\langle T_{\xi t}\rangle = 
			-\frac{(d+1)(1- \theta/D)}{r_c^{d+2-(d+1)\theta/D}} h_{t\xi}
			= -\frac{-3}{r_c^{-2}} h_{t\xi}  \;,  \\ 
		&\langle T_{ij}\rangle = -\frac{(d+1)(1- \theta/D)}{r_c^{d+2-(d+1)\theta/D}}  h_{ij}
		= -\frac{-3}{r_c^{-2}}  h_{ij}  \;.
	\end{align}   
The null energy condition is violated in this case. 
The stress energy tensor one point functions have opposite sign compared to the cases we consider above.

\smallskip 

{\it  Semiclassical propagators}

The static semiclassical propagator is given by 
	\begin{align}    \label{staticPropagatorKKMonopole}
		G(\Delta x_i) &\sim  \exp \left[2m \frac{D}{\theta} ~ \epsilon^{\theta/D}\right]  
		 \exp \left[-m \frac{D}{\theta} c_{\theta/D} |\Delta x_i|^{\theta/D}\right] \;, \nonumber \\
		 &\qquad c_{\theta/D} \equiv \left(\frac{2\sqrt{\pi}\,\Gamma\left(\frac{2-\theta/D}{2(1-\theta/D)} \right)}
		{\Gamma \left(\frac{1}{2(1-\theta/D)} \right)} \right)^{1-\theta/D} \;.
	\end{align}     
where $\theta/D=3/2$. The result is independent of $z$.

\smallskip  

{\it Outlook}

Seemingly the physical properties show some unusual properties, such as imaginary entanglement entropy, 
stress energy tensors with wrong signs. Do they signal that this background is physically not acceptable? 
Instead of answering the question directly. 
We would like to bring out a system with negative dynamical exponent \cite{Singh:2012un} 
along with the reference \cite{Cvetic:1998jf}. The latter discussed the Euclidean version 
of the metric similar to the one considered here (\ref{KKMonopoleMetricU}). 
It will be interesting to investigate the systems with negative dynamical exponent along the line.


\begin{thebibliography}{aaaa}


\def\hri#1#2{\href{http://arxiv.org/abs/#1}{[#1]#2}}
\def\hre#1#2{\href{http://arxiv.org/abs/#1/#2}{[#1/#2]}}


\bibitem{Sachdev:2012dq} 
  S.~Sachdev,
  {\em ``The Quantum phases of matter,''}
  25th Solvay Conference on Physics, ``The Theory of the Quantum World'', Brussels, Oct 2011,
  \hri{arXiv:1203.4565}{[hep-th]}.

\bibitem{Fisher} 
D.~S.~Fisher, 
{\em ``Scaling and critical slowing down in random-field Ising systems,''}
Phys. Rev. Lett. {\bf 56}, 416 (1986).

\bibitem{SachdevBook} 
S.~Sachdev, 
{\em "Quantum Phase Transitions,"}
2nd Ed., Cambridge University Press (2011). 

%
\bibitem{CGKKM} 
C.~Charmousis, B.~Gouteraux, B.~S.~Kim, E.~Kiritsis and R.~Meyer,
  {\em ``Effective Holographic Theories for low-temperature condensed matter systems,''}
  JHEP {\bf 1011}, 151 (2010)
  [\hri{arXiv:1005.4690}{[hep-th]}. 

\bibitem{Gouteraux:2011ce} 
  B.~Gouteraux and E.~Kiritsis,
  {\em ``Generalized Holographic Quantum Criticality at Finite Density,''}
  JHEP {\bf 1112}, 036 (2011)
  \hri{arXiv:1107.2116}{[hep-th]}.

\bibitem{Ogawa:2011bz} 
  N.~Ogawa, T.~Takayanagi and T.~Ugajin,
  {\em ``Holographic Fermi Surfaces and Entanglement Entropy,''}
  \hri{arXiv:1111.1023}{[hep-th]}.

\bibitem{Huijse:2011ef} 
  L.~Huijse, S.~Sachdev and B.~Swingle,
  {\em ``Hidden Fermi surfaces in compressible states of gauge-gravity duality,''}
  \hri{arXiv:1112.0573}{[cond-mat.str-el]}.
   
\bibitem{kachru}
  X.~Dong, S.~Harrison, S.~Kachru, G.~Torroba and H.~Wang,
  {\em ``Aspects of holography for theories with hyperscaling violation,''}
  \hri{arXiv:1201.1905}{[hep-th]}.
  
\bibitem{Shaghoulian:2011aa} 
  E.~Shaghoulian,
  {\em ``Holographic Entanglement Entropy and Fermi Surfaces,''}
  \hri{arXiv:1112.2702}{[hep-th]}.

\bibitem{Narayan:2012hk} 
  K.~Narayan,
  {\em ``On Lifshitz scaling and hyperscaling violation in string theory,''}
  Phys.\ Rev.\ D {\bf 85}, 106006 (2012)
  \hri{arXiv:1202.5935}{[hep-th]}.
  
\bibitem{kimHyperscaling}
  B.~S.~Kim,
  {\em ``Schr\"odinger Holography with and without Hyperscaling Violation,''}
  JHEP {\bf 1206}, 116 (2012)
  \hri{arXiv:1202.6062}{[hep-th]}.
  
\bibitem{Singh:2012un} 
  H.~Singh,
  {\em ``Lifshitz/Schr\"odinger Dp-branes and dynamical exponents,''}
  \hri{arXiv:1202.6533}{[hep-th]}.

\bibitem{Hartnoll:2012wm} 
  S.~A.~Hartnoll and E.~Shaghoulian,
  {\em ``Spectral weight in holographic scaling geometries,''}
  \hri{arXiv:1203.4236}{[hep-th]}.

\bibitem{Dey:2012tg} 
  P.~Dey and S.~Roy,
  {\em ``Lifshitz-like space-time from intersecting branes in string/M theory,''}
  \hri{arXiv:1203.5381}{[hep-th]}.

\bibitem{Myung:2012cb} 
  Y.~S.~Myung and T.~Moon,
  {\em ``Quasinormal frequencies and thermodynamic quantities for the Lifshitz black holes,''}
  Phys.\ Rev.\ D {\bf 86}, 024006 (2012)
  \hri{arXiv:1204.2116}{[hep-th]}.
    
\bibitem{Dey:2012rs} 
  P.~Dey and S.~Roy,
  {\em ``Intersecting D-branes and Lifshitz-like space-time,''}
  \hri{arXiv:1204.4858}{[hep-th]}.

\bibitem{Perlmutter:2012he} 
  E.~Perlmutter,
  {\em ``Hyperscaling violation from supergravity,''}
  JHEP {\bf 1206}, 165 (2012)
  \hri{arXiv:1205.0242}{[hep-th]}.

\bibitem{Cadoni:2012uf} 
  M.~Cadoni and S.~Mignemi,
  {\em ``Phase transition and hyperscaling violation for scalar Black Branes,''}
  JHEP {\bf 1206}, 056 (2012)
  \hri{arXiv:1205.0412}{[hep-th]}.

\bibitem{Charmousis:2012dw} 
  C.~Charmousis, B.~Gouteraux and E.~Kiritsis,
  {\em ``Higher-derivative scalar-vector-tensor theories: black holes, Galileons, singularity cloaking and holography,''}
  \hri{arXiv:1206.1499}{[hep-th]}.
  
\bibitem{Ammon:2012je} 
  M.~Ammon, M.~Kaminski and A.~Karch,
  {\em ``Hyperscaling-Violation on Probe D-Branes,''}
  \hri{arXiv:1207.1726}{[hep-th]}.

\bibitem{Kiritsis:2012ta} 
  E.~Kiritsis,
  {\em ``Lorentz violation, Gravity, Dissipation and Holography,''}
  \hri{arXiv:1207.2325}{[hep-th]}.

\bibitem{Bhattacharya:2012zu} 
  J.~Bhattacharya, S.~Cremonini and A.~Sinkovics,
  {\em ``On the IR completion of geometries with hyperscaling violation,''}
  \hri{arXiv:1208.1752}{[hep-th]}.
  
\bibitem{Dey:2012hf} 
  P.~Dey and S.~Roy,
  {\em ``Holographic entanglement entropy of the near horizon 1/4 BPS F-D$p$ bound states,''}
  \hri{arXiv:1208.1820}{[hep-th]}.

\bibitem{Kundu:2012jn} 
  N.~Kundu, P.~Narayan, N.~Sircar and S.~P.~Trivedi,
  {\em ``Entangled Dilaton Dyons,''}
  \hri{arXiv:1208.2008}{[hep-th]}.
  
\bibitem{Kulaxizi:2012gy} 
  M.~Kulaxizi, A.~Parnachev and K.~Schalm,
  {\em ``On Holographic Entanglement Entropy of Charged Matter,''}
  \hri{arXiv:1208.2937}{[hep-th]}.
  
\bibitem{Alishahiha:2012cm} 
  M.~Alishahiha and H.~Yavartanoo,
  {\em ``On Holography with Hyperscaling Violation,''}
  \hri{arXiv:1208.6197}{[hep-th]}.
  
\bibitem{Park:2012mc} 
  C.~Park,
  {\em ``Membrane paradigm in the Einstein-dilaton theory,''}
  \hri{arXiv:1209.0842}{[hep-th]}.

\bibitem{Dey:2012mc} 
  P.~Dey and S.~Roy,
  {\em ``Lifshitz metric with hyperscaling violation from NS5-Dp states in string theory,''}
  \hri{arXiv:1209.1049}{[hep-th]}.


\bibitem{Sadeghi:2012ix} 
  J.~Sadeghi, B.~Pourhassan and A.~Asadi,
  {\em ``Thermodynamics of string black hole with hyperscaling violation,''}
  \hri{arXiv:1209.1235}{[hep-th]}.

\bibitem{Sadeghi:2012vv} 
  J.~Sadeghi, B.~Pourhasan and F.~Pourasadollah,
  {\em ``Schr\"{o}dinger black holes with hyperscaling violation,''}
  \hri{arXiv:1209.1874}{[hep-th]}.

\bibitem{Pal:2012zn} 
  S.~S.~Pal,
  {\em ``Fermi-like Liquid From Einstein-DBI-Dilaton System,''}
  \hri{arXiv:1209.3559}{[hep-th]}.

\bibitem{Alishahiha:2012qu} 
  M.~Alishahiha, E.~OColgain and H.~Yavartanoo,
  {\em ``Charged Black Branes with Hyperscaling Violating Factor,''}
  \hri{arXiv:1209.3946}{[hep-th]}.
  
\bibitem{Bueno:2012sd} 
  P.~Bueno, W.~Chemissany, P.~Meessen, T.~Ortin and C.~S.~Shahbazi,
  {\em ``Lifshitz-like solutions with hyperscaling violation in ungauged supergravity,''}
  \hri{arXiv:1209.4047}{[hep-th]}.
  
\bibitem{Narayan:201209} 
  K.~Narayan,
  {\em ``AdS null deformations with inhomogeneities,''}
  \hri{arXiv:1209.4348}{[hep-th]}.

\bibitem{Cadoni:2012ea} 
  M.~Cadoni and M.~Serra,
  {\em ``Hyperscaling violation for scalar black branes in arbitrary dimensions,''}
  \hri{arXiv:1209.4484}{[hep-th]}.

\bibitem{Maldacena:1997re} 
  J.~M.~Maldacena,
  {\em ``The Large N limit of superconformal field theories and supergravity,''}
  Adv.\ Theor.\ Math.\ Phys.\  {\bf 2}, 231 (1998)
  \hre{arXiv:hep-th}{9711200}.

\bibitem{Aharony:1999ti}
  O.~Aharony, S.~S.~Gubser, J.~M.~Maldacena, H.~Ooguri and Y.~Oz,
  {\em ``Large N field theories, string theory and gravity,''}
  Phys.\ Rept.\  {\bf 323}, 183 (2000)
  \hre{arXiv:hep-th}{9905111}.

\bibitem{Huijse:2011hp} 
  L.~Huijse and S.~Sachdev,
  {\em ``Fermi surfaces and gauge-gravity duality,''}
  Phys.\ Rev.\ D {\bf 84}, 026001 (2011)
  \hri{arXiv:1104.5022}{[hep-th]}.
  
\bibitem{Sachdev:2012tj} 
  S.~Sachdev,
  {\em ``Compressible quantum phases from conformal field theories in 2+1 dimensions,''}
  \hri{arXiv:1209.1637}{[hep-th]}.

\bibitem{Lee:2008xf} 
  S.~-S.~Lee,
  {\em ``A Non-Fermi Liquid from a Charged Black Hole: A Critical Fermi Ball,''}
  Phys.\ Rev.\ D {\bf 79}, 086006 (2009)
  \hri{arXiv:0809.3402}{[hep-th]}.
  
\bibitem{Liu:2009dm} 
  H.~Liu, J.~McGreevy and D.~Vegh,
  {\em ``Non-Fermi liquids from holography,''}
  Phys.\ Rev.\ D {\bf 83}, 065029 (2011)
  \hri{arXiv:0903.2477}{[hep-th]}.
  
\bibitem{Cubrovic:2009ye} 
  M.~Cubrovic, J.~Zaanen and K.~Schalm,
  {\em ``String Theory, Quantum Phase Transitions and the Emergent Fermi-Liquid,''}
  Science {\bf 325}, 439 (2009)
  \hri{arXiv:0904.1993}{[hep-th]}.

\bibitem{Faulkner:2009wj} 
  T.~Faulkner, H.~Liu, J.~McGreevy and D.~Vegh,
  {\em ``Emergent quantum criticality, Fermi surfaces, and AdS(2),''}
  Phys.\ Rev.\ D {\bf 83}, 125002 (2011)
  \hri{arXiv:0907.2694}{[hep-th]}.
  
\bibitem{Doiron-Leyraud:2007}
  N. Doiron-Leyraud et. al., 
  {\em ``Quantum oscillations and the Fermi surface in an underdoped high-Tc superconductor,''}
  \href{http://www.nature.com/nature/journal/v447/n7144/abs/nature05872.html}{Nature 447, 565 (2007)}. 
  
\bibitem{Eisert:2008ur} 
  J.~Eisert, M.~Cramer and M.~B.~Plenio,
  {\em ``Area laws for the entanglement entropy - a review,''}
  Rev.\ Mod.\ Phys.\  {\bf 82}, 277 (2010)
  \hri{arXiv:0808.3773}{[quant-ph]}.
    
\bibitem{HoloEntanglement1} 
  S.~Ryu and T.~Takayanagi,
  {\em ``Holographic derivation of entanglement entropy from AdS/CFT,''}
  Phys.\ Rev.\ Lett.\  {\bf 96} (2006) 181602
  \hre{arXiv:hep-th}{0603001}.

\bibitem{HoloEntanglement2}
 S.~Ryu and T.~Takayanagi,
  {\em ``Aspects of holographic entanglement entropy,''}
  JHEP {\bf 0608} (2006) 045
  \hre{arXiv:hep-th}{0605073}.
  
\bibitem{Hubeny:2007xt} 
  V.~E.~Hubeny, M.~Rangamani and T.~Takayanagi,
  {\em ``A Covariant holographic entanglement entropy proposal,''}
  JHEP {\bf 0707}, 062 (2007)
  \hri{arXiv:0705.0016}{[hep-th]}.

\bibitem{Takayanagi:2012kg} 
  T.~Takayanagi,
  {\em ``Entanglement Entropy from a Holographic Viewpoint,''}
  Class.\ Quant.\ Grav.\  {\bf 29}, 153001 (2012)
  \hri{arXiv:1204.2450}{[gr-qc]}.
  
\bibitem{Balasubramanian:2009rx} 
  K.~Balasubramanian and J.~McGreevy,
  {\em ``An Analytic Lifshitz black hole,''}
  Phys.\ Rev.\ D {\bf 80}, 104039 (2009)
  \hri{arXiv:0909.0263}{[hep-th]}.

\bibitem{GKMReview} 
  B.~Gouteraux, B.~S.~Kim and R.~Meyer,
  {\em ``Charged Dilatonic Black Holes and their Transport Properties,''}
  Fortsch.\ Phys.\  {\bf 59}, 723 (2011)
  \hri{arXiv:1102.4440}{[hep-th]}.

\bibitem{GKMReview2} 
  R.~Meyer, B.~Gouteraux and B.~S.~Kim,
  {\em ``Strange Metallic Behaviour and the Thermodynamics of Charged Dilatonic Black Holes,''}
  Fortsch.\ Phys.\  {\bf 59}, 741 (2011)
  \hri{arXiv:1102.4433}{[hep-th]}.
  
\bibitem{Hashimoto:1999ut} 
  A.~Hashimoto and N.~Itzhaki,
  {\em ``Noncommutative Yang-Mills and the AdS / CFT correspondence,''}
  Phys.\ Lett.\ B {\bf 465}, 142 (1999)
  \hre{hep-th}{9907166}.

\bibitem{Maldacena:1999mh} 
  J.~M.~Maldacena and J.~G.~Russo,
  {\em ``Large N limit of noncommutative gauge theories,''}
  JHEP {\bf 9909}, 025 (1999)
  \hre{hep-th}{9908134}.

\bibitem{Alishahiha:2000pu} 
  M.~Alishahiha, Y.~Oz and J.~G.~Russo,
  {\em ``Supergravity and light - like noncommutativity,''}
  JHEP {\bf 0009}, 002 (2000)
  \hre{arXiv:hep-th}{0007215}. 
  
\bibitem{Hubeny:2005qu} 
  V.~E.~Hubeny, M.~Rangamani and S.~F.~Ross,
  {\em ``Causal structures and holography,''}
  JHEP {\bf 0507}, 037 (2005)
  \hre{arXiv:hep-th}{0504034}. 
  
\bibitem{Bergman:2001rw} 
  A.~Bergman, K.~Dasgupta, O.~J.~Ganor, J.~L.~Karczmarek and G.~Rajesh,
  {\em ``Nonlocal field theories and their gravity duals,''}
  Phys.\ Rev.\ D {\bf 65}, 066005 (2002)
  \hre{arXiv:hep-th}{0103090}.
  
\bibitem{Bergman:2000cw} 
  A.~Bergman and O.~J.~Ganor,
  {\em ``Dipoles, twists and noncommutative gauge theory,''}
  JHEP {\bf 0010}, 018 (2000)
  \hre{arXiv:hep-th}{0008030}.

\bibitem{Ganor:2006ub} 
  O.~J.~Ganor,
  {\em ``A New Lorentz violating nonlocal field theory from string-theory,''}  
  Phys.\  Rev.\  D {\bf 75}, 025002 (2007)
  \hri{arXiv:hep-th}{0609107}.
  
\bibitem{Ganor:2007qh} 
  O.~J.~Ganor, A.~Hashimoto, S.~Jue, B.~S.~Kim and A.~Ndirango,
  {\em ``Aspects of Puff Field Theory,''}
  JHEP {\bf 0708}, 035 (2007)
  \hre{arXiv:hep-th}{0702030}. 
  
\bibitem{Kachru:2008yh}
  S.~Kachru, X.~Liu and M.~Mulligan,
  {\em ``Gravity Duals of Lifshitz-like Fixed Points,''}
  Phys.\ Rev.\  D {\bf 78}, 106005 (2008)
  \hri{arXiv:0808.1725}{[hep-th]}.
  
\bibitem{Taylor:2008tg} 
  M.~Taylor,
  {\em ``Non-relativistic holography,''}
  \hri{arXiv:0812.0530}{[hep-th]}.

\bibitem{Charmousis:2009xr} 
  C.~Charmousis, B.~Gouteraux and J.~Soda,
  {\em ``Einstein-Maxwell-Dilaton theories with a Liouville potential,''}
  Phys.\ Rev.\ D {\bf 80}, 024028 (2009)
  \hri{arXiv:0905.3337}{[gr-qc]}.
  
\bibitem{Gubser}
S.~S.~Gubser and F.~D.~Rocha,
  {\em ``Peculiar properties of a charged dilatonic black hole in AdS$_5$,''}
  Phys.\ Rev.\ D {\bf 81}, 046001 (2010)
  \hri{arXiv:0911.2898}{[hep-th]}.

\bibitem{GKPT}
K.~Goldstein, S.~Kachru, S.~Prakash and S.~P.~Trivedi,
  {\em ``Holography of Charged Dilaton Black Holes,''}
  JHEP {\bf 1008}, 078 (2010)
  \hri{arXiv:0911.3586}{[hep-th]}.
  
\bibitem{CDP}
 M.~Cadoni, G.~D'Appollonio and P.~Pani,
  {\em ``Phase transitions between Reissner-Nordstrom and dilatonic black holes in 4D AdS spacetime,''}
  JHEP {\bf 1003}, 100 (2010)
  \hri{arXiv:0912.3520}{[hep-th]}.

\bibitem{CDP2}
  M.~Cadoni and P.~Pani,
  {\em ``Holography of charged dilatonic black branes at finite temperature,''}
  JHEP {\bf 1104}, 049 (2011)
  \hri{arXiv:1102.3820}{[hep-th]}.
  
\bibitem{BBPZ}
G.~Bertoldi, B.~A.~Burrington and A.~W.~Peet,
  {\em ``Thermal behavior of charged dilatonic black branes in AdS and UV completions of Lifshitz-like geometries,''}
  Phys.\ Rev.\ D {\bf 82}, 106013 (2010)
  \hri{arXiv:1007.1464}{[hep-th]}.
  
\bibitem{BBPZ2}  
G.~Bertoldi, B.~A.~Burrington, A.~W.~Peet and I.~G.~Zadeh,
  {\em ``Lifshitz-like black brane thermodynamics in higher dimensions,''}
  Phys.\ Rev.\ D {\bf 83}, 126006 (2011)
  \hri{arXiv:1101.1980}{[hep-th]}.

\bibitem{GKPTIW}
K.~Goldstein, N.~Iizuka, S.~Kachru, S.~Prakash, S.~P.~Trivedi and A.~Westphal,
  {\em ``Holography of Dyonic Dilaton Black Branes,''}
  JHEP {\bf 1010}, 027 (2010)
  \hri{arXiv:1007.2490}{[hep-th]}.

\bibitem{IKNT}
N.~Iizuka, N.~Kundu, P.~Narayan and S.~P.~Trivedi,
  {\em ``Holographic Fermi and Non-Fermi Liquids with Transitions in Dilaton Gravity,''}
  \hri{arXiv:1105.1162}{[hep-th]}.

\bibitem{Lee:2010ii} 
  B.~-H.~Lee, D.~-W.~Pang and C.~Park,
  {\em ``Strange Metallic Behavior in Anisotropic Background,''}
  JHEP {\bf 1007}, 057 (2010)
  \hri{arXiv:1006.1719}{[hep-th]}.
  
\bibitem{BBM}
P.~Berglund, J.~Bhattacharyya and D.~Mattingly,
  {\em ``Charged Dilatonic AdS Black Branes in Arbitrary Dimensions,''}
  \hri{arXiv:1107.3096}{[hep-th]}.

\bibitem{Gouteraux:2011qh} 
  B.~Gouteraux, J.~Smolic, M.~Smolic, K.~Skenderis and M.~Taylor,
  {\em ``Holography for Einstein-Maxwell-dilaton theories from generalized dimensional reduction,''}
  JHEP {\bf 1201}, 089 (2012)
  \hri{arXiv:1110.2320}{[hep-th]}.

\bibitem{Gubser:2000nd} 
  S.~S.~Gubser,
  {\em ``Curvature singularities: The Good, the bad, and the naked,''}
  Adv.\ Theor.\ Math.\ Phys.\  {\bf 4}, 679 (2000)
  \hre{hep-th}{0002160}.

\bibitem{Hoyos:2010at} 
  C.~Hoyos and P.~Koroteev,
  {\em ``On the Null Energy Condition and Causality in Lifshitz Holography,''}
  Phys.\ Rev.\ D {\bf 82}, 084002 (2010)
  [Erratum-ibid.\ D {\bf 82}, 109905 (2010)]
  \hri{arXiv:1007.1428}{[hep-th]}.

\bibitem{Swingle:2011mk} 
  B.~Swingle and T.~Senthil,
  {\em ``Universal crossovers between entanglement entropy and thermal entropy,''}
  \hri{arXiv:1112.1069}{[cond-mat.str-el]}.
  
\bibitem{Gursoy:2007er} 
  U.~Gursoy, E.~Kiritsis and F.~Nitti,
  {\em ``Exploring improved holographic theories for QCD: Part II,''}
  JHEP {\bf 0802}, 019 (2008)
  \hri{arXiv:0707.1349}{[hep-th]}.
  
\bibitem{Gursoy:2008za} 
  U.~Gursoy, E.~Kiritsis, L.~Mazzanti and F.~Nitti,
 {\em ``Holography and Thermodynamics of 5D Dilaton-gravity,''}
  JHEP {\bf 0905}, 033 (2009)
  \hri{arXiv:0812.0792}{[hep-th]}.
 
\bibitem{Charmousis:2001hg} 
  C.~Charmousis, R.~Emparan and R.~Gregory,
  {\em ``Selfgravity of brane worlds: A New hierarchy twist,''}
  JHEP {\bf 0105}, 026 (2001)
  \hre{hep-th}{0101198}.

\bibitem{Son:2008ye}
  D.~T.~Son,
  {\em ``Toward an AdS/cold atoms correspondence: a geometric realization of the
  Schroedinger symmetry,''}
  Phys.\ Rev.\  D {\bf 78}, 046003 (2008)
  \hri{arXiv:0804.3972}{[hep-th]}.
  
\bibitem{Balasubramanian:2008dm}
  K.~Balasubramanian and J.~McGreevy,
  {\em ``Gravity duals for non-relativistic CFTs,''}
  Phys.\ Rev.\ Lett.\  {\bf 101}, 061601 (2008)
  \hri{arXiv:0804.4053}{[hep-th]}.
  
\bibitem{Herzog:2008wg}
  C.~P.~Herzog, M.~Rangamani and S.~F.~Ross,
  {\em "Heating up Galilean holography,"}
  JHEP {\bf 0811}, 080 (2008)
  \hri{arXiv:0807.1099}{[hep-th]}.
    
\bibitem{Maldacena:2008wh}
  J.~Maldacena, D.~Martelli and Y.~Tachikawa,
  {\em "Comments on string theory backgrounds with non-relativistic conformal
  symmetry,"}
  JHEP {\bf 0810}, 072 (2008)
  \hri{arXiv:0807.1100}{[hep-th]}.
    
\bibitem{Adams:2008wt}
  A.~Adams, K.~Balasubramanian and J.~McGreevy,
  {\em "Hot Spacetimes for Cold Atoms,"}
  JHEP {\bf 0811}, 059 (2008)
  \hri{arXiv:0807.1111}{[hep-th]}. 

\bibitem{Yamada:2008if}
  D.~Yamada,
  {\em "Thermodynamics of Black Holes in Schroedinger Space,"}
  Class.\ Quant.\ Grav.\  {\bf 26}, 075006 (2009)
  \hri{arXiv:0809.4928}{[hep-th]}. 
  
\bibitem{Ammon:2010eq}
  M.~Ammon, C.~Hoyos, A.~O'Bannon and J.~M.~S.~Wu,
  {\em ``Holographic Flavor Transport in Schrodinger Spacetime,''}
  JHEP {\bf 1006}, 012 (2010)
  \hri{arXiv:1003.5913}{[hep-th]}.

\bibitem{Alishahiha:2003ru}
  M.~Alishahiha and O.~J.~Ganor,
  {\em ``Twisted backgrounds, PP waves and nonlocal field theories,''}
  JHEP {\bf 0303}, 006 (2003)
  \hre{arXiv:hep-th}{0301080}.
  
\bibitem{Gimon:2003xk}
  E.~G.~Gimon, A.~Hashimoto, V.~E.~Hubeny, O.~Lunin and M.~Rangamani,
  {\em ``Black strings in asymptotically plane wave geometries,''}
  JHEP {\bf 0308}, 035 (2003)
  \hre{arXiv:hep-th}{0306131}.

\bibitem{Mazzucato:2008tr}
  L.~Mazzucato, Y.~Oz and S.~Theisen,
  {\em ``Non-relativistic Branes,''}
  JHEP {\bf 0904}, 073 (2009)
  \hri{arXiv:0810.3673}{[hep-th]}.

\bibitem{Duval:2008jg}
  C.~Duval, M.~Hassaine and P.~A.~Horvathy,
  {\em ``The Geometry of Schrodinger symmetry in gravity background/non-relativistic
  CFT,''}
  Annals Phys.\  {\bf 324}, 1158 (2009)
  \hri{arXiv:0809.3128}{[hep-th]}.
  
\bibitem{Hartnoll:2008rs}
  S.~A.~Hartnoll and K.~Yoshida,
  {\em ``Families of IIB duals for nonrelativistic CFTs,''}
  JHEP {\bf 0812}, 071 (2008)
  \hri{arXiv:0810.0298}{[hep-th]}. 

\bibitem{Adams:2008zk} 
  A.~Adams, A.~Maloney, A.~Sinha and S.~E.~Vazquez,
  {\em ``1/N Effects in Non-Relativistic Gauge-Gravity Duality,''}
  JHEP {\bf 0903}, 097 (2009)
  \hri{arXiv:0812.0166}{[hep-th]}. 
  
\bibitem{Gauntlett:2009zw} 
  J.~P.~Gauntlett, S.~Kim, O.~Varela and D.~Waldram,
  {\em ``Consistent supersymmetric Kaluza-Klein truncations with massive modes,''}
  JHEP {\bf 0904}, 102 (2009)
  \hri{arXiv:0901.0676}{[hep-th]}.
  
\bibitem{Donos:2009en} 
  A.~Donos and J.~P.~Gauntlett,
  {\em ``Supersymmetric solutions for non-relativistic holography,''}
  JHEP {\bf 0903}, 138 (2009)
  \hri{arXiv:0901.0818}{[hep-th]}.
  
\bibitem{Pal:2009np} 
  S.~S.~Pal,
  {\em ``Non-relativistic supersymmetric Dp branes,''}
  Class.\ Quant.\ Grav.\  {\bf 26}, 245014 (2009)
  \hri{arXiv:0904.3620}{[hep-th]}.

\bibitem{Bobev:2009mw} 
  N.~Bobev, A.~Kundu and K.~Pilch,
  {\em ``Supersymmetric IIB Solutions with Schrodinger Symmetry,''}
  JHEP {\bf 0907}, 107 (2009)
  \hri{arXiv:0905.0673}{[hep-th]}.

\bibitem{Donos:2009xc} 
  A.~Donos and J.~P.~Gauntlett,
  {\em ``Solutions of type IIB and D=11 supergravity with Schrodinger(z) symmetry,''}
  JHEP {\bf 0907}, 042 (2009)
  \hri{arXiv:0905.1098}{[hep-th]}.
  
\bibitem{O'Colgain:2009yd} 
  E.~O. Colgain, O.~Varela and H.~Yavartanoo,
  {\em ``Non-relativistic M-Theory solutions based on Kaehler-Einstein spaces,''}
  JHEP {\bf 0907}, 081(2009)
  \hri{arXiv:0906.0261}{[hep-th]}.
  
\bibitem{Cremonesi:2009gy} 
  S.~Cremonesi, D.~Melnikov and Y.~Oz,
  {\em ``Stability of Asymptotically Schr\"odinger RN Black Hole and Superconductivity,''}
  JHEP {\bf 1004}, 048 (2010)
  \hri{arXiv:0911.3806}{[hep-th]}.

\bibitem{Jeong:2009aa}
  J.~Jeong, H.~C.~Kim, S.~Lee, E.~O Colgain and H.~Yavartanoo,
  {\em ``Schrodinger invariant solutions of M-theory with Enhanced Supersymmetry,''}
  JHEP {\bf 1003}, 034 (2010)
  \hri{arXiv:0911.5281}{[hep-th]}.
  
\bibitem{Banerjee:2011jb} 
  N.~Banerjee, S.~Dutta and D.~P.~Jatkar,
  {\em ``Geometry and Phase Structure of Non-Relativistic Branes,''}
  Class. Quant. Grav.{\bf 28},165002 (2011)
  \hri{arXiv:1102.0298}{[hep-th]}.
   
\bibitem{Kraus:2011pf} 
  P.~Kraus and E.~Perlmutter,
  {\em ``Universality and exactness of Schrodinger geometries in string and M-theory,''}
  JHEP {\bf 1105}, 045 (2011)
  \hri{arXiv:1102.1727}{[hep-th]}.

\bibitem{Kim:2011fb} 
  H.~-C.~Kim, S.~Kim, K.~Lee and J.~Park,
  {\em ``Emergent Schrodinger geometries from mass-deformed CFT,''}
  JHEP {\bf 1108}, 111 (2011)
  \hri{arXiv:1106.4309}{[hep-th]}.

\bibitem{Brown:2011av} 
  C.~M.~Brown and O.~DeWolfe,
  {\em ``The Godel-Schrodinger Spacetime and Stringy Chronology Protection,''}
  JHEP {\bf 1201}, 032 (2012)
  \hri{arXiv:1110.3840}{[hep-th]}.

\bibitem{Hyun:2011qj} 
  S.~Hyun, J.~Jeong and B.~S.~Kim,
  {\em ``Finite Temperature Aging Holography,''}
  JHEP {\bf 1203}, 010 (2012)
  \hri{arXiv:1108.5549}{[hep-th]}.

\bibitem{Hyun:2012fd} 
  S.~Hyun, J.~Jeong and B.~S.~Kim,
  {\em ``Aging Logarithmic Conformal Field Theory : a holographic view,''}
  \hri{arXiv:1209.2417}{[hep-th]}.

\bibitem{Guica:2010sw}
  M.~Guica, K.~Skenderis, M.~Taylor and B.~C.~van Rees,
  {\em ``Holography for Schrodinger backgrounds,''}
  JHEP {\bf 1102}, 056 (2011)
  \hri{arXiv:1008.1991}{[hep-th]}.

\bibitem{Balasubramanian:2010uw}
  K.~Balasubramanian and J.~McGreevy,
  {\em ``The Particle number in Galilean holography,''}
  JHEP {\bf 1101}, 137 (2011)
  \hri{arXiv:1007.2184}{[hep-th]}.

\bibitem{Itzhaki:1998dd}
N.~Itzhaki, J.~M. Maldacena, J.~Sonnenschein, and S.~Yankielowicz, 
	{\em ``Supergravity and the large N limit of theories with sixteen supercharges,''}
  Phys. \ Rev. \ {\bf D58} (1998) 046004,
  \hre{hep-th}{9802042}.
  
\bibitem{Boonstra:1998mp}
H.~Boonstra, K.~Skenderis, and P.~Townsend, 
	{\em ``The domain wall / QFT correspondence,''}  
	JHEP {\bf 9901} (1999) 003,
  	\hre{hep-th}{9807137}.
  
\bibitem{Kim:2010tf}
  B.~S.~Kim and D.~Yamada,
  {\em ``Properties of Schroedinger Black Holes from AdS Space,''}
  JHEP {\bf 1107}, 120 (2011)
  \hri{arXiv:1008.3286}{[hep-th]}.
  
\bibitem{Bobev:2011qx} 
  N.~Bobev and B.~C.~van Rees,
  {\em ``Schrodinger Deformations of $AdS_3 x S^3$,''}
  JHEP {\bf 1108}, 062 (2011)
 \hri{arXiv:1102.2877}{[hep-th]}.

\bibitem{Costa:2010cn} 
  R.~N.~Caldeira Costa and M.~Taylor,
  {\em ``Holography for chiral scale-invariant models,''}
  JHEP {\bf 1102}, 082 (2011)
 \hri{arXiv:1010.4800}{[hep-th]}.

\bibitem{Goldberger:2008vg}
  W.~D.~Goldberger,
  {\em ``AdS/CFT duality for non-relativistic field theory,''}
  JHEP {\bf 0903}, 069 (2009)
  \hri{arXiv:0806.2867}{[hep-th]}.
  
\bibitem{Barbon:2008bg}
  J.~L.~F.~Barbon and C.~A.~Fuertes,
  {\em ``On the spectrum of nonrelativistic AdS/CFT,''}
  JHEP {\bf 0809}, 030 (2008)
  \hri{arXiv:0806.3244}{[hep-th]}.

\bibitem{Kim:2010zq}
  B.~S.~Kim, E.~Kiritsis and C.~Panagopoulos,
  {\em ``Holographic quantum criticality and strange metal transport,''}
  New J.\ Phys.\  {\bf 14}, 043045 (2012)
  \hri{arXiv:1012.3464}{[cond-mat.str-el]}.

\bibitem{Kim:2011zd} 
  K.~-Y.~Kim and D.~-W.~Pang,
  {\em ``Holographic DC conductivities from the open string metric,''}
  JHEP {\bf 1109}, 051 (2011)
  \hri{arXiv:1108.3791}{[hep-th]}.
  
\bibitem{Fadafan:2012hr} 
  K.~B.~Fadafan,
  {\em ``Strange metals at finite 't Hooft coupling,''}
  \hri{arXiv:1208.1855}{[hep-th]}.
  
\bibitem{Horowitz:1996ay} 
  G.~T.~Horowitz, J.~M.~Maldacena and A.~Strominger,
  {\em ``Nonextremal black hole microstates and U duality,''}
  Phys.\ Lett.\ B {\bf 383}, 151 (1996)
  \hre{arXiv:hep-th}{9603109}.

\bibitem{Tseytlin:1997cs} 
  A.~A.~Tseytlin,
  {\em ``Composite BPS configurations of p-branes in ten-dimensions and eleven-dimensions,''}
  Class.\ Quant.\ Grav.\  {\bf 14}, 2085 (1997)
  \hre{arXiv:hep-th}{9702163}.

\bibitem{Maldacena:2001km} 
  J.~M.~Maldacena and H.~Ooguri,
  {\em ``Strings in AdS(3) and the SL(2,R) WZW model. Part 3. Correlation functions,''}
  Phys.\ Rev.\ D {\bf 65}, 106006 (2002)
  \hre{arXiv:hep-th}{0111180}.
  
\bibitem{Boonstra:1997dy} 
  H.~J.~Boonstra, B.~Peeters and K.~Skenderis,
  {\em ``Duality and asymptotic geometries,''}
  Phys.\ Lett.\ B {\bf 411}, 59 (1997)
  \hre{arXiv:hep-th}{9706192}.

\bibitem{Cvetic:1998jf} 
  M.~Cvetic, H.~Lu and C.~N.~Pope,
  {\em ``Space-times of boosted p-branes and CFT in infinite momentum frame,''}
  Nucl.\ Phys.\ B {\bf 545}, 309 (1999)
  \hre{arXiv:hep-th}{9810123}.


\end{thebibliography}
\end{document}